\documentclass[11pt,oneside]{article}

\usepackage[natbibapa]{apacite}
\usepackage{xcolor} 

\usepackage{soul}
\usepackage{amsmath, amssymb}
\usepackage{calrsfs}
\DeclareMathAlphabet{\pazocal}{OMS}{zplm}{m}{n}
\usepackage[shortlabels]{enumitem}
\usepackage{stmaryrd}			
\usepackage{bm}    				
\usepackage{amsthm}
\usepackage[nottoc]{tocbibind}
\usepackage{tocloft}
\DeclareMathOperator*{\esssup}{ess\,sup}

\newcommand{\R}{\mathbb{R}}
\newcommand{\N}{\mathbb{N}}
\newcommand{\U}{\mathbb{U}}

\def\E{\mathbb E}
\def\F{\mathcal F}
\def\P{\mathbb P}

\def\red{\textcolor{red}}

\usepackage{bbm}					

\numberwithin{equation}{section}

\newtheorem{thm}{Theorem}[section]
\newtheorem{lem}{Lemma}[section]
\newtheorem{prop}{Proposition}[section]
\newtheorem{cor}{Corollary}[section]

\theoremstyle{definition}
\newtheorem{defn}{Definition}[section]

\theoremstyle{remark}
\newtheorem{example}{Example}[section]

\theoremstyle{remark}
\newtheorem{rem}{Remark}[section]

\theoremstyle{theorem}
\newtheorem{assumption}{Assumption}[section]

\title{Arbitrage theory in a market of stochastic dimension}

\vspace{2mm}

\author{  
	\textsc{Erhan Bayraktar} 
	\thanks{
		Department of Mathematics, University of Michigan, Ann Arbor, MI, USA (E-mail: {\it erhan@umich.edu}). 
	}  
	\and
	\textsc{Donghan Kim} 
	\thanks{ 
		Department of Mathematics, University of Michigan, Ann Arbor, MI, USA (E-mail: {\it donghank@umich.edu}).
	}
	\and
	\textsc{Abhishek Tilva}
	\thanks{ 
		Department of Statistics, Columbia University, New York, NY, USA (E-mail: {\it akt2143@columbia.edu}).
	}
}

\begin{document}
	
	\maketitle
	
	\begin{abstract}
		\noindent
		This paper studies an equity market of stochastic dimension, where the number of assets fluctuates over time. In such a market, we develop the fundamental theorem of asset pricing, which provides the equivalence of the following statements: (i) there exists a supermartingale num\'eraire portfolio; (ii) each dissected market, which is of a fixed dimension between dimensional jumps, has locally finite growth; (iii) there is no arbitrage of the first kind; (iv) there exists a local martingale deflator; (v) the market is viable. We also present the optional decomposition theorem, which characterizes a given nonnegative process as the wealth process of some investment-consumption strategy. Furthermore, similar results still hold in an open market embedded in the entire market of stochastic dimension, where investors can only invest in a fixed number of large capitalization stocks. These results are developed in an equity market model where the price process is given by a piecewise continuous semimartingale of stochastic dimension. Without the continuity assumption on the price process, we present similar results but without explicit characterization of the num\'eraire portfolio.
	\end{abstract}
	
	\smallskip
	
	{\it MSC 2020 subject classifications:} Primary 60G48, 91G10; Secondary 60H05, 60H30. 
	
	\smallskip
	{\it Keywords and phrases:} piecewise semimartingale, fundamental theorem of asset pricing, num\'eraire portfolio, local martingale deflator, market viability, optional decomposition theorem, open market, superhedging.
	
	\smallskip
	
	\setcounter{tocdepth}{2}
	\tableofcontents
	
	\input amssym.def
	\input amssym
	
	\section{Introduction}
	\label{sec: introduction}
	
	Equity markets are usually modeled as \textit{closed}, that is, the total number of assets is assumed to be constant~(say $n \in \N$) over time such that entrance to or exit from the market of an individual asset is prohibited. The evolution of individual assets is then modeled by an $n$-dimensional stochastic process, for example, geometric Brownian motion, It\^o process, or more general semimartingale. However, this equity market model compels investors to put their money into the immutable space of investable assets, even though the number and composition of companies in actual markets fluctuate over time. In fact, the U.S. stock exchanges experienced dimensional changes quite frequently over the recent decades due to IPOs, bankruptcies, privatizations, etc; New York Stock Exchange~(NYSE) and American Stock Exchange~(AMEX) underwent dimensional changes every $1.68$ and $2.57$ trading days, respectively, on average, over the last 40 years from $1982$ to $2021$ (see Section 4.3 of \cite{BKT2} for more details).
	
	In order better to simulate investors' behavior in real equity markets than in the closed market model, the concept of \textit{open} markets is introduced in \cite{Fernholz:2018}, and widely studied in \cite{Karatzas:Kim2}. An open market consists of a fixed number~(say $m < n$) of higher capitalization stocks within a wider equity universe. Though the dimension of the open market is also assumed to be fixed, it is open in the sense that the constituents change over time according to their capitalizations; it is similar to high-capitalization indexes, in which a stock is replaced when its capitalization falls too low.
	
	However, both the closed and open market models fail to capture the intrinsic property of the real stock market, where new stocks enter as a result of IPOs or spin-offs, and old stocks exit following bankruptcies, privatization, or mergers and acquisitions. To this end, this paper and our companion paper, \cite{BKT2}, study a stock market model of changing dimensions, using the concept of piecewise semimartingales introduced in \cite{Strong2}. Whereas the companion paper focuses on how dimensional changes of the market impact performances of self-financing stock portfolios with empirical results, this paper develops more theoretical aspects of arbitrage theory in the setting of equity market with changing number of assets (thus we refer to \cite{BKT2} for those who are more interested in practical aspects).
	
	\cite{Strong2} defines the piecewise semimartingale of stochastic dimension and its stochastic integration, by patching several pieces~(or \textit{dissections}) of finite-dimensional semimartingales of different dimensions. Two fundamental theorems of asset pricing~(FTAPs) are then developed in the market of stochastic dimension; the equivalence of no free lunch with vanishing risk~(NFLVR) to the existence of an equivalent sigma-martingale measure~(E$\sigma$MM) for the price process~(\cite{DS_1998}), and the equivalence of no arbitrage of the first kind~($\text{NA}_1$) to the existence of a local martingale deflator~(ELMD) for the set of nonnegative wealth processes~(\cite{Kardaras_2012_viability}).
	
	Our paper uses the same notion of piecewise semimartingale to model an equity market, but further extends the latter FTAP~($\text{NA}_1 \iff \text{ELMD}$) by establishing most of the results presented in the recent monograph by \cite{KK2} under two different settings.
	
	First, we assume that each component of the price process is continuous and strictly positive between each dimensional jump. Under this assumption, we present the so-called fundamental theorem~(see Theorem~\ref{thm : fundamental theorem}), which is a complete version of the aforementioned FTAP. The central part of this result is to show the existence of the num\'eraire portfolio~(roughly speaking, a portfolio which cannot be outperformed) under $\text{NA}_1$ or ELMD. If it exists, the num\'eraire portfolio can be characterized in terms of local drift and covariation rates of each \textit{dissected market}, that is, each `piece' of the market with a fixed dimension between the dimensional changes. From this characterization, which we call the \textit{structural condition} of the dissected market, we derive that the num\'eraire portfolio is optimal, in the sense that it attains maximal growth rate and has relative log-optimal property. In particular, the maximal growth rate of the num\'eraire portfolio implies that each dissected market must have finite local growth if such a portfolio exists. This growth condition on every dissected market is then connected with the market viability, another notion of no-arbitrage; funding a nontrivial cumulative capital withdrawal stream should not be possible from an arbitrarily small amount of initial wealth.
	
	In addition to the fundamental theorem, we also show the optional decomposition theorem~(ODT) under the same assumption on the price process. It provides a local martingale~(supermartingale) formulation for a nonnegative process, which can be expressed as a stochastic integral~(minus a nondecreasing component, respectively) of some investment strategy with respect to the price process of stochastic dimension, under the equivalent conditions of the fundamental theorem.
	
	Moreover, we study an open market embedded in the entire market of stochastic dimension. When investors are only allowed to invest in a fixed number of large capitalization stocks, while the total number of assets in the entire market fluctuates, both the fundamental theorem and the ODT are presented. In order to handle the embedded open market, we combine the idea of \textit{dissecting} the market with the idea of \textit{censoring} the return process from \cite{Karatzas:Kim2}.
	
	The assumption of continuity in the price process is useful to have a complete and explicit treatment of arbitrage theory, including an open market within a larger market of stochastic dimension. It allows us to carry out the arbitrage theory by means of portfolios and leads to explicit characterizations of the num\'eraire portfolio, the set of local martingale deflators, and the investment strategy which appears in the ODT, in a pedagogically illustrative way.
	
	Subsequently, in the second main part of the paper, we consider a more general market of stochastic dimension, relaxing the continuity condition on the price process such that each component is only assumed to be a right-continuous with left limits~(RCLL) semimartingale between the dimensional changes. When jumps are present in the dynamics of asset prices, two notions of num\'eraire~(supermartingale num\'eraire and local martingale num\'eraire) portfolios may not coincide~(see \cite{kabanov2016no} and \cite{Takaoka}), hence one is not able to exploit the explicit characterization of the num\'eraire portfolio anymore in developing the arbitrage theory. Thus, the results obtained in the second part, while being more general, are less explicit than the ones in the first part of the paper.
	
	Moreover, \cite{KK:OD} noted that the same proof technique does not extend in a nontrivial way to prove the ODT if the price process contains jumps. Hence, we adopt a different method to prove the ODT first, relying on the classical version of ODT from \cite{Stricker:Yan}. We then apply the ODT for establishing the fundamental theorem; the essential part is again to prove the existence of a wealth process having the supermartingale num\'eraire property, under the other equivalent conditions~(e.g. the existence of a local martingale deflator). Finally, as further applications of the ODT, we present in full generality the superhedging duality and the second fundamental theorem of asset pricing, which provides an equivalent condition for the completeness of the market.
	
	\smallskip
	
	\textit{Preview: } This paper is organized as follows. Section~\ref{sec : piecewise semimartingales} reviews the concept of piecewise semimartingale and its stochastic integration from \cite{Strong2}. Section~\ref{sec : continuous price} studies a market of stochastic dimension, where the asset price is assumed to be continuous and strictly positive piecewise semimartingale; once we introduce relevant definitions and preliminary results in succession, the fundamental theorem is presented. We then provide the optional decomposition theorem and study the open market embedded in the entire market. In Section~\ref{sec : RCLL price}, under a more general equity market model with a right-continuous price process, we develop similar results of the ODT and the fundamental theorem, but with different arguments. We also present the superhedging duality and the second fundamental theorem of asset pricing under the general model.
	
	\bigskip
	
	\section{Piecewise semimartingales}	\label{sec : piecewise semimartingales}
	This section reviews the notion of piecewise semimartingales of stochastic dimension, introduced by \cite{Strong2}. We provide the definitions, notations, and summary of the results from Section~2 of \cite{Strong2}, which are useful to develop a full-fledged arbitrage theory for a market with stochastic dimension in the later sections.
	
	We consider a state space $\U:= \cup_{n=1}^{\infty} \R^n$, equipped with the topology generated by the union of the standard topologies of $\R^n$. Besides the $n$-dimensional zeros $0^{(n)} \in \R^n$ for each $n \in \N$, we define an additive identity element $\odot$, a topologically isolated point in $\hat{\U} := \U \cup \{\odot\}$ satisfying $\odot + x = x + \odot = x$ and $\odot x = x \odot = \odot$ for each $x \in \hat{\U}$. We define the modified indicator
	\begin{equation}	\label{def : modified indicator}
	\hat{\mathbbm{1}}_A(t, \omega) :=
	\begin{cases}
	1 \in \R \quad &\text{for } (t, \omega) \in A \subset [0, \infty) \times \Omega,
	\\
	\odot &\text{otherwise},
	\end{cases}
	\end{equation}
	which will be useful for dissecting stochastic processes. In order for expressions involving $\hat{\mathbbm{1}}$ to have the correct dimension in $\U$, we shall add a zero vector $0^{(n)}$ of an appropriate dimension $n$. Moreover, $1^{(n)}$ denotes the $n$-dimensional vector of ones, and $\mathbbm{1}_A$ is the usual indicator function for set $A$. 
	
	We use the notations $\R_+ := [0, \infty)$ and $B^{\top}$ the transpose of a matrix $B$. For any stochastic processes $Y$ and $Z$, the identity $Y=Z$ means that $Y$ and $Z$ are indistinguishable, $Y^{\alpha}(\cdot) := Y(\alpha \wedge \cdot)$ denotes the process stopped at a random time $\alpha$. All relationships among random variables are understood to hold almost surely. We shall denote the set $[n] := \{1, \cdots, n\}$ for every $n \in \N$.
	
	On a filtered probability space $(\Omega, \mathcal{F}, (\mathcal{F}_t)_{t \ge 0}, \mathbb{P})$ satisfying the usual conditions, let $X$ be a $\U$-valued progressive process having paths with left and right limits at all times. We denote $N := \dim X$ the dimension process of $X$ also having paths with left and right limits at all times. The following definition characterizes time instants of dimensional jumps for a given $\U$-valued process $X$, as a sequence of stopping times.
	
	\smallskip
	
	\begin{defn} [Reset sequence]
		A sequence of stopping times $(\tau_k)_{k \ge 0}$ is called a \textit{reset sequence} for a progressive $\U$-valued process $X$, if the following hold for $\mathbb{P}$-a.e. $\omega$:
		\begin{enumerate} [(i)]
			\item $\tau_0(\omega) = 0$, $\tau_{k-1}(\omega) \le \tau_k(\omega)$ for all $k \in \N$, and $\lim_{k \rightarrow \infty} \tau_k(\omega) = \infty$;
			\item $N(t, \omega) = N(\tau_{k-1}+, \omega)$ for every $t \in (\tau_{k-1}(\omega), \tau_k(\omega)]$ and $k \in \N$;
			\item $t \mapsto X(t, \omega)$ is right-continuous on $(\tau_{k-1}(\omega), \tau_k(
			\omega))$ for every $k \in \N$.
		\end{enumerate}
	\end{defn}
	
	When $X$ has a reset sequence $(\tau_k)_{k \ge 0}$, we shall always consider the minimal one $(\hat{\tau}_k)_{k \ge 0}$ in the sense of the fewest resets by a given time:
	\begin{equation*}
	\hat{\tau}_0 := 0, \qquad \hat{\tau}_k := \inf\{t > \hat{\tau}_{k-1} \, : \, X(t+) \neq X(t) \}, \quad k \in \N,
	\end{equation*}
	and assume that the initial dimension is deterministic, i.e., $\dim(X(0)) = N_0 \in \N$.
	
	We also emphasize here that there is no restriction on the size of each dimensional change, in the sense that the quantity $\vert N(\tau_k+) - N(\tau_{k-1}+)\vert$ can be bigger than or equal to one for each $k \in \N$. This allows us to include several equity market models with a changing number of assets in our setting. For example, the diverse market model of \cite{Karatzas:Sarantsev} considers a particular form of splits and mergers between companies; the largest company is split into two companies~(modeling a regulatory breakup) and any two of the existent companies merge into one at random times. In their model, the size of each dimensional jump is always one.
	
	In what follows, we fix such $\U$-valued process $X$ with the reset sequence $(\tau_k)_{k \ge 0}$, and define the \textit{dissections} of $\Omega$ and $X$ for every pair $(k, n) \in \N^2$
	\begin{align}
	\Omega^{k, n} &:= \{ \tau_{k-1} < \infty, ~ N(\tau_{k-1}+) = n \} \subset \Omega,	\label{def : Omega dissect}
	\\
	X^{k, n} &:= \big(X^{\tau_k} - X(\tau_{k-1}+)\big) \hat{\mathbbm{1}}_{\rrbracket \tau_{k-1}, \infty \llbracket \cap (\R_+ \times \Omega^{k, n})} + 0^{(n)}.		\label{def : X dissect}
	\end{align}
	
	We now introduce piecewise semimartingales as integrators of stochastic integrals, with an appropriate class of integrands. The $(k, n)$-dissection $X^{k, n}$ of $X$, defined in \eqref{def : X dissect}, is appropriate when $X$ plays the role of integrator. For integrands, a different definition of dissection is necessary.
	
	\smallskip
	
	\begin{defn} [Piecewise semimartingale]	\label{Def : piecewise semimartingale}
		A \textit{piecewise semimartingale} $X$ is a $\U$-valued progressive process having paths with left and right limits at all times, and possessing a reset sequence $(\tau_k)_{k \ge 0}$ such that $X^{k, n}$ is an $\R^n$-valued semimartingale for every $(k, n) \in \N^2$.
		
		A piecewise semimartingale $X$ is called \textit{piecewise continuous~(RCLL) semimartingale}, if each dissection $X^{k, n}$ is an $n$-dimensional continuous~(RCLL, respectively) semimartingale for every $(k, n) \in \N^2$.
	\end{defn}

    \smallskip
    
    We note that even though the piecewise semimartingale $X$ is continuous, an existing component $X_i$ can experience a right-discontinuous jump \textit{at} a reset time $\tau_k$, i.e., $X_i(\tau_k) \neq X_i(\tau_k+)$, but every component of $X$ should be continuous \textit{between} two consecutive reset times. We also note that any right-discontinuities of $X$ do not affect the value of the stochastic integral, defined in \eqref{def : stochastic integral} below.
    
	\smallskip
	
	\begin{defn} [Stochastic integral]
		For a piecewise semimartingale $X$ and its reset sequence $(\tau_k)_{k \ge 0}$, let $H$ be a $\U$-valued predictable process satisfying $\dim H = N = \dim X$. We dissect $H$ in the following manner
		\begin{equation}	\label{def : H dissect}
		H^{(k, n)} := H\hat{\mathbbm{1}}_{\rrbracket \tau_{k-1}, \tau_k \rrbracket \cap (\R_+ \times \Omega^{k, n})} + 0^{(n)}, \qquad \forall (k, n) \in \N^2,
		\end{equation}
		and define
		\begin{align}
		&\mathcal{L}(X) := \{ H \text{ predictable} \, : \, \dim H = N \text{ and } H^{(k, n)} \text{ is } X^{k, n} \text{-integrable, } \forall (k, n) \in \N^2 \},	\label{def : space of integrand}
		\\
		&\mathcal{L}_0(X) := \{ H \in \mathcal{L}(X) \, : \, H_0 = 0^{(N_0)} \}. 	\nonumber
		\end{align}
		For $H \in \mathcal{L}(X)$, the stochastic integral $H \cdot X$ is defined as 
		\begin{equation}	\label{def : stochastic integral}
		H \cdot X := H_0^{\top}X_0 + \sum_{k=1}^{\infty} \sum_{n=1}^{\infty} (H^{(k, n)} \cdot X^{k, n}).
		\end{equation}
	\end{defn}
	
	Note that each dissection $H^{(k, n)}$ of \eqref{def : H dissect} is predictable, since the process $H$ and the $(k, n$)-dissection set $\rrbracket \tau_{k-1}, \tau_k \rrbracket \cap (\R_+ \times \Omega^{k, n})$ are predictable. We also note that \cite{Strong2} uses the same notation for two dissections $X^{k, n}$ in \eqref{def : X dissect} and $H^{(k, n)}$ in \eqref{def : H dissect}, whereas we shall differentiate the two notations throughout this paper. The stochastic integral of \eqref{def : stochastic integral} generalizes the usual $\R^n$-valued semimartingale stochastic integration, since any increasing sequence $\tau_k \uparrow \infty$ of stopping times is a reset sequence for any semimartingales of fixed dimension. We use interchangeably the notations
	\begin{equation*}
	H \cdot X = \int H dX, \qquad H^{(k, n)} \cdot X^{k, n} = \int H^{(k, n)} \, dX^{k, n} = \int \sum_{i=1}^n H^{(k, n)}_i \, dX^{k, n}_i,
	\end{equation*}
	for the stochastic integrals of \eqref{def : stochastic integral}. Although we shall not make direct use of the following result, we summarize some of the properties of the stochastic integral from \cite{Strong2}.
	
	\smallskip
	
	\begin{prop}
		Let $X$ be a piecewise semimartingale with reset sequence $(\tau_k)_{k \ge 0}$ and $H, G \in \mathcal{L}(X)$.
		\begin{enumerate} [(i)]
			\item If $X$ has another reset sequence $(\tilde{\tau}_k)_{k \ge 0}$, then $\tilde{X}^{k, n}$ defined via \eqref{def : X dissect} is an $\R^n$-valued semimartingale for every $(k, n) \in \N^2$. The class $\mathcal{L}(X)$ and the process $H \cdot X$ do not depend on the choice of reset sequence of $X$.
			\item The stochastic integral $H \cdot X$ is an $\R$-valued semimartingale.
			\item $\mathcal{L}(X)$ is a vector space; $H \cdot X + G \cdot X = (H+G) \cdot X$.
			\item For any stopping time $\alpha$, the stopped process $X^{\alpha}$ is a piecewise semimartingale and the identity $(X^{\alpha})^{k, n} = (X^{k, n})^{\alpha}$ holds for every $(k, n) \in \N^2$.
		\end{enumerate}
	\end{prop}
	
	\bigskip
	
	\section{Market with continuous, positive price process}	\label{sec : continuous price}
	
	In this section, we develop a full-fledged arbitrage theory in a market of stochastic dimension, when a stock price is modeled as a $\U$-valued piecewise \textit{continuous} semimartingale. This means that between two consecutive dimensional jumps $\tau_{k-1}$ and $\tau_k$, each $n$-dimensional piece~(or dissection) of the price process remains \textit{continuous} for every $(k, n) \in \N^2$. Under this setting, we ultimately present the fundamental theorem~(Section~\ref{subsec : fundamental theorem}) and the optional decomposition theorem~(Section~\ref{subsec : ODT}).
	
	\medskip
	
	\subsection{Price process, return process, and portfolio}	\label{subsec : basic processes}
	
	Using the concept of piecewise semimartingale from Section~\ref{sec : piecewise semimartingales}, we describe in this subsection an equity market with a stochastic number of investable assets.
	
	\smallskip
	
	\begin{defn} [Price process]	\label{Def : price process}
		A $\U$-valued piecewise continuous semimartingale $S$ is called a \textit{price process}~(of stochastic dimension), if every component of $S$ is strictly positive on each dissection set $\rrbracket \tau_{k-1}, \tau_k \rrbracket \cap (\R_+ \times \Omega^{k, n})$ for every $(k, n) \in \N^2$. 
	\end{defn}
	
	The dimension process $N = \dim(S)$ of $S$ represents the number of companies present in the market, and the $n$ components of $S$ on the set $\rrbracket \tau_{k-1}, \tau_k \rrbracket \cap (\R_+ \times \Omega^{k, n})$ represent the discounted stock prices~(or the capitalizations) of the $n$ existent companies, for every $(k, n) \in \N^2$. In order to simplify the model, we assume that every stock has a single outstanding share, so that the price of a stock is equal to its capitalization. By definition, every $\cup_{n=1}^{\infty}(0, \infty)^n$-valued piecewise continuous semimartingale can be a price process. We note that the strict positivity of the price process is essential to define the return process in the following.
	
	\smallskip
	
	\begin{defn} [Return process]	\label{Def : return process}
		For a given price process $S$, we call $R$ a \textit{return process}, if $R$ is a $\U$-valued piecewise continuous semimartingale with $\dim(R) = N = \dim(S)$, and its dissection is given by
		\begin{equation}	\label{def : return process}
		R^{k, n}_i(t) = \int_0^{t} \bigg( \frac{1}{S_i(u)} \hat{\mathbbm{1}}_{\rrbracket \tau_{k-1}, \tau_k \rrbracket \cap (\R_+ \times \Omega^{k, n})} + 0^{(1)} \bigg) \, dS^{k, n}_i(u), \quad t \ge 0,
		\end{equation}
		for every $i \in [n]$ and $(k, n) \in \N^2$.
	\end{defn}
	
	Since $S^{k, n}$ is constant out of the interval $\rrbracket \tau_{k-1}, \tau_k \rrbracket$, and the increments $dS^{k, n}_i$ and $dS_i$ coincide on the dissection set ${\rrbracket \tau_{k-1}, \tau_k \rrbracket \cap (\R_+ \times \Omega^{k, n})}$ from the definition \eqref{def : X dissect}, the identity of \eqref{def : return process} can be loosely rewritten as
	\begin{equation*}
	R^{k, n}_i(t) = \int_0^{t} \frac{1}{S_i(u)} \, dS^{k, n}_i(u) = \int_0^{t} \bigg( \frac{1}{S_i(u)} \hat{\mathbbm{1}}_{\rrbracket \tau_{k-1}, \tau_k \rrbracket \cap (\R_+ \times \Omega^{k, n})} + 0^{(1)} \bigg) \, dS_i(u),
	\end{equation*}
	which is reminiscent of the fact that the return of a stock is defined as a stochastic logarithm of its price process. 
	
	The quantity $R^{k, n}_i(t)$ can be interpreted as the cumulative return of the $i$-th stock until time $t$ when there are $n$ stocks extant between $(k-1)$-th and $k$-th dimensional changes of the market. We note that each value of $R^{k, n}$ is accumulated only on the interval $\rrbracket \tau_{k-1}, \tau_k \rrbracket$ by definition, thus $R^{k, n}_i \equiv 0$ on $\llbracket 0, \tau_{k-1} \rrbracket$ and $R^{k, n}_i \equiv R^{k, n}_i(\tau_k)$ on $\llbracket \tau_k, \infty \rrbracket$. Moreover, $R^{k, n}_i$ and $R^{k+1, m}_i$ may indicate cumulative returns of different companies for $i = 1, \cdots, \min(n, m)$, because the indexing of the stocks in the market can be inconsistent between each dimensional change; for example, if the $i$-th company exits the market, the $(i+1)$-st company inherits the index $i$ from the next epoch, and so on.
 
    We consider the canonical decomposition of each $n$-dimensional semimartingale 
	\begin{equation}	\label{R decomposition}
	R^{k, n}_i := A^{k, n}_i + M^{k, n}_i
	\end{equation}
	such that $A^{k, n}_i$ is of finite variation, $M^{k, n}_i$ is a local martingale for every $i \in [n]$ and $(k, n) \in \N^2$.
	Recalling the notation \eqref{def : space of integrand}, we define portfolios in this market.
	
	\smallskip
	
	\begin{defn} [Portfolio]
		For a given return process $R$, we call $\pi$ a \textit{portfolio}, if $\pi \in \mathcal{L}(R)$. The (cumulative) return process of $\pi$ is defined by $R_{\pi} := \pi \cdot R$, and the wealth process of $\pi$ is given by
		\begin{equation}	\label{def : X pi}
		X_{\pi} := \mathcal{E}(R_{\pi}),
		\end{equation}
		where $\mathcal{E}(Z) := \exp\big(Z - \frac{1}{2} [ Z, Z ] \big)$ denotes the stochastic exponential of a scalar continuous semimartingale $Z$ with $Z(0) = 0$. A portfolio $\pi$ is called \textit{null portfolio} if its return process is a zero process, i.e., $R_{\pi} \equiv 0$.
	\end{defn}
	
	In Definition~\ref{Def : return process}, there was no condition on the initial vector $R_0$ for the return process, but from now on we shall assume $R_0 = 0^{(N_0)}$, in order to have the identity $X_{\pi}(0) = 1$, thus the wealth of any portfolio $\pi$ is always normalized at the initial time. For each $(k, n) \in \N^2$, the component $\pi^{(k, n)}_i$ of dissection of $\pi$ represents the proportion of wealth invested in the $i$-th company among $n$ companies, and
	\begin{equation}	\label{def : pi0}
	\pi^{(k, n)}_0 = 1 - \sum_{j=1}^n \pi^{(k, n)}_j,
	\end{equation}
	is the proportion of capital invested in the money market. Here, we assume that the money market earns no interest. Moreover, when the dimension of the market changes from $n$~(in the $k$-th epoch) to $m$~(in the $(k+1)$-st epoch), re-distribution of wealth from $\pi^{(k, n)}$ to $\pi^{(k+1, m)}$ is financed by the money market such that the corresponding portion of wealth changes from $\pi^{(k, n)}_0$ to $\pi^{(k+1, m)}_0$.
	
	Recalling \eqref{def : stochastic integral}, the return process of $\pi$ can be dissected as
	\begin{equation}	\label{def : R pi}
	R_{\pi} = \sum_{k=1}^{\infty} \sum_{n=1}^{\infty} \pi^{(k, n)} \cdot R^{k, n}
	= \sum_{k=1}^{\infty} \sum_{n=1}^{\infty} R^{k, n}_{\pi}, \qquad \text{where} \quad 
	R^{k, n}_{\pi} := \pi^{(k, n)} \cdot R^{k, n}.
	\end{equation}
    We note that the double sum in \eqref{def : R pi} is a finite sum for each time point and the same is true for the sums of this type that follow.
    
	Given two portfolios $\pi, \rho \in \mathcal{L}(R)$, we denote $C_{\pi \rho}$ the covariation process between the cumulative returns $R_{\pi}$, $R_{\rho}$
	\begin{equation}	\label{def : C pi rho}
	C_{\pi \rho} \equiv [ R_{\pi}, R_{\rho} ] := \sum_{k=1}^{\infty} \sum_{n=1}^{\infty} C^{k, n}_{\pi \rho},
	\end{equation} 
	where
	\begin{equation}	\label{def : C pi rho dissection}
	C^{k, n}_{\pi \rho} := [ R^{k, n}_{\pi}, R^{k, n}_{\rho} ]
	= \int_0^{\cdot} \sum_{i=1}^n \sum_{j=1}^n \pi^{(k, n)}_i(s) \rho^{(k, n)}_j(s) \, d [ R^{k, n}_i, R^{k, n}_j ](s).
	\end{equation}
	Moreover, we write for every $i \in [n]$ and $(k, n) \in \N^2$
	\begin{equation}	\label{def : C i rho}
	C^{k, n}_{i \rho} := [ R^{k, n}_i, R^{k, n}_{\rho} ].
	\end{equation}
	This notation is justified as $C^{k, n}_{i \rho} \equiv C^{k, n}_{\nu \rho}$ when a portfolio $\nu$ invests all its wealth in the $i$-th stock, if at least $i$ stocks are present in the market; i.e., $\nu$ is defined via its dissection for a fixed $i \in \N$
	\begin{align}
	\nu &:= 0^{(N_0)} + \sum_{k=1}^{\infty} \sum_{n=1}^{\infty} \hat{\mathbbm{1}}_{\rrbracket \tau_{k-1}, \tau_k \rrbracket \cap (\R_+ \times \Omega^{k, n})} \nu^{(k, n)}, \qquad \text{where}	\label{def : unit portfolio}
	\\
	\nu^{(k, n)} &:= e^i \hat{\mathbbm{1}}_{\rrbracket \tau_{k-1}, \tau_k \rrbracket \cap (\R_+ \times \Omega^{k, n}) \cap \{i \le n\}} + 0^{(n)}		\nonumber
	\end{align}
	where $e^i$ denotes the $n$-dimensional unit vector with the $i$-th entry equal to one. It is straightforward to check $R^{k, n}_{\nu} = R^{k, n}_i$ and $C^{k, n}_{\nu \rho} = [R^{k, n}_i, R^{k, n}_{\rho}]$ whenever $i \le n$.
	
	\medskip
	
	\subsection{Num\'eraire portfolios}	\label{subsec : numeraire portfolio}
	
	For any given portfolios $\pi, \rho \in \mathcal{L}(R)$, we compare the relative performance of $\pi$ with respect to the other `baseline' portfolio $\rho$ by computing the ratio
	\begin{equation}	\label{def : relative wealth}
	X^{\rho}_{\pi} := \frac{X_{\pi}}{X_{\rho}}.
	\end{equation}
	
	The following result gives a representation of $X^{\rho}_{\pi}$ as the stochastic exponential of $R^{\rho}_{\pi}$, which we call \textit{relative (cumulative) return of $\pi$ with respect to $\rho$}.
	
	\smallskip
	
	\begin{lem}	\label{lem : X rho pi representation}
		For every $(k, n) \in \N^2$, recalling the notations \eqref{def : R pi}, \eqref{def : C pi rho dissection} and \eqref{def : C i rho}, we define
		\begin{align}
		R^{\rho, k, n}_0 &:= C^{k, n}_{\rho \rho} - R^{k, n}_{\rho},	\label{def : R0}
		\\
		R^{\rho, k, n}_i &:= R^{\rho, k, n}_0 + (R^{k, n}_i - C^{k, n}_{i \rho}), \qquad i \in [n],			\label{def : Ri}
		\end{align}
		and
		\begin{equation}	\label{def : R rho pi}
		R^{\rho}_{\pi} := \sum_{k=1}^{\infty} \sum_{n=1}^{\infty} \int_0^{\cdot} \sum_{i=0}^n \pi^{(k, n)}_i(s) \, dR^{\rho, k, n}_i(s).
		\end{equation}
		Then, we have the representation of $X^{\rho}_{\pi}$ in \eqref{def : relative wealth}
		\begin{equation}	\label{eq : X rho pi}
		X^{\rho}_{\pi} = \mathcal{E}(R^{\rho}_{\pi}).
		\end{equation}
	\end{lem}
	
	\begin{proof}
		Using the definitions \eqref{def : R0}-\eqref{eq : X rho pi}, \eqref{def : C i rho}, and \eqref{def : C pi rho dissection}, the integral on the right-hand side of \eqref{def : R rho pi} can be written as
		\begin{align}
		\int_0^{\cdot} \sum_{i=0}^n \pi^{(k, n)}_i(s) \, dR^{\rho, k, n}_i(s)
		&= R^{\rho, k, n}_0(\cdot) + \int_0^{\cdot} \sum_{i=1}^n \pi^{(k, n)}_i(s) \, dR^{k, n}_i(s) - \int_0^{\cdot} \sum_{i=1}^n \pi^{(k, n)}_i(s) \, dC^{k, n}_{i \rho}(s) \nonumber
		\\
		& = R^{\rho, k, n}_0 + R^{k, n}_{\pi}-C^{k, n}_{\pi \rho}
		= C^{k, n}_{\rho\rho} - R^{k, n}_{\rho} + R^{k, n}_{\pi} - C^{k, n}_{\pi \rho}           \nonumber
		\\
		&= C^{k, n}_{(\rho-\pi)\rho} - R^{k, n}_{\rho-\pi}.    \label{eq : R rho pi expression}
		\end{align}
		Thus, we derive
		\begin{align*}
		\log \big( \mathcal{E}(R^{\rho}_{\pi}) \big) =R^{\rho}_{\pi} - \frac{1}{2} [R^{\rho}_{\pi}, R^{\rho}_{\pi}]
		&= \sum_{k=1}^{\infty}\sum_{n=1}^{\infty} \Big( C^{k, n}_{(\rho-\pi)\rho} - R^{k, n}_{\rho-\pi} - \frac{1}{2}[R^{k, n}_{\rho-\pi}, R^{k, n}_{\rho-\pi}] \Big)
		\\
		&=\sum_{k=1}^{\infty}\sum_{n=1}^{\infty} \Big( R^{k, n}_{\pi-\rho} - \frac{1}{2}C^{k, n}_{\pi\pi} + \frac{1}{2}C^{k, n}_{\rho\rho} \Big) = R_{\pi-\rho} - \frac{1}{2}C_{\pi\pi} + \frac{1}{2}C_{\rho\rho}.
		\end{align*}
		Here, the second identity uses the fact $[R^{k, n}_{\rho}, R^{\ell, m}_{\rho}] = [R^{k, n}_{\pi}, R^{\ell, m}_{\pi}] = [R^{k, n}_{\rho}, R^{\ell, m}_{\pi}] \equiv 0$ whenever $(k, n) \neq (\ell, m)$, from the property of dissection defined in \eqref{def : R pi}.
		
		On the other hand, we have
		\begin{align*}
		\log (X^{\rho}_{\pi}) = \log \big( \mathcal{E}(R_{\pi}) \big) - \log \big( \mathcal{E}(R_{\rho}) \big) = \Big( R_{\pi} - \frac{1}{2}C_{\pi\pi} \Big) - \Big( R_{\rho} - \frac{1}{2}C_{\rho\rho} \Big),
		\end{align*}
		thus the result follows.
	\end{proof}
	
	\smallskip
	
	\begin{defn} [Num\'eraire portfolio]	\label{Def : numeraire portfolio}
		A portfolio $\rho \in \mathcal{L}(R)$ is called \textit{supermartingale~(local martingale) num\'eraire portfolio}, if the relative wealth process $X^{\rho}_{\pi}$ is a supermartingale~(local martingale, respectively) for every portfolio $\pi \in \mathcal{L}(R)$ in the market.
	\end{defn}
	
	If a supermartingale~(or local martingale) num\'eraire portfolio $\rho$ exists, then it is unique modulo null portfolios~(Lemma~3.3 of \cite{Karatzas:Kim2}). In this case, the wealth process $X_{\rho}$ is called a supermartingale~(or local martingale) num\'eraire. The next result further shows that two num\'eraire portfolios are actually the same, whenever they exist, and gives an equivalent characterization.
	
	\smallskip
	
	\begin{prop}	\label{prop : numeraire portfolio}
		For a portfolio $\rho \in \mathcal{L}(R)$, the following statements are equivalent:
		\begin{enumerate} [(i)]
			\item $\rho$ is a supermartingale num\'eraire portfolio.
			\item $\rho$ is a local martingale num\'eraire portfolio.
			\item $A^{k, n}_i = C^{k, n}_{i \rho}$ holds for every $i \in [n]$ and $(k, n) \in \N^2$. 
		\end{enumerate}
	\end{prop}
	
	\begin{proof}
		We first show that $(iii)$ implies $(ii)$. Thanks to the decomposition \eqref{R decomposition}, $R^{k, n}_i - C^{k, n}_{i \rho} = M^{k, n}_i$ is a local martingale for every $i \in [n]$ and $(k, n) \in \N^2$. Moreover, we deduce from \eqref{def : R0} that
		\begin{align}
		R^{\rho, k, n}_0 
		&= C^{k, n}_{\rho\rho} - R^{k, n}_{\rho}
		= [R^{k, n}_{\rho}, R^{k, n}_{\rho}] - R^{k, n}_{\rho}
		= \bigg[\int_0^{\cdot} \sum_{i=1}^n \rho^{(k, n)}_i(s) \, dR^{k, n}_i(s), R^{k, n}_{\rho} \bigg] - R^{k, n}_{\rho}	\label{eq : R rho, k, n, 0 expression}
		\\
		&= \int_0^{\cdot} \sum_{i=1}^n \rho^{(k, n)}_i(s) \, dC^{k, n}_{i \rho}(s) - \int_0^{\cdot} \sum_{i=1}^n \rho^{(k, n)}_i(s) \, dR^{k, n}_i(s)
		= - \int_0^{\cdot} \sum_{i=1}^n \rho^{(k, n)}_i(s) \, dM^{k, n}_i(s).	\nonumber
		\end{align}
		Thus, $R^{\rho, k, n}_0$ and every $R^{\rho, k, n}_i$ of \eqref{def : Ri}, are local martingales. Every integral of \eqref{def : R rho pi} is then a local martingale, and we apply Lemma~2.12 of \cite{Strong2}~(as in the proof of Theorem~2.15, Corollary~2.16 of \cite{Strong2}) to conclude that $R^{\rho}_{\pi}$, and also $X^{\rho}_{\pi}$, are local martingales.
		
		The implication $(ii) \Longrightarrow (i)$ is trivial, since every nonnegative local martingale is a supermartingale by Fatou's lemma.
		
		We now assume $(i)$. Let us fix $(i, j) \in \N^2$ satisfying $i \le j$, and construct a portfolio $\nu_{(i, j)}$ via dissection
		\begin{align}
		\nu_{(i, j)} &:= 0^{(N_0)} + \sum_{k=1}^{\infty} \sum_{n=1}^{\infty} \hat{\mathbbm{1}}_{\rrbracket \tau_{k-1}, \tau_k \rrbracket \cap (\R_+ \times \Omega^{k, n})} \nu^{(k, n)}_{(i, j)}, \qquad \text{where}	\nonumber
		\\
		\nu^{(k, n)}_{(i, j)} &:= e^i \hat{\mathbbm{1}}_{\rrbracket \tau_{k-1}, \tau_k \rrbracket \cap (\R_+ \times \Omega^{k, n}) \cap \{j \le n\}} + 0^{(n)}.	\label{def : nu i j}
		\end{align}
		This portfolio $\nu_{(i, j)}$ is a generalization of the portfolio $\nu$ depending on a single index $i \in \N$, defined in \eqref{def : unit portfolio}. It allocates all wealth to the $i$-th stock, if there exist more than or equal to $j~(\ge i)$ stocks in the market; otherwise, it invests all wealth into the money market.
		
		From the assumption $(i)$, $X^{\rho}_{\rho+\nu_{(i, j)}}$ and $X^{\rho}_{\rho-\nu_{(i, j)}}$ are supermartingales and their stochastic logarithms $R^{\rho}_{\rho+\nu_{(i, j)}}$ and $R^{\rho}_{\rho-\nu_{(i, j)}}$ in Lemma~\ref{lem : X rho pi representation} are then local supermartingales. We compute
		\begin{equation*}
		R^{\rho}_{\rho+\nu_{(i, j)}} = \sum_{k=1}^{\infty} \sum_{n=1}^{\infty} \int_0^{\cdot} \sum_{\ell=0}^n \big( \rho^{(k, n)}_{\ell}(s) + \nu^{(k, n)}_{(i, j), \ell}(s) \big) \, dR^{\rho, k, n}_{\ell}(s)
		\end{equation*}
		via \eqref{def : R rho pi}.	Using the definitions~\eqref{def : R0}-\eqref{def : R rho pi}, it is easy to show that
		\begin{equation*}
		\int_0^{\cdot} \sum_{\ell=0}^n \rho^{(k, n)}_{\ell}(s) \, dR^{\rho, k, n}_{\ell}(s) = 0
		\end{equation*}
		holds, thus we have $R^{\rho}_{\rho-\nu_{(i, j)}} = - R^{\rho}_{\rho+\nu_{(i, j)}}$, and both quantities are local martingales. Moreover, we obtain
		\begin{align*}
		R^{\rho}_{\rho+\nu_{(i, j)}} 
		= \sum_{k=1}^{\infty} \sum_{n=1}^{\infty} \int_0^{\cdot} \sum_{\ell=0}^n \nu^{(k, n)}_{(i, j), \ell}(s) \, dR^{\rho, k, n}_{\ell}(s)
		&= \sum_{k=1}^{\infty} \sum_{n=1}^{j-1} R^{\rho, k, n}_0 +  
		\sum_{k=1}^{\infty} \sum_{n=j}^{\infty} R^{\rho, k, n}_i
		\\
		&=\sum_{k=1}^{\infty} \sum_{n=1}^{\infty} R^{\rho, k, n}_0 +  
		\sum_{k=1}^{\infty} \sum_{n=j}^{\infty} \big( R^{k, n}_i - C^{k, n}_{i \rho} \big).
		\end{align*}
		The stopped local martingale $(R^{\rho}_{\rho+\nu_{(i, j)}})^{\tau_k}$ is again a local martingale for every $k \in \N$, therefore the difference
		\begin{equation*}
		(R^{\rho}_{\rho+\nu_{(i, j)}})^{\tau_k} - (R^{\rho}_{\rho+\nu_{(i, j)}})^{\tau_{k-1}}
		= \sum_{n=1}^{\infty} R^{\rho, k, n}_0 + \sum_{n=j}^{\infty} \big( R^{k, n}_i - C^{k, n}_{i \rho} \big)
		=: R(i, j, k)
		\end{equation*}
		is also a local martingale for every $k \in \N$. Since we fixed $(i, j) \in \N^2$ satisfying $i \le j$ arbitrarily, we now choose any $n \in \N$, $i \in [n]$, and set $j = n$ and $j = n+1$ to conclude that
		\begin{equation*}
		R(i, n, k) - R(i, n+1, k) = R^{k, n}_i - C^{k, n}_{i \rho}
		\end{equation*}
		is a local martingale and the condition (iii) holds.	
	\end{proof}
	
	\medskip
	
	\subsection{Structural condition of each dissected market}	\label{subsec : structural condition}
	
	The condition $(iii)$ of Proposition~\ref{prop : numeraire portfolio} provides an important characterization of the num\'eraire portfolio $\rho$; for every dissection $(R^{k, n})_{(k, n) \in \N^2}$ of the return process, the identity $A^{k, n}_i = C^{k, n}_{i \rho}$, derived by means of \eqref{R decomposition} and \eqref{def : C i rho}, must hold for each component $i = 1, \cdots, n$. This condition is called \textit{structural condition}, and it is known to be connected to the maximal growth rate of the market, in the spirit of Sections 2.1.2 and 2.1.3 of \cite{KK2}. This subsection demonstrates such connection for every \textit{dissected market}. First, we can reformulate the structural condition in terms of \textit{local rates} as in the following.
	
	Let us fix $(k, n) \in \N^2$, recall the decomposition \eqref{R decomposition}, and denote $C^{k, n}$ the $(n \times n)$ matrix-valued process with entries 
	\begin{equation}	\label{def : C entries}
	C^{k, n}_{i, j} := [M^{k, n}_i, M^{k, n}_j].
	\end{equation}
	We define the \textit{dissected operational clock} by a real-valued process
	\begin{equation}
	O^{k, n} := \sum_{i=1}^n \int_0^{\cdot} \Big( \vert d A^{k, n}_i(t) \vert + dC^{k, n}_{i, i}(t) \Big),
	\end{equation}
	where $\int \vert dA^{k, n}_i(t) \vert$ denotes the total variation of $A^{k, n}_i$. With respect to this nondecreasing process, we consider the Radon-Nikod\'ym derivatives $\alpha^{k, n} = (\alpha^{k, n}_i)_{1 \le i \le n}$ and $c^{k, n} = (c^{k, n}_{i, j})_{1 \le i, j \le n}$ of $A^{k, n} = (A^{k, n}_i)_{1 \le i \le n}$ and $C^{k, n}$, respectively:
	\begin{equation}	\label{def : alpha, c}
	A^{k, n} = \int_0^{\cdot} \alpha^{k, n}(t) \, dO^{k, n}(t), \qquad C^{k, n} = \int_0^{\cdot} c^{k, n}(t) \, dO^{k, n}(t).
	\end{equation}
	These predictable processes $\alpha^{k, n}$ and $c^{k, n}$ are called \textit{local return rate} and \textit{local covariation rate} of the dissected market, since they are derived from the \textit{dissected return} $R^{k, n}$. Here, we note that an $n$-dimensional process $\nu$ is $R^{k, n}$-integrable, if and only if,
	\begin{equation}	\label{con : integrability}
	\int_0^T  \big( \vert \nu^{\top} \alpha^{k, n} \vert + \nu^{\top}c^{k, n} \nu \big) (t) \, dO^{k, n}(t) < \infty \qquad \text{holds for any } T \ge 0.
	\end{equation}
	
	We then obtain the representation
	\begin{align*}
	C^{k, n}_{i\rho} = \Big[ R^{k, n}_i, \int_0^{\cdot} \sum_{j=1}^n \rho^{(k, n)}_j(t) \, dR^{k, n}_j(t) \Big] 
	&= \int_0^{\cdot} \sum_{j=1}^n \rho^{(k, n)}_j(t) \, dC^{k, n}_{i,j}(t)
	\\
	&= \int_0^{\cdot} \sum_{j=1}^n \rho^{(k, n)}_j(t) c^{k, n}_{i, j}(t) \, dO^{k, n}(t),
	\end{align*}
	thus the condition $(iii)$ of Proposition~\ref{prop : numeraire portfolio} is equivalent to
	\begin{equation}	\label{eq : dissected structural condition}
	\alpha^{k, n} = c^{k, n} \rho^{(k, n)}, \qquad (\mathbb{P} \otimes O^{k, n})-\text{a.e.}
	\end{equation}
	for every $(k, n) \in \N^2$. In other words, the existence of num\'eraire portfolio $\rho$ is equivalent to the structural condition \eqref{eq : dissected structural condition} of the \textit{$(k, n)$-dissected market} for every $(k, n) \in \N^2$.
	
	In order to show that the num\'eraire portfolio $\rho$ has the maximal growth rate, we first need to define the growth~(both cumulative and local rate) of portfolios in the market. For fixed $(k, n) \in \N^2$ and a portfolio $\pi \in \mathcal{L}(R)$, recalling the decomposition \eqref{R decomposition}, we consider the decomposition of $R^{k, n}_{\pi}$ in \eqref{def : R pi}
	\begin{equation*}
	A^{k, n}_{\pi} := \pi^{(k, n)} \cdot A^{k, n} = \int_0^{\cdot} \sum_{i=1}^n \pi^{(k, n)}_i(t) \, dA^{k, n}_i(t),
	\qquad
	M^{k, n}_{\pi} := \pi^{(k, n)} \cdot M^{k, n},
	\end{equation*}
	and define
	\begin{equation}	\label{def : Gamma pi}
	\Gamma^{k, n}_{\pi} := A^{k, n}_{\pi} - \frac{1}{2}C^{k, n}_{\pi\pi}.
	\end{equation}
	It is then easy to derive
	\begin{equation*}
	\log \big( \mathcal{E}(R^{k, n}_{\pi}) \big) = R^{k, n}_{\pi} - \frac{1}{2} C^{k, n}_{\pi\pi} = \Gamma^{k, n}_{\pi} + M^{k, n}_{\pi},
	\end{equation*}
	and moreover, using the property that $R^{k, n}_{\pi}$ and $R^{\ell, m}_{\pi}$ are orthogonal, i.e., $[R^{k, n}_{\pi}, R^{\ell, m}_{\pi}] \equiv 0$, whenever $(k, n) \neq (\ell, m)$, we obtain
	\begin{align*}
	X_{\pi} = \mathcal{E} \Big( \sum_{k=1}^{\infty} \sum_{n=1}^{\infty} R^{k, n}_{\pi} \Big)
	= \exp \Big( \sum_{k=1}^{\infty} \sum_{n=1}^{\infty} R^{k, n}_{\pi} - \frac{1}{2} \sum_{k=1}^{\infty} \sum_{n=1}^{\infty} [R^{k, n}_{\pi}, R^{k, n}_{\pi}] \Big)
	= \prod_{k=1}^{\infty} \prod_{n=1}^{\infty} \mathcal{E} \big( R^{k, n}_{\pi} \big).
	\end{align*}
	Combining the last identities, we have the following representation of the log-wealth process
	\begin{equation}	\label{eq : log wealth}
	\log (X_{\pi}) = \sum_{k=1}^{\infty} \sum_{n=1}^{\infty} \big( \Gamma^{k, n}_{\pi} + M^{k, n}_{\pi} \big).
	\end{equation}
	We call the finite variation process $\Gamma_{\pi} := \sum_{k=1}^{\infty} \sum_{n=1}^{\infty} \Gamma^{k, n}_{\pi}$ the \textit{cumulative growth} of $\pi$ with dissection $\Gamma^{k, n}_{\pi}$ for each $(k, n) \in \N^2$. Furthermore, if we denote $\gamma^{k, n}_{\pi}$ the Radon-Nikod\'ym derivative of $\Gamma^{k, n}_{\pi}$ with respect to $O^{k, n}$, it is easy to verify the following relationship from \eqref{def : Gamma pi}
	\begin{equation}	\label{def : gamma pi dissection}
	\gamma^{k, n}_{\pi} = (\alpha^{k, n})^{\top} \pi^{(k, n)} - \frac{1}{2} (\pi^{(k, n)})^{\top} c^{k, n} \pi^{(k, n)}, \qquad (\mathbb{P} \otimes O^{k, n})-\text{a.e.}
	\end{equation}
	We call $\gamma^{k, n}_{\pi}$ the \textit{local growth rate} of the portfolio $\pi$ in the $(k, n)$-dissected market.
	
	We now define the \textit{maximal growth rate}
	\begin{equation}	\label{def : maximal growth rate}
	g^{k, n} := \sup_{p \in \R^n} \Big( (\alpha^{k, n})^{\top}p - \frac{1}{2} p^{\top}c^{k, n}p \Big),
	\end{equation}
	achievable in the $(k, n)$-dissected market. Note that $g^{k, n}$ is predictable, since the supremum can be taken over a countable dense subset of $\R^n$. Before presenting the following result, we define \textit{pseudo-inverse} of an $(n \times n)$ matrix-valued process $c$ by
	\begin{equation}
	c^{\dagger} := \lim_{m \rightarrow \infty} \bigg( \Big(c+ \frac{\textbf{id}_n}{m}\Big)^{-2} c \bigg),
	\end{equation}
	where $\textbf{id}_n$ denotes the $n \times n$ identity matrix.
	
	\smallskip
	
	\begin{prop}	\label{prop : structural conditions}
		For a fixed, arbitrary pair $(k, n) \in \N^2$, the following statements are equivalent:
		\begin{enumerate} [(i)]
			\item There exists an $n$-dimensional $R^{k, n}$-integrable process $\rho^{(k, n)}$ with $\rho^{(k, n)}(0) = 0^{(n)}$ such that $\alpha^{k, n} = c^{k, n} \rho^{(k, n)}$ holds $(\mathbb{P} \otimes O^{k, n})$-a.e.
			\item $\alpha^{k, n} \in \textbf{range}(c^{k, n})$ for $(\mathbb{P} \otimes O^{k, n})$-a.e., and $\int_0^T \big(\alpha^{k, n}(t)\big)^{\top} \big(c^{k, n}(t)\big)^{\dagger}\alpha^{k, n}(t) \, dO^{k, n}(t) < \infty$ for every $T \ge 0$.
			\item The $(k, n)$-dissected market has locally finite growth, that means, for every $T \ge 0$, we have $G^{k, n}(T) := \int_0^T g^{k, n}(t) \, dO^{k, n}(t) < \infty$.
		\end{enumerate}
	\end{prop}
	
	We refer to Sections 2.1.2 and 2.1.3 of \cite{KK2} for the detailed proof of Proposition~\ref{prop : structural conditions}, because the same argument can be applied, if the symbols~$\alpha, c, \rho, O, g, G$ there, are replaced with the ones with the superscript ${k, n}$ here. To provide the idea of the proof, the process 
	\begin{equation}	\label{def : numeraire portfolio dissection}
	\rho^{(k, n)} := (c^{k, n})^{\dagger}\alpha^{k, n}, \qquad \text{(modulo null portfolio)}	
	\end{equation}
	satisfies the conditions $(i)-(iii)$ of Proposition~\ref{prop : structural conditions}. In this case, $\rho^{(k, n)}$ attains the maximal growth rate $g^{k, n}$ of \eqref{def : maximal growth rate}, which is equal to $(1/2) (\alpha^{k, n})^{\top} (c^{k, n})^{\dagger} \alpha^{k, n}$. Generally, $g^{k, n}$ of \eqref{def : maximal growth rate} can be expressed as 
	\begin{equation}	\label{eq : g dissection}
	g^{k, n} = \frac{1}{2} \Big( (\alpha^{k, n})^{\top} (c^{k, n})^{\dagger} \alpha^{k, n} \Big) \mathbbm{1}_{\{\alpha^{k, n} \in \textbf{range}(c^{k, n})\}} + \infty \mathbbm{1}_{\{\alpha^{k, n} \notin \textbf{range}(c^{k, n})\}}
	\end{equation}
	and the finiteness of $(iii)$ guarantees the $R^{k, n}$-integrability of the process $\rho^{(k, n)}$ above.
	
	\smallskip
	
	Thanks to Proposition~\ref{prop : numeraire portfolio}, the num\'eraire portfolio exists if and only if the structural conditions of Proposition~\ref{prop : structural conditions} hold for every $(k, n)$-dissected market for $(k, n) \in \N^2$. Then, the num\'eraire portfolio $\rho$ can be constructed as
	\begin{equation}	\label{def : numeraire portfolio}
	\rho := 0^{(N_0)} + \sum_{k=1}^{\infty} \sum_{n=1}^{\infty} \hat{\mathbbm{1}}_{\rrbracket \tau_{k-1}, \tau_k \rrbracket \cap (\R_+ \times \Omega^{k, n})} \rho^{(k, n)},
	\end{equation}
	where every dissection $\rho^{(k, n)}$ satisfies the condition $(i)$ of Proposition~\ref{prop : structural conditions}.
	
	\medskip
	
	\subsection{Arbitrage of the first kind and local martingale deflators}	\label{subsec : NA1}
	
	According to the fundamental result~(Theorem~2.31 of \cite{KK2}) of arbitrage theory in equity markets~(with fixed number of stocks), there are more concepts related to the equivalent conditions of Propositions~\ref{prop : numeraire portfolio} and \ref{prop : structural conditions}, namely the market viability~(or lack of arbitrage of the first kind), and the existence of local martingale deflators. The equivalence between the last two notions is actually proven in \cite{Strong2} in the setting of market with a stochastic number of assets. In this subsection, we state their result, supplement it with related concepts, and provide connections with the other results from the previous subsections.
	
	We first define investment strategies, which play the role of integrand for the price process, as opposed to portfolios acting as an integrand for the return process.
	
	\smallskip
	
	\begin{defn} [Investment strategy] \label{Def: investment strategy}
		For a given price process $S$ of Definition~\ref{Def : price process}, we call $\vartheta$ an \textit{investment strategy}, if $\vartheta \in \mathcal{L}_0(S)$. The wealth process of $\vartheta$ with initial capital $x$ is defined by
		\begin{equation}	\label{def : wealth of vartheta}
		X(\cdot ; x, \vartheta) := x + \vartheta \cdot S = x + \sum_{k=1}^{\infty} \sum_{n=1}^{\infty} \int_0^{\cdot} \sum_{i=1}^n \vartheta^{(k, n)}_i(u) \, dS^{k, n}_i(u).
		\end{equation}
		The investment strategy $\vartheta$ with initial capital $x > 0$ is said to be \textit{admissible~(strictly admissible)}, if its wealth process is nonnegative~(strictly positive) at all times, i.e.,
		\begin{equation*}
		X(\cdot ; x, \vartheta) \ge 0 \qquad \big(X(\cdot ; x, \vartheta) > 0, \text{ respectively}\big).
		\end{equation*}
	\end{defn}
	
	For a given strictly admissible investment strategy $\vartheta$ with initial capital $x = 1$, we can construct corresponding portfolio
	\begin{equation}
	\pi := 0^{(N_0)} + \sum_{k=1}^{\infty} \sum_{n=1}^{\infty} \hat{\mathbbm{1}}_{\rrbracket \tau_{k-1}, \tau_k \rrbracket \cap (\R_+ \times \Omega^{k, n})} \pi^{(k, n)},
	\end{equation}
	where for every $(k, n) \in \N^2$
	\begin{equation}	\label{def : corresponding pi dissection}
	\pi^{(k, n)}_i := \frac{S_i \vartheta^{(k, n)}_i}{X(\cdot ; 1, \vartheta)} \hat{\mathbbm{1}}_{\rrbracket \tau_{k-1}, \tau_k \rrbracket \cap (\R_+ \times \Omega^{k, n})} + 0^{(1)}, \qquad i \in [n].
	\end{equation}
	Here, the strict admissibility of $\vartheta$ is necessary as the wealth process appears in the denominator. The wealth $X_{\pi}$ of this portfolio $\pi$ is then equal to the wealth $X(\cdot; 1, \vartheta)$ of $\vartheta$:
	\begin{align}
	X_{\pi}(t) &= \mathcal{E} \big(R_{\pi}(t) \big)
	= \mathcal{E} \bigg(\int_0^t \sum_{k=1}^{\infty} \sum_{n=1}^{\infty}  \sum_{i=1}^n \frac{S_i(u) \vartheta^{(k, n)}_i(u)}{X(u ; 1, \vartheta)} \hat{\mathbbm{1}}_{\rrbracket \tau_{k-1}, \tau_k \rrbracket \cap (\R_+ \times \Omega^{k, n})} \, dR^{k, n}_i(u) \bigg)            \label{eq : corresponding wealth of portfolio}
	\\
	&= \mathcal{E} \bigg(\int_0^t \sum_{k=1}^{\infty} \sum_{n=1}^{\infty}  \sum_{i=1}^n \frac{\vartheta^{(k, n)}_i(u)}{X(u ; 1, \vartheta)} \hat{\mathbbm{1}}_{\rrbracket \tau_{k-1}, \tau_k \rrbracket \cap (\R_+ \times \Omega^{k, n})} \, dS^{k, n}_i(u) \bigg)                                                          \nonumber
	\\
	&= \mathcal{E} \bigg( \int_0^t \frac{dX(u ; 1, \vartheta)}{X(u ; 1, \vartheta)} \bigg)
	= X(t; 1, \vartheta), \qquad \forall \, t \ge 0,                \nonumber
	\end{align}
	after plugging in \eqref{def : corresponding pi dissection}, \eqref{def : return process}, \eqref{def : wealth of vartheta}, and using the property of stochastic exponential.
	
	Conversely, for a given portfolio $\pi \in \mathcal{L}(R)$~(note that its wealth $X_{\pi}$ is always positive as a stochastic exponential), we can define corresponding investment strategy
	\begin{equation*}
	\vartheta := 0^{(N_0)} + \sum_{k=1}^{\infty} \sum_{n=1}^{\infty} \hat{\mathbbm{1}}_{\rrbracket \tau_{k-1}, \tau_k \rrbracket \cap (\R_+ \times \Omega^{k, n})} \vartheta^{(k, n)},
	\end{equation*}
	where for every $(k, n) \in \N^2$
	\begin{equation}	\label{def : corresponding vartheta dissection}
	\vartheta^{(k, n)}_i := \frac{(X_{\pi}) \pi^{(k, n)}_i}{S_i} \hat{\mathbbm{1}}_{\rrbracket \tau_{k-1}, \tau_k \rrbracket \cap (\R_+ \times \Omega^{k, n})} + 0^{(1)}, \qquad i \in [n].
	\end{equation}
	It is also easy to check that the wealth $X(t; 1, \vartheta)$ of $\vartheta$ is equal to $X_{\pi}(t)$ for every $t \ge 0$, by plugging in \eqref{def : corresponding vartheta dissection} into \eqref{def : wealth of vartheta}, and using the identity $dS^{k, n}_i(u) = dS_i(u)$ on the set $\rrbracket \tau_{k-1}, \tau_k \rrbracket \cap (\R_+ \times \Omega^{k, n})$, together with the property of stochastic exponential.
	
	\smallskip
	
	\begin{defn} [Arbitrage of the first kind] \label{def: NA1}
		An \textit{arbitrage of the first kind} for horizon $T > 0$ is an $\mathcal{F}_{T}$-measurable random variable $h$ satisfying $\mathbb{P}(h \ge 0) = 1$, $\mathbb{P}(h > 0) > 0$, such that for every $x > 0$, there exists an admissible investment strategy $\vartheta$ satisfying $X(T; x, \vartheta) \ge h$. If there are no arbitrages of the first kind, we say $\textit{NA}_1$ holds.
	\end{defn}
	
	Since this weak notion of arbitrage was introduced by \cite{I}, it has appeared under different names; \textit{no asymptotic arbitrage with first kind} in \cite{Kabanov:Kramkov}, \textit{bounded in probability~(or BK)} in \cite{Kabanov}, \textit{cheap thrills} in \cite{LW_snack}, and \textit{no unbounded profit with bounded risk~(NUPBR)} in \cite{KK}. It is also known to be equivalent to other conditions which we provide in the following.
	
	\smallskip
	
	\begin{defn} [Local martingale deflator]	\label{Def : deflator}
		An adapted, strictly positive process $Y$ is called \textit{local martingale deflator}, if $Y(0)=1$ and $YX_{\pi}$ is a local martingale for every portfolio $\pi \in \mathcal{L}(R)$.
	\end{defn}
	
	\smallskip
	
	\begin{defn} [Market viability] \label{Def : market viability}
		A nondecreasing, adapted, and right-continuous process $K$ with $K(0) = 0$ is called a \textit{cumulative withdrawal stream}. For any given cumulative withdrawal stream $K$, we define the \textit{superhedging capital associated with $K$} by
		\begin{equation}    \label{def : superhedging capital}
		x(K) := \inf \{ x \ge 0 \, \vert \, \exists \, \vartheta \in \mathcal{L}_0(S) \text{ such that } X(\cdot ; x, \vartheta) \ge K\},
		\end{equation}
		representing the smallest initial capital starting from which the process $K$ can be financed or hedged in the market. 
		We say that market is \textit{viable}, if $x(K) = 0$ implies $K \equiv 0$.
	\end{defn}
	
	\smallskip
	
	\begin{prop} [\cite{Strong2} and \cite{KK2}]	\label{prop : NA1}
		The following statements are equivalent:
		\begin{enumerate} [(i)]
			\item $\textit{NA}_1$ holds.
			\item There exists a local martingale deflator.
			\item The market is viable.
			\item The collection of portfolio wealth processes is bounded in probability, i.e.,
			\begin{equation}	\label{con : bounded in prob}
			\lim_{m \rightarrow \infty} \sup_{\pi \in \mathcal{L}(R)} \mathbb{P} \big[ X_{\pi}(T) > m \big] = 0 \text{ holds for any } T \ge 0.
			\end{equation}
		\end{enumerate}
	\end{prop}
	
	\begin{proof}
		Theorem~3.5 of \cite{Strong2} proves that the existence of local martingale deflator is equivalent to $\textit{NA}_1$ in the market of stochastic dimension. The equivalences (i)$\iff$(iii) and (iii)$\iff$(iv) follow from Exercise~2.21 and Proposition~2.22 of \cite{KK2}; the argument can be applied to any market~(of either fixed or stochastic dimension).
	\end{proof}
	
	We also refer to Section 1.4 of \cite{KS2} for some ramifications of this concept of arbitrage, in a classical model of the equity market.
	
	We conclude this subsection with the following explicit expression of local martingale deflators in terms of a local martingale num\'eraire portfolio. This characterization of local martingale deflators is originally studied by \cite{Schweizer:1995}, also appears in Propositions~2.3, 3.2 of \cite{Lar:Zit:2007}, and Exercise~2.27 of \cite{KK2}.
	
	Recalling the decomposition \eqref{R decomposition} of each dissected return, let us denote $\mathcal{Y}$ the collection of local martingale deflators in Definition~\ref{Def : deflator}, and $\mathcal{M}^{\bot}_{loc}(M)$ the collection of scalar local martingales $L$ with RCLL paths, satisfying $L(0)=0$ and the strong orthogonality condition $[L, M^{k, n}_i] = 0$ for every $(k, n) \in \N^2$ and $i \in [n]$.
	
	\smallskip
	
	\begin{prop}	\label{prop : form of deflators}
		If there exists a local martingale num\'eraire portfolio $\rho$, then we have
		\begin{equation}
		\mathcal{Y} = \bigg\{ \frac{\mathcal{E}(L)}{X_{\rho}} \, \Big\vert \, L \in \mathcal{M}^{\bot}_{loc}(M), \quad \Delta L > -1 \bigg\}.
		\end{equation}
	\end{prop}
	
	\begin{proof}
		Let $Z$ be a local martingale deflator. Recalling the notation \eqref{def : relative wealth}, $ZX_{\rho}$ and $ZX_{\pi} = ZX_{\rho} X^{\rho}_{\pi}$ are strictly positive local martingales for every $\pi \in \mathcal{L}(R)$. Then, it follows that $Q := ZX_{\rho}$ is strongly orthogonal to $X^{\rho}_{\pi}$. Since we can express $Q = \mathcal{E}(L)$, where
		\begin{equation*}
		L = \int_0^{\cdot} \frac{dQ(t)}{Q(t-)}, \qquad \text{with} \qquad \Delta L(t) = \frac{Q(t) - Q(t-)}{Q(t-)} > -1,
		\end{equation*}
		and $X^{\rho}_{\pi} = \mathcal{E}(R^{\rho}_{\pi})$ from \eqref{eq : X rho pi}, we conclude from the strict positivity of $Q$ and $X^{\rho}_{\pi}$, that $L$ is strongly orthogonal to $R^{\rho}_{\pi}$.
		
		Now that $\rho$ satisfies the condition $(iii)$ of Proposition~\ref{prop : numeraire portfolio}, the notations \eqref{def : R0}, \eqref{def : Ri} of Lemma~\ref{lem : X rho pi representation} have alternative representations from \eqref{eq : R rho, k, n, 0 expression}
		\begin{align*}
		R^{\rho, k, n}_0 &= - \int_0^{\cdot} \sum_{i=1}^n \rho^{(k, n)}_i \, dM^{k, n}_i,
		\\
		R^{\rho, k, n}_i &= R^{\rho, k, n}_0 + M^{k, n}_i, \qquad \qquad \qquad i \in [n],
		\end{align*}
		and a straightforward computation gives for every portfolio $\pi$
		\begin{equation}	\label{eq : R rho pi representation}
		R^{\rho}_{\pi} = \sum_{k=1}^{\infty} \sum_{n=1}^{\infty} \int_0^{\cdot} \sum_{i=1}^n (\pi^{(k, n)}_i - \rho^{(k, n)}_i) \, dM^{k, n}_i.
		\end{equation}
		For an arbitrary pair $(k, n) \in \N^2$ and number $i \in [n]$, we define a portfolio $\pi$ via the following recipe
		\begin{align*}
		\pi^{(\ell, p)} &= \rho^{(\ell, p)}, \qquad \text{if } (\ell, p) \neq (k, n),
		\\
		\pi^{(k, n)}_j &= \rho^{(k, n)}_j \qquad \text{ if } j \neq i,
		\\
		\pi^{(k, n)}_i &= \rho^{(k, n)}_i+1,
		\end{align*}
		such that $R^{\rho}_{\pi} = M^{k, n}_i$. In other words, $L$ is strongly orthogonal to every $M^{k, n}_i$, thus it belongs to $\mathcal{M}^{\bot}_{loc}(M)$, which establishes the representation $Z = \mathcal{E}(L)/X_{\rho}$.
		
		Conversely, for any $L \in \mathcal{M}^{\bot}_{loc}(M)$ satisfying $\Delta L > -1$, it is enough to show that $\mathcal{E}(L) X^{\rho}_{\pi}$ is a local martingale for every portfolio $\pi$. The identities \eqref{eq : X rho pi}, \eqref{eq : R rho pi representation}, the strong orthogonality condition on $L$, and Yor's formula prove that
		\begin{equation*}
		\mathcal{E}(L)X^{\rho}_{\pi} = \mathcal{E}(L)\mathcal{E}(R^{\rho}_{\pi})
		= \mathcal{E}(L+ R^{\rho}_{\pi})
		\end{equation*}
		is a local martingale.
	\end{proof}
	
	\medskip
	
	\subsection{The fundamental theorem}	\label{subsec : fundamental theorem}
	
	We are now ready to connect all results~(Propositions~\ref{prop : numeraire portfolio}, \ref{prop : structural conditions}, and \ref{prop : NA1}) in the previous subsections, and state the following cornerstone theorem of arbitrage theory in the market of stochastic dimension.
	
	\smallskip
	
	\begin{thm}	\label{thm : fundamental theorem}
		The following statements are equivalent:
		\begin{enumerate} [(i)]
			\item There exists a supermartingale~(also a local martingale) num\'eraire portfolio.
			\item There exists a portfolio $\rho$ satisfying $A^{k, n}_i = C^{k, n}_{i\rho}$ for every $i \in [n]$ and $(k, n) \in \N^2$. 
			\item Each $(k, n)$-dissected market has locally finite growth, i.e., $G^{k, n}(T) < \infty$ for every $T \ge 0$ and $(k, n) \in \N^2$.
			\item $\text{NA}_1$ holds.
			\item There exists a local martingale deflator.
			\item The market is viable.
		\end{enumerate}
	\end{thm}
	
	\begin{proof}
		The equivalences between $(i)-(iii)$ are explained in Sections~\ref{subsec : numeraire portfolio}, \ref{subsec : structural condition}, and the equivalences between $(iv)-(vi)$ are from Proposition~\ref{prop : NA1}. If a local martingale num\'eraire portfolio $\rho$ exists, then the reciprocal of its wealth $1/X_{\rho}$ is a local martingale deflator by definition, thus $(i)$ implies $(v)$. We show in the following the implication $(vi) \Longrightarrow (iii)$, which shows the result.
		
		We assume that there exists a pair of natural numbers $(\ell, p)$ such that the $(\ell, p)$-dissected market fails to have locally finite growth, or the condition $(ii)$ of Proposition~\ref{prop : structural conditions} is violated. Following the proof of Theorem~2.31 of \cite{KK2}, we shall treat two cases: either
		\begin{enumerate} [(A)]
			\item the set $\{\alpha^{\ell, p} \notin \textbf{range}(c^{\ell, p})\}$ fails to be $(\mathbb{P} \otimes O^{\ell, p})$-null, or
			\item the set $\{\alpha^{\ell, p} \notin \textbf{range}(c^{\ell, p})\}$ is $(\mathbb{P} \otimes O^{\ell, p})$-null, but there exists $T \ge 0$ such that 
			\begin{equation}	\label{con : G infinity}
			\mathbb{P} \Big[ \int_0^T \Big( (\alpha^{\ell, p})^{\top} (c^{\ell, p})^{\dagger} \alpha^{\ell, p}\Big) (t) \, dO^{\ell, p}(t) = \infty \Big] > 0, \qquad \text{holds.}
			\end{equation}
		\end{enumerate}
		
		For case (A), we define a $p$-dimensional predictable process
		\begin{equation}    \label{def : varphi ell p}
		\varphi^{(\ell, p)} := \frac{1}{\Vert \alpha^{\ell, p}-c^{\ell, p}(c^{\ell, p})^{\dagger}\alpha^{\ell, p}\Vert^2} \big(\alpha^{\ell, p} - c^{\ell, p}(c^{\ell, p})^{\dagger}\alpha^{\ell, p} \big) \mathbbm{1}_{\{\alpha^{\ell, p} \notin \textbf{range}(c^{\ell, p})\}}.
		\end{equation}
		Recalling the fact that $c^{\ell, p}(c^{\ell, p})^{\dagger}$ is the projection operator on $\textbf{range}(c^{\ell, p})$, we can deduce that $\varphi^{(\ell, p)}$ is well-defined, i.e., the denominator is nonzero on the set $\{\alpha^{\ell, p} \notin \textbf{range}(c^{\ell, p})\}$, and the identities $c^{\ell, p} \varphi^{(\ell, p)} \equiv 0$ and $(\varphi^{(\ell, p)})^{\top} \alpha^{\ell, p} = \mathbbm{1}_{\{\alpha^{\ell, p} \notin \textbf{range}(c^{\ell, p}) \}}$ hold. From \eqref{con : integrability}, $\varphi^{(\ell, p)}$ is $R^{\ell, p}$-integrable, however, $c^{\ell, p} \varphi^{(\ell, p)} \equiv 0$ yields $[\int_0^{\cdot} (\varphi^{(\ell, p)})^{\top} dM^{\ell, p}] \equiv 0$, thus 
		\begin{equation}    \label{eq : varphi M zero}
		\int_0^{\cdot} (\varphi^{(\ell, p)})^{\top} dM^{\ell, p} \equiv 0.
		\end{equation}
		Therefore, we set nondecreasing process
		\begin{equation}
		K := \int_0^{\cdot} (\varphi^{(\ell, p)})^{\top} dR^{\ell, p} 
		= \int_0^{\cdot} (\varphi^{(\ell, p)})^{\top} dA^{\ell, p}
		= \int_0^{\cdot} \mathbbm{1}_{\{\alpha^{\ell, p} \notin \textbf{range}(c^{\ell, p})\}} dO^{\ell, p},
		\end{equation}
		and the assumption (A) implies $\mathbb{P} [K(\infty) > 0] > 0$. We now define an investment strategy $\vartheta$ via dissection
		\begin{align*}
		\vartheta^{(\ell, p)}_i &:= \frac{\varphi^{(\ell, p)}_i}{S_i} \hat{\mathbbm{1}}_{\rrbracket \tau_{\ell-1}, \tau_{\ell} \rrbracket \cap (\R_+ \times \Omega^{\ell, p})} + 0^{(1)}, \qquad i = 1, \cdots, p,
		\\
		\vartheta^{(k, n)} &:= 0^{(n)}, \qquad \qquad \qquad \qquad \qquad \qquad \text{if } (k, n) \neq (\ell, p),
		\end{align*}
		then its wealth process is equal to $K$:
		\begin{equation*}
		X(\cdot ; 0, \vartheta) = 0 + \vartheta \cdot S
		= \int_0^{\cdot} (\vartheta^{(\ell, p)})^{\top} dS^{\ell, p}
		= \int_0^{\cdot} (\varphi^{(\ell, p)})^{\top} dR^{\ell, p}
		= K.
		\end{equation*}
		This violates the market viability, since we can ``finance'' nonzero process $K$ from zero initial capital.
		
		For case (B), we assume \eqref{con : G infinity} and we shall deduce that the market is not viable, by showing that the condition \eqref{con : bounded in prob} is violated. We define a portfolio $\rho$ via dissection
		\begin{align*}
		\rho^{(\ell, p)} &:= (c^{\ell, p})^{\dagger}\alpha^{\ell, p} \hat{\mathbbm{1}}_{\rrbracket \tau_{\ell-1}, \tau_{\ell} \rrbracket \cap (\R_+ \times \Omega^{\ell, p})} + 0^{(p)},
		\\
		\rho^{(k, n)} &:= 0^{(n)}, \qquad \qquad \qquad \qquad \qquad \qquad \qquad \text{if } (k, n) \neq (\ell, p),
		\end{align*}
		as well as a sequence of portfolios $(\rho_{\eta})_{\eta \in \N}$ such that
		\begin{equation}
		\rho_{\eta}^{(k, n)} := \rho^{(k, n)} \mathbbm{1}_{\{\Vert \rho^{(k, n)} \Vert < \eta\}}, \qquad \text{for every } (k, n) \in \N^2.
		\end{equation}
		Recalling \eqref{eq : log wealth} and \eqref{def : gamma pi dissection}, it is straightforward to deduce for every $\eta \in \N$
		\begin{align*}
		\log(X_{\rho_{\eta}}) = \frac{1}{2} \int_0^{\cdot} \mathbbm{1}_{\{\Vert \rho^{(\ell, p)} \Vert < \eta\}} (\rho^{(\ell, p)})^{\top} c^{\ell, p} \rho^{(\ell, p)} \, dO^{\ell, p} + \int_0^{\cdot} \mathbbm{1}_{\{\Vert \rho^{(\ell, p)} \Vert < \eta\}} (\rho^{(\ell, p)})^{\top} \, dM^{\ell, p}.
		\end{align*}
		Since the first integral
		\begin{equation*}
		2G^{\eta} := \int_0^{\cdot} \mathbbm{1}_{\{\Vert \rho^{(\ell, p)} \Vert < \eta\}} (\rho^{(\ell, p)})^{\top} c^{\ell, p} \rho^{(\ell, p)} \, dO^{\ell, p}
		\end{equation*}
		is the quadratic variation of the last integral $\int_0^{\cdot} \mathbbm{1}_{\{\Vert \rho^{(\ell, p)} \Vert < \eta\}} (\rho^{(\ell, p)})^{\top} \, dM^{\ell, p}$, which is a local martingale, the Dambis-Dubins-Schwarz representation~(Theorem~3.4.6 of \cite{KS1}) with the scaling property of Brownian motion shows that there exists a Brownian motion $W^{\eta}$, possibly on an enlarged probability space, satisfying
		\begin{equation*}
		\log(X_{\rho_{\eta}}) = G^{\eta} + \sqrt{2}W^{\eta}(G^{\eta}), \qquad \text{for every } \eta \in \N.
		\end{equation*}
		Now that $\big(G^{\eta}(T)\big)_{\eta \in \N}$ is nondecreasing and converges as $\eta \rightarrow \infty$ to 
		\begin{equation*}
		\frac{1}{2} \int_0^T (\rho^{(\ell, p)})^{\top} c^{\ell, p} \rho^{(\ell, p)} \, dO^{\ell, p}
		= \frac{1}{2} \int_0^T (\alpha^{\ell, p})^{\top} (c^{\ell, p})^{\dagger} \alpha^{\ell, p} \, dO^{\ell, p} =: G(T), 
		\end{equation*}
		the condition \eqref{con : G infinity} and the strong law of large numbers imply
		\begin{equation*}
		\lim_{\eta \rightarrow \infty} \mathbb{P} \bigg[ \frac{W^{\eta}\big( G^{\eta}(T) \big) }{G^{\eta}(T)} \le -\frac{1}{2\sqrt{2}}, \quad G(T) = \infty \bigg] = 0,
		\end{equation*}
		thus
		\begin{equation*}
		\lim_{\eta \rightarrow \infty} \mathbb{P} \bigg[ \frac{\log \big(X_{\rho_{\eta}}(T) \big)}{G^{\eta}(T)} \le \frac{1}{2}, \quad G(T) = \infty \bigg] = 0.
		\end{equation*}
		Therefore, the sequence of random variables $\big(X_{\rho_{\eta}}(T)\big)_{\eta \in \N}$ fails to be bounded in probability under the assumption \eqref{con : G infinity}, which violates the condition \eqref{con : bounded in prob}.
	\end{proof}

        \smallskip

        Example 2.11 of \cite{KK2} provides a simple example of a stock market (of a fixed dimension) that is not viable. We can easily generalize it in our setting to give the following example of a market in which any condition of Theorem \ref{thm : fundamental theorem} fails to hold.
        
        \begin{example}
            For a fixed $n \in \mathbb{N}$, consider $n$ independent copies of Brownian motions $W_1, \cdots, W_n$ and the $n$-dimensional vector $B := (|W_1|, \cdots, |W_n|)$. Note from Tanaka's formula that the component $B_i$ of $B$ can be decomposed as $B_i = A_i + M_i$, where $A_i$ is the local time process of $W_i$ at the origin, and $M_i = \int_0^{\cdot} \text{sign}(W_i(s))dW_i(s)$ is a Brownian motion. Moreover, $A_i$ is singular with respect to the Lebesgue measure, and $[M_i,M_j](t) \equiv \delta_{i, j}t$ holds for every $i, j \in [n]$. 
            Suppose that the return process of the $(1, n)$-dissected market follows the dynamics of $B$, i.e.,
            \begin{equation}    \label{ex: not viable}
                R^{1, n}_i(t) = B^{\tau_1}_i(t), \qquad \text{for every } t \ge 0, \quad i \in [n].
            \end{equation}
            Then, we have $A^{1, n}_i = A^{\tau_1}_i$, $M^{1, n}_i = M^{\tau_1}_i$, and 
            \begin{equation*}
                C^{1, n}_{i \rho}(t) = \int_0^{t \wedge \tau_1} \sum_{j=1}^n \rho^{(1, n)}_j(s) d[M^{1, n}_i, M^{1, n}_j](s) = \int_0^{t \wedge \tau_1} \rho^{(1, n)}_i(s) ds.
            \end{equation*}
            However, the singularity of $A_i$ with respect to the Lebesgue measure concludes that a portfolio $\rho$ satisfying $A^{1, n}_i = C^{1, n}_{i\rho}$ does not exist. Since condition (ii) of Theorem \ref{thm : fundamental theorem} is violated, this market is not viable.
            
            For any $k\in\mathbb{N}$, we can also make the return process of $(k, n)$-dissected market follow $B$ by setting
            \begin{equation*}
                R^{k, n}_i(t + \tau_{k-1}) = B^{\tau_k}_i(t), \qquad \text{for every } t \ge 0, \quad i \in [n],
            \end{equation*}
            instead of \eqref{ex: not viable}, then the same argument concludes that this market violates condition (ii) of Theorem \ref{thm : fundamental theorem} for the pair $(k, n) \in \mathbb{N}^2$.
        \end{example}
        
	\medskip
	
	\subsection{Optimal properties of the num\'eraire portfolio}	\label{subsec : log-optimality}

    In Section \ref{subsec : structural condition}, we showed that the (supermartingale) num\'eraire portfolio attains the maximal growth rate of \eqref{def : maximal growth rate} for every dissected market. This `optimality' in the growth can be formally defined as follows.
    \begin{defn} [Growth optimal portfolio]
        A portfolio $\rho \in \mathcal{L}(R)$ is called \textit{growth optimal portfolio}, if
        \begin{equation}    \label{con : growth optimal}
            \gamma^{k, n}_{\rho} \ge \gamma^{k, n}_{\pi}
        \end{equation}
        holds for every $(k, n) \in \mathbb{N}^2$ and every portfolio $\pi \in \mathcal{L}(R)$.
    \end{defn}

    Recalling the cumulative growth $\Gamma^{k, n}_{\pi}$ of $\pi$ in \eqref{def : Gamma pi} and the fact that $\gamma^{k, n}_{\pi}$ is a Radon-Nikod\'ym derivative of $\Gamma^{k, n}_{\pi}$ with respect to $O^{k, n}$, we note that the inequality of \eqref{con : growth optimal} is equivalent to the condition that the process
    \begin{equation*}
        (\Gamma^{\rho}_{\pi})^{k, n} := \Gamma^{k, n}_{\pi} - \Gamma^{k, n}_{\rho}
    \end{equation*}
    is non-increasing.
        
    Furthermore, as described in Section 2.3.3 of \cite{KK2}, the (supermartingale) num\'eraire portfolio (in a market of fixed dimension) has the relative log-optimality, which is related to the maximization of expected logarithmic utility of the wealth process. We generalize this result in our setting of market with a changing number of assets.
	
	\smallskip
 
	\begin{defn} [Relative log-optimal portfolio]
		A portfolio $\rho \in \mathcal{L}(R)$ is called \textit{relatively log-optimal portfolio}, if 
		\begin{equation}    \label{con : log-optimal}
		\E \Big[ \big(\log X^{\rho}_{\pi}(\tau) \big)^+ \Big] < \infty \qquad \text{and} \qquad
		\E \big[ \log X^{\rho}_{\pi}(\tau) \big] \le 0
		\end{equation}
		hold for every stopping time $\tau$ and every portfolio $\pi \in \mathcal{L}(R)$.
	\end{defn}
	
	\smallskip
	
	\begin{prop}	\label{prop : log-optimality}
		For a portfolio $\rho \in \mathcal{L}(R)$, the following statements are equivalent:
		\begin{enumerate}[(i)]
			\item $\rho$ is a supermartingale num\'eraire portfolio.
                \item $\rho$ is a growth optimal portfolio.
			\item $\rho$ is a relatively log-optimal portfolio.
		\end{enumerate}
	\end{prop}
	
	The conditions $(ii)$ and $(iii)$ of Proposition~\ref{prop : log-optimality} can be appended to Proposition~\ref{prop : numeraire portfolio}, hence the existence of the growth optimal portfolio or the relative log-optimal portfolio can also be added as an equivalent statement in the fundamental theorem~(Theorem~\ref{thm : fundamental theorem}).
	
	\begin{proof}
        The equivalence (i) $\iff$ (ii) can be proven by applying the same argument in the proof of Proposition 2.41 of \cite{KK2} to each dissected market (with the notations $\pi^{k, n}, \rho^{k, n}, c^{k, n}, \alpha^{k, n}$, etc) for every $(k, n) \in \mathbb{N}^2$.
        
		For the equivalence (i) $\iff$ (iii), suppose that $\rho$ is a supermartingale num\'eraire portfolio. From a trivial inequality $(\log x)^+ < x$, together with the Optional Sampling Theorem and Fatou's lemma applied to the nonnegative supermartingale $X^{\rho}_{\pi}$, we obtain
		\begin{equation*}
		\E \Big[ \big(\log X^{\rho}_{\pi}(\tau) \big)^+ \Big]
		< \E \big[ X^{\rho}_{\pi}(\tau) \big]
		\le 1
		\end{equation*}
		for any stopping time $\tau$ and $\pi \in \mathcal{L}(R)$. Jensen's inequality applied to the last inequality also yields the second condition $\E \big[ \log X^{\rho}_{\pi}(\tau) \big] \le 0$ of \eqref{con : log-optimal}.
		
		In order to show the reverse implication, suppose that $\rho$ is a relative log-optimal portfolio. Recalling the proof of Theorem~\ref{thm : fundamental theorem}, we first claim that the set $\{\alpha^{\ell, p} \notin \textbf{range}(c^{\ell, p})\}$ is $(\mathbb{P} \otimes O^{\ell, p})$-null for every $(\ell, p) \in \N^2$.
		
		For each pair $(\ell, p) \in \N^2$, let us bring back the $p$-dimensional process $\varphi^{(\ell, p)}$, defined in \eqref{def : varphi ell p}, which satisfies the identities
		\begin{equation}    \label{eq : identities of varphi}
		c^{\ell, p} \varphi^{(\ell, p)} \equiv 0, \qquad (\varphi^{(\ell, p)})^{\top} \alpha^{\ell, p} = \mathbbm{1}_{\{\alpha^{\ell, p} \notin \textbf{range}(c^{\ell, p})\}}.
		\end{equation}
		We construct a portfolio $\varphi$ from these dissections
		\begin{equation*}
		\varphi := 0^{(N_0)} + \sum_{\ell=1}^{\infty} \sum_{p=1}^{\infty} \hat{\mathbbm{1}}_{\rrbracket \tau_{\ell-1}, \tau_{\ell} \rrbracket \cap (\R_+ \times \Omega^{\ell, p})} \varphi^{(\ell, p)}.
		\end{equation*}
		From the representation \eqref{eq : log wealth} of the log-wealth process, we obtain
		\begin{align*}
		\log X^{\rho}_{\rho + \varphi}
		&= \sum_{\ell=1}^{\infty} \sum_{p=1}^{\infty} \Big( \Gamma^{\ell, p}_{\rho+\varphi} - \Gamma^{\ell, p}_{\rho} + M^{\ell, p}_{\rho+\varphi} - M^{\ell, p}_{\rho} \Big)
		\\
		&= \sum_{\ell=1}^{\infty} \sum_{p=1}^{\infty} \bigg( \int_0^{\cdot} (\gamma^{\ell, p}_{\rho+\varphi} - \gamma^{\ell, p}_{\rho})^{\top} \, dO^{\ell, p} + \int_0^{\cdot} (\varphi^{(\ell, p)})^{\top} dM^{\ell, p} \bigg)
		\\
		&= \sum_{\ell=1}^{\infty} \sum_{p=1}^{\infty} \int_0^{\cdot} \mathbbm{1}_{\{\alpha^{\ell, p} \notin \textbf{range}(c^{\ell, p}) \}} dO^{\ell, p}.
		\end{align*}
		Here, the last equality follows from the definition \eqref{def : gamma pi dissection} of the local growth rate, the identities \eqref{eq : identities of varphi}, and \eqref{eq : varphi M zero}. The log-optimal property $\E[\log X^{\rho}_{\rho + \varphi}] \le 0$ now yields that each set $\{\alpha^{\ell, p} \notin \textbf{range}(c^{\ell, p})\}$ should be $(\mathbb{P} \otimes O^{\ell, p})$-null for every $(\ell, p) \in \N^2$.
		
		Next, we recall from \eqref{def : numeraire portfolio dissection} that the process $\nu^{(k, n)} := (c^{k, n})^{\dagger} \alpha^{k, n}$ is a candidate for (a $(k, n)$-dissection of) the num\'eraire portfolio. Let us fix an arbitrary $m \in \N$ and define for every $(k, n) \in \N^2$
		\begin{align*}
		\rho^{(k, n), m} := \rho^{(k, n)} \mathbbm{1}_{\{ c^{k, n}\rho^{(k, n)} = \alpha^{k, n} \}} 
		&+ \rho^{(k, n)} \mathbbm{1}_{\{ c^{k, n}\rho^{(k, n)} \neq \alpha^{k, n}, \, \Vert \nu^{(k, n)} \Vert > m \}} 
		\\
		&+ \nu^{(k, n)} \mathbbm{1}_{\{ c^{k, n}\rho^{(k, n)} \neq \alpha^{k, n}, \, \Vert \nu^{(k, n)} \Vert \le m \}}.
		\end{align*}
		Since the last term is bounded by $m$ and $\rho^{(k, n)}$ is $R^{k, n}$-integrable, the process $\rho^{(k, n), m}$ is also $R^{k, n}$-integrable. We also define a portfolio $\rho^m$ by collecting the dissections
		\begin{equation*}
		\rho^m := 0^{(N_0)} + \sum_{k=1}^{\infty} \sum_{n=1}^{\infty} \hat{\mathbbm{1}}_{\rrbracket \tau_{k-1}, \tau_{k} \rrbracket \cap (\R_+ \times \Omega^{k, n})} \rho^{(k, n), m}.
		\end{equation*}
		In what follows, we shall show that $X^{\rho^m}_{\rho} = \mathcal{E}(R^{\rho^m}_{\rho})$ is a local martingale. 
		From \eqref{def : R rho pi} and \eqref{eq : R rho pi expression}, we have the representation
		\begin{equation}    \label{eq : R rho m rho}
		R^{\rho^m}_{\rho} = \sum_{k=1}^{\infty} \sum_{n=1}^{\infty} \int_0^{\cdot} \sum_{i=0}^n \pi^{(k, n)}_i(s) \, dR^{\rho^m, k, n}_i(s)
		= \sum_{k=1}^{\infty} \sum_{n=1}^{\infty} \Big( R^{k, n}_{\rho-\rho^m} - C^{k, n}_{(\rho-\rho^m)\rho^m} \Big).
		\end{equation}
		Let us fix a pair $(k, n) \in \N^2$. On the set $\Xi^{k, n, m} := \{ c^{k, n}\rho^{(k, n)} \neq \alpha^{k, n}, \, \Vert \nu^{(k, n)} \Vert \le m \}$, we have $\rho^{(k, n), m} = \nu^{(k, n)}$, which implies $c^{k, n}\rho^{(k, n), m} = \alpha^{k, n}$, hence also $A^{k, n}_i = C^{k, n}_{i \rho^m}$ for every $i \in [n]$ from \eqref{eq : dissected structural condition}. Then, the expression $R^{k, n}_i - C^{k, n}_{i \rho^m} = M^{k, n}_i$ is a local martingale for each $i \in [n]$. Moreover, from \eqref{def : R0}, \eqref{def : Ri}, and \eqref{eq : R rho, k, n, 0 expression}, all integrators $R^{\rho^m, k, n}_0, \cdots, R^{\rho^m, k, n}_n$ are local martingales, hence the integral $\int_0^{\cdot} \sum_{i=0}^{\infty} \pi^{(k, n)}_i dR^{\rho^m}_{i}$ is a local martingale. On the complement set $(\Xi^{k, n, m})^c$, we have $\rho^{(k, n), m} = \rho^{(k, n)}$, thus each summand $R^{k, n}_{\rho-\rho^m} - C^{k, n}_{(\rho-\rho^m)\rho^m}$ of \eqref{eq : R rho m rho} vanishes. Therefore, every summand is a local martingale, and Lemma~2.12 of \cite{Strong2} concludes that $R^{\rho^m}_{\rho}$ is also a local martingale. This proves the claim that $X^{\rho^m}_{\rho}$ is a (nonnegative) local martingale, hence also a supermartingale.
		
		If we have $\P[X_{\rho}(T) \neq X_{\rho^m}(T)] > 0$ for some $T>0$, Jensen's inequality and the Optional Sampling Theorem give
		\begin{equation*}
		\E \big[ \log X^{\rho^m}_{\rho}(T) \big] < \log \E \big[ X^{\rho^m}_{\rho}(T) \big] \le 0
		\end{equation*}
		which contradicts the relative log-optimality of $\rho$. Thanks to the continuity of $X_{\rho}$ and $X_{\rho^m}$, we conclude that $X_{\rho} \equiv X_{\rho^m}$, thus $\rho-\rho^m$ is a null portfolio for every $m \in \N$. Then, we have $c^{k, n}\nu^{(k, n)} = \alpha^{k, n} = c^{k, n}\rho^{(k, n), m} = c^{k, n}\rho^{(k, n)}$ on the set $\Xi^{k, n, m}$, which implies that $\Xi^{k, n, m}$ should be a $(\P \times O^{k, n})$-null set for every $m \in \N$. Therefore,
		\begin{equation*}
		\bigcup_{m=1}^{\infty} \Xi^{k, n, m} = \{ c^{k, n}\rho^{(k, n)} \neq \alpha^{k, n} \}
		\end{equation*}
		is also a $(\P \times O^{k, n})$-null set, and $c^{k, n}\rho^{(k, n)} = \alpha^{k, n}$ holds $(\P \times O^{k, n})$-a.e. for each $(k, n) \in \N^2$. The supermartingale num\'eraire property of $\rho$ now follows from the structural condition \eqref{eq : dissected structural condition}.
	\end{proof}
	
	\medskip
	
	\subsection{The optional decomposition theorem}	\label{subsec : ODT}
	
	When the num\'eraire~(either supermartingale or local martingale) exists in the market, the wealth of any portfolio~(or corresponding investment strategy), divided by the num\'eraire, is a local martingale. The following result shows that this local martingale can be represented as a stochastic integral with respect to the $M^{k, n}$ of \eqref{R decomposition}, on each dissection set.
	
	\smallskip
	
	\begin{lem}	\label{lem : X / X rho representation}
		Let $\rho$ be a num\'eraire portfolio. For any given initial capital $x \in \R$ and an investment strategy $\vartheta \in \mathcal{L}_0(S)$ with a notation $X(\cdot) \equiv X(\cdot ; x, \vartheta) = x + \vartheta \cdot S$, there exists a $\U$-valued process $\eta$ such that each $\eta^{(k, n)}$ is $M^{k, n}$-integrable for every $(k, n) \in \N^2$, and the ratio of its wealth process $X$ to the num\'eraire $X_{\rho}$ can be represented as
		\begin{equation}	\label{eq : X / X rho representation}
		\frac{X}{X_{\rho}} = x + \sum_{k=1}^{\infty} \sum_{n=1}^{\infty} \int_0^{\cdot} \sum_{i=1}^n \eta^{(k, n)}_i(u) \, dM^{k, n}_i(u).
		\end{equation}
	\end{lem}
	
	\begin{proof}
		Using \eqref{def : H dissect}, \eqref{def : return process}, and \eqref{R decomposition}, we derive that
		\begin{align*}
		dX(t) &= \sum_{k=1}^{\infty} \sum_{n=1}^{\infty} \sum_{i=1}^n \vartheta^{(k, n)}_i(t) \hat{\mathbbm{1}}_{\rrbracket \tau_{k-1}, \tau_k \rrbracket \cap (\R_+ \times \Omega^{k, n})} \, dS^{k, n}_i(t)
		\\
		&= \sum_{k=1}^{\infty} \sum_{n=1}^{\infty} \sum_{i=1}^n S_i(t) \vartheta^{(k, n)}_i(t) \hat{\mathbbm{1}}_{\rrbracket \tau_{k-1}, \tau_k \rrbracket \cap (\R_+ \times \Omega^{k, n})} \, dR^{k, n}_i(t)
		\\
		&= \sum_{k=1}^{\infty} \sum_{n=1}^{\infty} \sum_{i=1}^n S_i(t) \vartheta^{(k, n)}_i(t) \Big( dA^{k, n}_i(t) + dM^{k, n}_i(t) \Big).
		\end{align*}
		On the other hand, let us consider a portfolio $\pi$ investing all its wealth in the money market at all times, i.e., $\pi^{(k, n)} \equiv 0^{(n)}$ for every $(k, n) \in \N^2$, such that $X_{\pi} \equiv 1$. Lemma~\ref{lem : X rho pi representation} and the identity \eqref{eq : R rho, k, n, 0 expression} show that
		\begin{equation*}
		\frac{1}{X_{\rho}} = \frac{X_{\pi}}{X_{\rho}} = \mathcal{E}(R^{\rho}_{\pi}) 
		= \mathcal{E}\bigg(\sum_{k=1}^{\infty} \sum_{n=1}^{\infty} R^{\rho, k, n}_0\bigg)
		= \mathcal{E}\bigg(-\sum_{k=1}^{\infty} \sum_{n=1}^{\infty} \int_0^{\cdot} \sum_{i=1}^n \rho^{(k, n)}_i(u) \, dM^{k, n}_i(u) \bigg),
		\end{equation*}
		thus
		\begin{equation*}
		d\bigg( \frac{1}{X_{\rho}} \bigg)(t) = - \sum_{k=1}^{\infty} \sum_{n=1}^{\infty} \sum_{i=1}^n \frac{\rho^{(k, n)}_i(t)}{X_{\rho}(t)} \, dM^{k, n}_i(t).
		\end{equation*}
		Applying the product rule, we have
		\begin{align*}
		d\bigg(\frac{X}{X_{\rho}}\bigg)(t) &= \frac{1}{X_{\rho}(t)} \, dX(t) + X(t) \, d\bigg(\frac{1}{X_{\rho}}\bigg)(t) + d \Big[X, \frac{1}{X_{\rho}}\Big](t)
		\\
		&= \sum_{k=1}^{\infty} \sum_{n=1}^{\infty} \Bigg[ \sum_{i=1}^n \frac{S_i(t) \vartheta^{(k, n)}_i(t)}{X_{\rho}(t)} \Big( dA^{k, n}_i(t) + dM^{k, n}_i(t) \Big)
		-\sum_{i=1}^n \frac{X(t)\rho^{(k, n)}_i(t)}{X_{\rho}(t)}\, dM^{k, n}_i(t)
		\\
		& \qquad \qquad \qquad \qquad \qquad \qquad \qquad - \sum_{i=1}^n \sum_{j=1}^n \frac{S_i(t)\vartheta^{(k, n)}_i(t) \rho^{(k, n)}_j(t)}{X_{\rho}(t)} \, d[M^{k, n}_i, M^{k, n}_j](t) \Bigg]
		\\
		&= \sum_{k=1}^{\infty} \sum_{n=1}^{\infty} \sum_{i=1}^n \frac{S_i(t) \vartheta^{(k, n)}_i(t) - X(t) \rho^{(k, n)}_i(t)}{X_{\rho}(t)} \, dM^{k, n}_i(t).
		\end{align*}
		Here, the finite variation terms vanish in the last equality, thanks to the property of the num\'eraire portfolio $\rho$ from Proposition~\ref{prop : numeraire portfolio} $(iii)$. Setting
		\begin{equation*}
		\eta^{(k, n)}_i(t) := \frac{S_i(t) \vartheta^{(k, n)}_i(t) - X(t) \rho^{(k, n)}_i(t)}{X_{\rho}(t)}, \qquad \forall \, (k, n) \in \N^2,
		\end{equation*}
		yields the result \eqref{eq : X / X rho representation}.
	\end{proof}
	
	\smallskip
	
	The converse of Lemma~\ref{lem : X / X rho representation} also holds, i.e., for a given $x \in \R$ and a $\U$-valued process $\eta$ such that each $\eta^{(k, n)}$ is $M^{k,n}$-integrable for every $(k, n) \in \N^2$, we can construct an investment strategy $\vartheta \in \mathcal{L}_0(S)$ satisfying \eqref{eq : X / X rho representation}, by reversing the proof.
	
	The left-hand side of \eqref{eq : X / X rho representation} can be written as $X(1/X_{\rho})$, and the expression $1/X_{\rho}$ is an example of local martingale deflator. Since we know the general form of deflators from Proposition~\ref{prop : form of deflators}, we can expect a generalization of the result by replacing $1/X_{\rho}$ in \eqref{eq : X / X rho representation} by any local martingale deflator. This gives rise to the following so-called optional decomposition Theorem.
	
	\smallskip
	
	\begin{thm}	\label{thm : optional decomposition}
		Suppose that the market is viable. For a nonnegative, adapted RCLL process $X$ with $X(0)= x \ge 0$, the following statements are equivalent:
		\begin{enumerate} [(i)]
			\item For every local martingale deflator $Y \in \mathcal{Y}$, the process $YX$ is a supermartingale.
			\item There exist an investment strategy $\vartheta \in \mathcal{L}_0(S)$ and an adapted, nondecreasing process $K$ satisfying $K(0)=0$ with right-continuous paths, such that
			\begin{equation}	\label{eq : X decomposition}
			X = x + \vartheta \cdot S - K = x + \sum_{k=1}^{\infty} \sum_{n=1}^{\infty} \int_0^{\cdot} \sum_{i=1}^n \vartheta^{(k, n)}_i(t) \, dS^{k, n}_i(t) - K.
			\end{equation}
		\end{enumerate}
	\end{thm}
	
	\begin{proof}
		We first prove the implication $(ii) \Longrightarrow (i)$. From Proposition~\ref{prop : form of deflators}, any local martingale deflator $Y$ admits the representation $\mathcal{E}(L)/X_{\rho}$, where $L \in \mathcal{M}^{\bot}_{loc}(M)$, $\Delta L > -1$, and $\rho$ is the local martingale num\'eraire portfolio. Using the representation from Lemma~\ref{lem : X / X rho representation}, we obtain
		\begin{align*}
		Y(X+K) = Y X(\cdot ; x, \vartheta) = \mathcal{E}(L)\frac{X(\cdot; x, \vartheta)}{X_{\rho}} = \mathcal{E}(L) \bigg( x + \sum_{k=1}^{\infty} \sum_{n=1}^{\infty} \int_0^{\cdot} \sum_{i=1}^n \eta^{(k, n)}_i(u) \, dM^{k, n}_i(u) \bigg).
		\end{align*}
		As a product of two orthogonal local martingales, the last expression is also a local martingale. Thus, $YX$, being a nonnegative local supermatingale, is a supermartingale.
		
		The proof of the reverse implication $(i) \Longrightarrow (ii)$ is divided into 4 parts. Following the argument of Sections~3.1.2, 3.1.3 of \cite{KK2}, we first impose Assumption~\ref{assump : conti} in the first three parts A-C, and then give the proof for the general case in the last part D.
		\begin{assumption}	\label{assump : conti}
			All local martingales on the filtered probability space $(\Omega, \mathcal{F}, (\mathcal{F}_t)_{t \ge 0}, \mathbb{P})$ have continuous paths.
		\end{assumption}
		\noindent
		\textbf{Part A:} Since the market is viable, there exists a local martingale num\'eraire portfolio $\rho$ by Theorem~\ref{thm : fundamental theorem}. From Proposition~\ref{prop : form of deflators}, the process $\mathcal{E}(L)X/X_{\rho}$ is a supermartingale for any element $L \in \mathcal{M}^{\bot}_{loc}(M)$ satisfying $\Delta L > -1$. The ratio $X/X_{\rho}$ is then also a supermartingale, thus the Doob-Meyer decomposition yields
		\begin{equation}	\label{eq : Doob-Meyer decomp}
		\frac{X}{X_{\rho}} = x + V - B,
		\end{equation}
		where $V$ is a local martingale~(which is continuous from Assumption~\ref{assump : conti}) and $B$ is an adapted, nondecreasing process with right-continuous paths, satisfying $V(0) = B(0)= 0$.
		
		\smallskip
		
		\noindent
		\textbf{Part B:} We shall prove two claims. We first show that there exist a collection of processes $\eta^{(k, n)}_i$ for every $(k, n) \in \N^2$, $i = [n]$ and a process $L \in \mathcal{M}^{\bot}_{loc}(M)$ such that each $\eta^{(k, n)}_i$ is $M^{k, n}_i$-integrable and the local martingale $V$ admits a decomposition
		\begin{equation}	\label{eq : decomposition of V}
		V = \sum_{k=1}^{\infty} \sum_{n=1}^{\infty} \int_0^{\cdot} \sum_{i=1}^n \eta^{(k, n)}_i(t) \, dM^{k, n}_i(t) + L = \eta \cdot M + L.
		\end{equation}
		In order to prove this claim, we shall use the Kunita-Watanabe decomposition~(KWD) repeatedly in a manner of `double induction'. First, we take a localizing sequence of stopping times $(\tau_{k})_{k \in \mathbb{N}}$ for the local martingale $V$, and apply KWD to the stopped local martingale $V^{\tau_1}$ such that there exist a $M^{1, 1}_1$-integrable process $\eta^{(1, 1)}_1$ and a local martingale $L^{\{1, 1\}}$ with $L^{\{1, 1\}}(0) = 0$ satisfying the identities
		\begin{equation*}
		V^{\tau_1} = \int_0^{\cdot} \eta^{(1, 1)}_1(t) \, dM^{1, 1}_1(t) + L^{\{1, 1\}}, \qquad [M^{1, 1}_1, \, L^{\{1, 1\}}] \equiv 0.
		\end{equation*}
		We next apply KWD to the local martingale $L^{\{1, 1\}}$ to obtain the identities
		\begin{equation*}
		L^{\{1, 1\}} = \int_0^{\cdot} \sum_{i=1}^2 \eta^{(1, 2)}_i(t) \, dM^{1, 2}_i(t) + L^{\{1, 2\}} = \eta^{(1, 2)} \cdot M^{1, 2} + L^{\{1, 2\}}, \quad [M^{1, 2}_i, \, L^{\{1, 2\}}] \equiv 0 \text{ for } i = 1, 2,
		\end{equation*}
		where $L^{\{1, 2\}}$ is a local martingale and $\eta^{(1, 2)}$ is a $2$-dimensional $M^{1, 2}$-integrable process.	Note that the components of $M^{1, n}$ are flat off the dissection set $\mathrm{diss}^{1, n} := \rrbracket \tau_{0}, \tau_1 \rrbracket \cap (\R_+ \times \Omega^{1, n})$, and each element of $(\mathrm{diss}^{1, n})_{n \in \N}$ is disjoint with each other, we have $[M^{1, 1}_1, \, M^{1, 2}_1] = [M^{1, 1}_1, \, M^{1, 2}_2] \equiv 0$, thus $[M^{1, 1}_1, \, L^{\{1, 2\}}] \equiv 0$ holds. Continuing this procedure, we obtain a sequence of processes $(\eta^{(1, n)})_{n \in \N}$ such that each $\eta^{(1, n)}$ is $n$-dimensional and $M^{1, n}$-integrable, along with a sequence of local martingales $(L^{\{1, n\}})_{n \in \N}$, satisfying for every $p \in \N$
		\begin{equation*}
		V^{\tau_1} = \sum_{n=1}^p \Big(\eta^{(1, n)} \cdot M^{1, n}\Big) + L^{\{1, p\}}, \qquad 
		[M^{1, n}_i, \, L^{\{1, p\}}] \equiv 0, \qquad  \text{for each } n \in [p], ~i \in [n].
		\end{equation*}
		We now define
		\begin{equation*}
		L^{\{1\}} := V^{\tau_1} - \sum_{n=1}^{\infty} \Big( \eta^{(1, n)} \cdot M^{1, n} \Big),
		\end{equation*}
		then it is easy to show that $L^{\{1\}}$ is a well-defined local martingale with the property 
		\begin{equation*}
		[M^{1, n}_i, \, L^{\{1\}}] \equiv 0 \text{ for each } n \in \N, ~i \in [n],
		\end{equation*}
		from the disjoint property of the set $(\mathrm{diss}^{1, n})_{n \in \N}$.
		
		For the second stopping time $\tau_2$, we consider the local martingale $V^{\tau_2} - V^{\tau_1}$ and apply the same argument to obtain the representation
		\begin{equation*}
		V^{\tau_2} - V^{\tau_1} = \sum_{n=1}^{\infty} \Big( \eta^{(2, n)} \cdot M^{2, n} \Big) + L^{\{2\}},
		\end{equation*}
		where $L^{\{2\}}$ is a local martingale satisfying $[M^{2, n}_i, \, L^{\{2\}}] \equiv 0$ for every $n \in \N$, $i \in [n]$, and each $\eta^{(2, n)}$ is an $n$-dimensional $M^{2, n}$-integrable process. Inductively, we obtain for every $k \in \N$
		\begin{equation*}
		V^{\tau_{k}} - V^{\tau_{k-1}} = \sum_{n=1}^{\infty} \Big( \eta^{(k, n)} \cdot M^{k, n} \Big) + L^{\{k\}},
		\end{equation*}
		where each $L^{\{k\}}$ is a local martingale satisfying $[M^{k, n}_i, \, L^{\{k\}}] \equiv 0$ for every $n \in \N$, $i \in [n]$, and each $\eta^{(k, n)}$ is an $n$-dimensional $M^{k, n}$-integrable process. Therefore, we arrive at the desired expression \eqref{eq : decomposition of V}
		\begin{equation*}
		V = \sum_{k=1}^{\infty} \Big( V^{\tau_{k}} - V^{\tau_{k-1}} \Big)
		= \sum_{k=1}^{\infty} \sum_{n=1}^{\infty} \Big( \eta^{(k, n)} \cdot M^{k, n} \Big) + L,
		\end{equation*}
		by setting $L := \sum_{k=1}^{\infty} L^{\{k\}}$. Since $L^{\{k\}}$ is flat off the interval $\rrbracket \tau_{k-1}, \tau_k \rrbracket$ for each $k \in \N$, we have for each fixed $(\ell, n) \in \N^2$ and $i \in [n]$
		\begin{equation*}
		[M^{\ell, n}_i, \, L] = [M^{\ell, n}_i, \, L^{\{\ell\}}] \equiv 0,
		\end{equation*}
		thus $L$ belongs to $\mathcal{M}^{\bot}_{loc}(M)$. This proves the first claim.
		
		We next claim that the process $L$ is actually a zero process, i.e., $L \equiv 0$, so that the identities \eqref{eq : Doob-Meyer decomp}, \eqref{eq : decomposition of V} reduce to
		\begin{equation}	\label{eq : DM decomp 0}
		\frac{X}{X_{\rho}} = x + \eta \cdot M - B.
		\end{equation}
		For a fixed arbitrary $m \in \N$, the process $mL$ belongs to $\mathcal{M}^{\bot}_{loc}(M)$, thus Proposition~\ref{prop : form of deflators} together with Assumption~\ref{assump : conti} implies that $\mathcal{E}(mL)/X_{\rho}$ is a local martingale deflator and $X \mathcal{E}(mL)/X_{\rho}$ is a supermartingale by condition (i). Moreover, since $\mathcal{E}(mL)$ also belongs to $\mathcal{M}^{\bot}_{loc}(M)$, we obtain $[\mathcal{E}(mL), \, \eta \cdot M] \equiv 0$, and $\mathcal{E}(mL) (\eta \cdot M)$ is a local martingale.
		
		Combining \eqref{eq : Doob-Meyer decomp} with \eqref{eq : decomposition of V} and multiplying $\mathcal{E}(mL)$, we derive
		\begin{equation*}
		\mathcal{E}(mL)(L-B) = \frac{X \mathcal{E}(mL)}{X_{\rho}} - \mathcal{E}(mL) \big( x + \eta \cdot M \big)
		\end{equation*}
		and the left-hand side is a local supermartingale from the above observations. By the product rule, the left-hand side can be expressed as
		\begin{equation*}
		\int_0^{\cdot} (L-B)(t-) \, d \mathcal{E}(mL)(t) + \int_0^{\cdot} \mathcal{E}(mL)(t) \, dL(t) + \int_0^{\cdot} \mathcal{E}(mL)(t) \, d\big( [mL, \, L] - B \big)(t).
		\end{equation*}
		The first two integrals are local martingales, so the last integrator $m[L, \, L] - B$ should be a local supermartingale for every $m \in N$. We conclude that $[L, \, L] \equiv 0$, thus $L \equiv 0$.
		
		\smallskip
		
		\noindent
		\textbf{Part C:} The product rule applied to $X = X_{\rho}(X/X_{\rho})$, along with \eqref{def : X pi} and \eqref{eq : DM decomp 0}, yields
		\begin{equation}	\label{eq : X expression}
		X = x + \int_0^{\cdot} X(t-) \, dR_{\rho}(t) - \int_0^{\cdot} X_{\rho}(t) \, dB(t) + \int_0^{\cdot} X_{\rho}(t) \, d(\eta \cdot M)(t) + [X_{\rho}, \, \eta \cdot M].
		\end{equation}
		Here, recalling the notations \eqref{def : C entries} and \eqref{def : C i rho}, the last term of the right-hand side is equal to
		\begin{align*}
		&\sum_{k=1}^{\infty} \sum_{n=1}^{\infty} \bigg[ \int_0^{\cdot} X_{\rho}(t) \, dR_{\rho}(t), \, \int_0^{\cdot} \sum_{i=1}^n \eta^{(k, n)}_i(t) \, dM^{k, n}_i(t) \bigg]
		\\
		&= \sum_{k=1}^{\infty} \sum_{n=1}^{\infty} \int_0^{\cdot} X_{\rho}(t) \sum_{i=1}^n \sum_{j=1}^n \eta^{(k, n)}_i(t) \rho^{(k, n)}_j(t) \, dC^{k, n}_{ij}(t)
		= \sum_{k=1}^{\infty} \sum_{n=1}^{\infty} \int_0^{\cdot} X_{\rho}(t) \sum_{i=1}^n \eta^{(k, n)}_i(t) \, dC^{k, n}_{i\rho}(t),
		\end{align*}
		whereas the second-last term is expressed as
		\begin{equation*}
		\sum_{k=1}^{\infty} \sum_{n=1}^{\infty} \int_0^{\cdot} X_{\rho}(t) \sum_{i=1}^n \eta^{(k, n)}_i(t) \, dM^{k, n}_i(t).
		\end{equation*}
		From the structural condition~(Proposition~\ref{prop : numeraire portfolio} (iii)) of the local martingale num\'eraire portfolio $\rho$, the sum of the last two expressions is
		\begin{equation*}
		\sum_{k=1}^{\infty} \sum_{n=1}^{\infty} \int_0^{\cdot} X_{\rho}(t) \sum_{i=1}^n \eta^{(k, n)}_i(t) \, dR^{k, n}_i(t),
		\end{equation*}
		thus the identity \eqref{eq : X expression} becomes
		\begin{align*}
		X &+ \int_0^{\cdot} X_{\rho}(t) \, dB(t) = x + \sum_{k=1}^{\infty} \sum_{n=1}^{\infty} \int_0^{\cdot} \sum_{i=1}^n \Big( X(t-) \rho^{(k, n)}_i(t) + X_{\rho}(t) \eta^{(k, n)}_i(t) \Big) \, dR^{k, n}_i(t)
		\\
		&= x + \sum_{k=1}^{\infty} \sum_{n=1}^{\infty} \int_0^{\cdot} \sum_{i=1}^n \bigg( \frac{X(t-) \rho^{(k, n)}_i(t) + X_{\rho}(t) \eta^{(k, n)}_i(t)}{S_i(t)} \hat{\mathbbm{1}}_{\rrbracket \tau_{k-1}, \tau_k \rrbracket \cap (\R_+ \times \Omega^{k, n})} + 0^{(1)} \bigg) \, dS^{k, n}_i(t)
		\end{align*}
		with the help of the definition~\eqref{def : return process}. We now construct the investment strategy $\vartheta$ via dissection
		\begin{equation}	\label{def : vartheta ODT}
		\vartheta^{(k, n)}_i := \frac{X(t-) \rho^{(k, n)}_i(t) + X_{\rho}(t) \eta^{(k, n)}_i(t)}{S_i(t)} \hat{\mathbbm{1}}_{\rrbracket \tau_{k-1}, \tau_k \rrbracket \cap (\R_+ \times \Omega^{k, n})} + 0^{(1)}
		\end{equation}
		and setting $K := \int_0^{\cdot} X_{\rho}(t) \, dB(t)$, we arrive at the identity \eqref{eq : X decomposition} of the condition $(ii)$.
		
		\smallskip
		
		\noindent
		\textbf{Part D:} Finally, we provide the proof of $(i) \Longrightarrow (ii)$ without Assumption~\ref{assump : conti}. Since any local martingale can be decomposed (uniquely up to indistinguishability) as the sum of the continuous part and the purely discontinuous part~(Theorem~I.4.18 of \cite{JacodS}), the joint decomposition \eqref{eq : Doob-Meyer decomp}, \eqref{eq : decomposition of V} is transformed into
		\begin{equation}  \label{eq : DM decomp 0 L}
		\frac{X}{X_{\rho}} = x + \eta \cdot M  + L^c + L^d - B.
		\end{equation}
		Here, the processes $\eta$, $B$ have the same properties as before, $L^c$ is a continuous local martingale in $\mathcal{M}^{\bot}_{loc}(M)$, and $L^d$ is a purely discontinuous local martingale, which is orthogonal to every continuous local martingale. We now define $\widetilde{B} := B-L^d$ then $\widetilde{B}$ is again orthogonal to every continuous local martingale. Using the argument of Section~3.1.3 of \cite{KK2}, we can show that $\widetilde{B}$ is a nondecreasing process. From the argument in part B, we have $L^c \equiv 0$, thus the identities \eqref{eq : DM decomp 0} and \eqref{eq : DM decomp 0 L} become
		\begin{equation*}
		\frac{X}{X_{\rho}} = x + \eta \cdot M - \widetilde{B}.
		\end{equation*}
		As in part C, defining $\vartheta$ in \eqref{def : vartheta ODT} and $K := \int_0^{\cdot} X_{\rho}(t) \, d\widetilde{B}(t)$ proves the condition (ii).
	\end{proof}
	
	\smallskip
	
	The name of this decomposition theorem is given from the fact that the process $K$ is optional~(being an adapted, right-continuous process). This version of theorem is close to the formulation of Theorem~3.1 of \cite{KK2}~(or originally from \cite{KK:OD}), in a closed market with fixed number of stocks modeled by continuous semimartingales under a general right-continuous filtration. We refer to \cite{KarouiQ1995}, \cite{Kram_OD}, \cite{FKram_OD}, and \cite{Stricker:Yan} for earlier studies of the optional decomposition.
	
	The following corollary presents the local martingale version of the optional decomposition theorem.
	
	\smallskip
	
	\begin{cor}	\label{cor : optional decomposition}
		Suppose that the market is viable. For a nonnegative, adapted RCLL process $X$ with $X(0)= x \ge 0$, the following statements are equivalent:
		\begin{enumerate} [(i)]
			\item For every local martingale deflator $Y \in \mathcal{Y}$, the process $YX$ is a local martingale.
			\item There exists an investment strategy $\vartheta \in \mathcal{L}_0(S)$ satisfying $X \equiv X(\cdot; x, \vartheta)$, that is,
			\begin{equation*}
			X = x + \vartheta \cdot S = x + \sum_{k=1}^{\infty} \sum_{n=1}^{\infty} \int_0^{\cdot} \sum_{i=1}^n \vartheta^{(k, n)}_i(t) \, dS^{k, n}_i(t).
			\end{equation*}
		\end{enumerate}
	\end{cor}
	
	\begin{proof}
		The implication $(ii) \Longrightarrow (i)$ is easy; as in the proof of Theorem~\ref{thm : optional decomposition}, $YX$ can be represented as the product of two orthogonal local martingales by Proposition~\ref{prop : form of deflators} and Lemma~\ref{lem : X / X rho representation}.
		
		We now assume $(i)$, then $YX$ is also a supermartingale~(being a nonnegative local martingale). From Theorem~\ref{thm : optional decomposition}, there exist $\vartheta \in \mathcal{L}_0(S)$ and an adapted, nondecreasing process $K$ with right-continuous paths satisfying \eqref{eq : X decomposition}. In what follows, we prove $K \equiv 0$. Let us choose $Y = 1/X_{\rho}$, the reciprocal of the local martingale num\'eraire, then 
		\begin{equation*}
		YX = \frac{X}{X_{\rho}} = \frac{X(\cdot; x, \vartheta)}{X_{\rho}} - \frac{K}{X_{\rho}}
		\end{equation*}
		is a local martingale. Moreover, we know from Lemma~\ref{lem : X / X rho representation} that $X(\cdot; x, \vartheta)/X_{\rho}$ is also a local martingale. Thus, $K/X_{\rho}$ is a local martingale and the product rule gives
		\begin{equation*}
		\frac{K}{X_{\rho}} = \int_0^{\cdot} K(t-) \, d\bigg(\frac{1}{X_{\rho}}\bigg)(t) + \int_0^{\cdot} \frac{1}{X_{\rho}(t)} \, dK(t).
		\end{equation*}
		Since the integrator of the first integral on the right-hand side is a local martingale, the last integral is also a local martingale. Being a nonnegative local martingale, the last term is also a supermartingale, and nondecreasing. This shows $K \equiv 0$, since the integrand $1/X_{\rho}$ is strictly positive.
	\end{proof}

	\medskip
	
	\subsection{Open market embedded in a market of stochastic dimension}	\label{subsec : open markets}
	
	The concept of \textit{open market} was introduced in \cite{Fernholz:2018}. When the equity market consists of a fixed number $M \in \N$ of stocks, the investors are allowed to invest only in the top $m$ capitalization stocks at all times for $m \le M$. The fundamental result of arbitrage theory, in the sense of Theorem~\ref{thm : fundamental theorem}, is then proven in \cite{Karatzas:Kim2} for this top $m$ open market. More recently, some empirical examples of numerically optimized portfolios under open market constraints are given in \cite{Campbell:Wong}, and the work of \cite{Itkin:Open_market} presents a parametric family of market weight models in a slightly generalized setting of an open market. We explore in this subsection the notion of open markets in the setting of investing universe with a stochastic dimension.
	
	Throughout this subsection, we shall fix a positive integer $m \in \N$ for the size of the top open market. One may impose an assumption that the number of assets in the market is always bigger than or equal to $m$, i.e., $N = \dim(S) \ge m$. In this case, only the $(k, n)$-dissections for dimension $n \ge m$ of $\U$-valued processes would be necessary. Without this assumption, when the number of stocks in the market is less than $m$, the top open market boils down to the usual \textit{closed} market of the previous subsections, and the same theory can be applied. Therefore, we shall not impose any assumption in the following on $m$, the size of the open market.
	
	For any $n$-dimensional vector $v = (v_1, \cdots, v_n)$ for every $n \in \N$, we define the $k$-th ranked component $v_{(k)}$ by
	\begin{equation}	\label{def : ranked component}
	v_{(k)} := \max_{1 \le i_1 \cdots \le i_k \le n} \min \{ v_{i_1}, \cdots, v_{i_k}\},
	\end{equation}
	satisfying
	\begin{equation*}
	\max_{i=1, \cdots, n} v_i = v_{(1)} \ge v_{(2)} \ge \cdots \ge v_{(n)} = \min_{i=1, \cdots, n} v_i.
	\end{equation*}
	We shall use a lexicographic rule for breaking ties that always assigns a higher rank~(smaller $(k)$) to a smaller index $i$.
	
	Moreover, for any $(k, n) \in \N^2$, we recall the notation $[n] := \{1, \cdots, n\}$, and define a process $[n] \times [0, \infty) \ni (i, t) \mapsto u^{k, n}_i(t) \in [n]$ such that each $u^{k, n}_i(\cdot)$ is predictable and satisfies
	\begin{align}	
	S_i(t) &= S_{\big(u^{k, n}_i(t)\big)}(t), \qquad \text{on the dissection set } \rrbracket \tau_{k-1}, \tau_k \rrbracket \cap (\R_+ \times \Omega^{k, n}),		\nonumber
	\\
	u^{k, n}_i(t) &= i, \qquad \qquad \qquad ~~  \text{otherwise},	\label{def : u k, n}
	\end{align}
	for every $i \in [n]$. In other words, on the dissection set, $u^{k, n}_i(t)$ represents the rank of the $i$-th stock in terms of capitalization among $n$ stocks at time $t$. Since $u^{k, n}_i$ shall act only on the dissection set, the choice $u^{k, n}_i \equiv i$ on the complement set, is not important. Here, both index and rank of the stocks may shift due to the dimensional change~(at each reset sequence $\tau_k$), and we assume that appropriate relabeling of index~(and corresponding rank) is performed whenever necessary to inherit each company's dynamics after every dimensional change.  
	
	\smallskip
	
	\begin{defn} [Censored return process] \label{Def: censored R}
		For a given return process $R$ of Definition~\ref{Def : return process}, we define \textit{censored return process} $\widetilde{R}$ via dissection
		\begin{equation*}
		\widetilde{R}^{k, n}_i(t) := \int_0^t \mathbbm{1}_{\{u^{k, n}_i(s) \le m\}} dR^{k, n}_i(s), \qquad i \in [n], \quad k, n \in \N, \quad t \ge 0.
		\end{equation*}
	\end{defn}
	The process $\widetilde{R}^{k, n}_i(t)$ represents the cumulative return of the $i$-th stock of the $(k, n)$-dissected market, accumulated over $[0, t]$ only when this stock ranks among the top $m$ by capitalization, out of $n$ companies present in the market. When $n < m$, i.e., the number of stocks existent in the market is less than the size $m$ of the open market, we note the identity $\widetilde{R}^{k, n} \equiv R^{k, n}$.
	
	Imposing an additional condition that restricts investing in the $i$-th stock, whenever the rank of the stock is bigger than $m$, we have the following definitions of portfolio and investment strategy in the top $m$ open market.
	
	\smallskip
	
	\begin{defn} [Portfolio and investment strategy in the top $m$ open market]    \label{Def: portfolio in open market}
		A portfolio $\pi \in \mathcal{L}(R)$ is called a \textit{portfolio among the top $m$ stocks}, if it satisfies for each $(k, n) \in \N^2$ and $i \in [n]$
		\begin{equation}	\label{def : top m portfolio}
		\pi^{(k, n)}_i(t) \mathbbm{1}_{\{ u^{k, n}_i(t) > m \}} = 0.
		\end{equation}
		We denote $\mathcal{L}(R) \cap \mathcal{T}(m)$ the collection of portfolios among the top $m$ stocks. Similarly, an \textit{investment strategy among the top $m$ stocks} is an investment strategy $\vartheta \in \mathcal{L}_0(S)$ satisfying
		\begin{equation*}
		\vartheta^{(k, n)}_i(t) \mathbbm{1}_{\{ u^{k, n}_i(t) > m \}} = 0,
		\end{equation*}
		and we denote $\mathcal{L}_0(S) \cap \mathcal{T}(m)$ the collection of investment strategies among the top $m$ stocks.
	\end{defn}
	
	We note that the condition \eqref{def : top m portfolio} is equivalent to $\pi^{(k, n)}_i(t) \mathbbm{1}_{\{ u^{k, n}_i(t) \le m \}} = \pi^{(k, n)}_i(t)$. From \eqref{def : X pi} and \eqref{def : R pi}, we have the similar representation of the wealth $X_{\pi}$ of $\pi \in \mathcal{L}(R) \cap \mathcal{T}(m)$
	\begin{align}
	X_{\pi} = \mathcal{E}(R_{\pi}) 
	&= \mathcal{E}\bigg(\sum_{k=1}^{\infty} \sum_{n=1}^{\infty} R^{k, n}_{\pi} \bigg), 	\nonumber
	\\
	R^{k, n}_{\pi} = \sum_{i=1}^n \int_0^{\cdot} \pi^{(k, n)}_i(s) \, dR^{k, n}_i(s) 
	&= \sum_{i=1}^n \int_0^{\cdot} \pi^{(k, n)}_i(s) \mathbbm{1}_{\{ u^{k, n}_i(s) \le m \}} \, dR^{k, n}_i(s)	\nonumber
	\\
	&= \sum_{i=1}^n \int_0^{\cdot} \pi^{(k, n)}_i(s) \, d\widetilde{R}^{k, n}_i(s) = \pi^{(k, n)} \cdot \widetilde{R}^{k, n}	\label{eq : R tilde representation}
	\end{align}
	where $R^{k, n}$ is replaced by $\widetilde{R}^{k, n}$.
	
	In what follows, we define censored version of the other processes from Sections~\ref{subsec : basic processes}-\ref{subsec : structural condition}, which are relevant for the open market. We consider the semimartingale decomposition $\widetilde{R}^{k, n}_i := \widetilde{A}^{k, n}_i + \widetilde{M}^{k, n}_i$ and define the cumulative covariation and the local rates
	\begin{equation*}
	\widetilde{C}^{k, n}_{i, j} := [\widetilde{M}^{k, n}_i, \widetilde{M}^{k, n}_j],
	\end{equation*}
	\begin{equation*}
	\widetilde{\alpha}^{k, n}_i := \mathbbm{1}_{\{ u^{k, n}_i \le m \}} \alpha^{k, n}_i, \qquad
	\widetilde{c}^{k, n}_{i, j} := \mathbbm{1}_{\{ u^{k, n}_i \le m \}} \mathbbm{1}_{\{ u^{k, n}_j \le m \}} c^{k, n}_{i, j}
	\end{equation*}
	for each $(i, j) \in [n]^2$ and $(k, n) \in \N^2$~(cf. \eqref{def : C entries}, \eqref{def : alpha, c}). Introducing the diagonal-matrix-valued predictable process $D^{k, n} \equiv (D^{k, n}_{i, j})_{i, j \in [n]}$ with entries
	\begin{equation*}
	D^{k, n}_{i, j}(t) := 
	\begin{cases}
	\mathbbm{1}_{\{ u^{k, n}_i(t) \le m \}}, \qquad &i \neq j,
	\\
	\qquad 0, \qquad &i = j,
	\end{cases}
	\qquad \qquad t \ge 0,
	\end{equation*}
	for each pair $(k, n) \in \N^2$, we have the alternative matrix form
	\begin{equation*}
	d\widetilde{C}^{k, n}(t) = D^{k, n}(t) \, dC^{k, n}(t) \, D^{k, n}(t), \qquad \widetilde{\alpha}^{k, n} = D^{k, n}\alpha^{k, n}, \qquad \widetilde{c}^{k, n} = D^{k, n} c^{k, n} D^{k, n}.
	\end{equation*}
	As in \eqref{eq : R tilde representation}, these definitions give rise to the following representations
	\begin{align*}
	\widetilde{A}^{k, n}_i = \int_0^{\cdot} \mathbbm{1}_{\{ u^{k, n}_i(t) \le m \}} dA^{k, n}_i(t) 
	&= \int_0^{\cdot} \widetilde{\alpha}^{k, n}_i(t) \, dO^{k, n}(t),
	\qquad
	\widetilde{C}^{k, n}_{i, j} = \int_0^{\cdot} \widetilde{c}^{k, n}_{i, j}(t) \, dO^{k, n}(t),
	\\
	dC^{k, n}_{\pi \rho}(t) &= d[R^{k, n}_{\pi}, R^{k, n}_{\rho}](t) = \pi^{(k, n)}(t) \, d\widetilde{C}^{k, n}(t) \, \rho^{(k, n)}(t),
	\end{align*}
	for every $(k, n) \in \N^2$ and $\pi, \rho \in \mathcal{L}(R) \cap \mathcal{T}(m)$~(cf. \eqref{def : alpha, c}, \eqref{def : C pi rho dissection}). Moreover, we denote
	\begin{equation}	\label{def : tilde C i rho}
	\widetilde{C}^{k, n}_{i \rho} := [ \widetilde{R}^{k, n}_i, R^{k, n}_{\rho} ],
	\end{equation}
	for any $\rho \in \mathcal{L}(R) \cap \mathcal{T}(m)$~(cf. \eqref{def : C i rho}). 
	
	Combining the argument of \cite{Karatzas:Kim2} and of the previous subsections, we can derive analogous results, ultimately the fundamental theorem for the top $m$ open market without much effort, by replacing the regular symbols with the corresponding tilde~(censored) symbols. We illustrate such results in the following under the framework of the open market.
	
	\smallskip
	
	\begin{defn} [Num\'eraire portfolio in the top $m$ open market]	\label{Def : numeraire portfolio open market}
		A portfolio $\rho \in \mathcal{L}(R) \cap \mathcal{T}(m)$ is called \textit{supermartingale~(local martingale) num\'eraire portfolio among the top $m$ stocks}, if the relative wealth process $X^{\rho}_{\pi}$ is a supermartingale~(local martingale, respectively) for every portfolio $\pi \in \mathcal{L}(R) \cap \mathcal{T}(m)$ among the top $m$ stocks.
	\end{defn}
	
	\smallskip
	
	\begin{prop}	\label{prop : numeraire portfolio open market}
		For a portfolio $\rho \in \mathcal{L}(R) \cap \mathcal{T}(m)$ among the top $m$ stocks, the following statements are equivalent:
		\begin{enumerate} [(i)]
			\item $\rho$ is a supermartingale num\'eraire portfolio among the top $m$ stocks.
			\item $\rho$ is a local martingale num\'eraire portfolio ampng the top $m$ stocks.
			\item $\widetilde{A}^{k, n}_i = \widetilde{C}^{k, n}_{i \rho}$ holds for every $i \in [n]$ and $k, n \in \N$. 
		\end{enumerate}
	\end{prop}
	
	The statement \eqref{eq : X rho pi} of Lemma~\ref{lem : X rho pi representation} remains true for $\rho, \pi \in \mathcal{L}(R) \cap \mathcal{T}(m)$, if we replace the symbols in the parenthesis of \eqref{def : Ri} with the censored dissected return $\widetilde{R}^{k, n}$ and the expression $\widetilde{C}^{k, n}_{i \rho}$ of \eqref{def : tilde C i rho}. For the proof of Proposition~\ref{prop : numeraire portfolio open market}, we define a new portfolio $\widetilde{\nu}_{(i, j)}$ among the top $m$ stocks for every fixed $i \le j$, defined via dissection from \eqref{def : nu i j}
	\begin{equation*}
	\widetilde{\nu}^{(k, n)}_{(i, j)}(\cdot) := \nu^{(k, n)}_{(i, j)}(\cdot) \mathbbm{1}_{\{u_i(\cdot) \le m\}}.
	\end{equation*}
	Using the aforementioned tilde symbols instead of the regular ones in the proof of Proposition~\ref{prop : numeraire portfolio} establishes Proposition~\ref{prop : numeraire portfolio open market}.
	
	Since the condition (iii) of Proposition~\ref{prop : numeraire portfolio open market} can be reformulated in terms of local rates~(cf. \eqref{eq : dissected structural condition})
	\begin{equation}
	\widetilde{\alpha}^{k, n} = \widetilde{c}^{k, n} \rho^{(k, n)}, \qquad (\mathbb{P} \otimes O^{k, n})- \text{a.e.},
	\end{equation}
	we expect the similar result of Proposition~\ref{prop : structural conditions} for the top $m$ open market. We first define the \textit{maximal growth rate}
	\begin{equation}	\label{def : maximal growth rate open market}
	\widetilde{g}^{k, n} := \sup_{p \in \R^n \cap \mathcal{T}(m)} \Big( (\widetilde{\alpha}^{k, n})^{\top}p - \frac{1}{2} p^{\top}\widetilde{c}^{k, n}p \Big),
	\end{equation}
	achievable in the $(k, n)$-dissected, top $m$ open market~(cf. \eqref{def : maximal growth rate}). Here, $\R^n \cap \mathcal{T}(m)$ denotes the subset of elements in $\R^n$ such that at most $m$ components are nonzero. Applying the same computational technique used in Section~3.3 of \cite{Karatzas:Kim2} to the tilde symbols of each dissected market, we can derive the following version of Proposition~\ref{prop : structural conditions}.
	
	\smallskip
	
	\begin{prop}	\label{prop : structural conditions open market}
		For a fixed, arbitrary pair $(k, n) \in \N^2$, the following statements are equivalent:
		\begin{enumerate} [(i)]
			\item There exists an $n$-dimensional $\widetilde{R}^{k, n}$-integrable process $\rho^{(k, n)}$ having values in $\R^n \cap \mathcal{T}(m)$ with $\rho^{(k, n)}(0) = 0^{(n)}$ such that $\widetilde{\alpha}^{k, n} = \widetilde{c}^{k, n} \rho^{(k, n)}$ holds $(\mathbb{P} \otimes O^{k, n})$-a.e.
			\item $\widetilde{\alpha}^{k, n} \in \textbf{range}(\widetilde{c}^{k, n})$ for $(\mathbb{P} \otimes O^{k, n})$-a.e., and $\int_0^T \big(\widetilde{\alpha}^{k, n}(t)\big)^{\top} \big(\widetilde{c}^{k, n}(t)\big)^{\dagger}\widetilde{\alpha}^{k, n}(t) \, dO^{k, n}(t) < \infty$ for every $T \ge 0$.
			\item The $(k, n)$-dissected top $m$ open market has locally finite growth, that means, for every $T \ge 0$, we have $\widetilde{G}^{k, n}(T) := \int_0^T \widetilde{g}^{k, n}(t) \, dO^{k, n}(t) < \infty$.
		\end{enumerate}
	\end{prop}
	
	Again, the process given by
	\begin{equation}	\label{def : numeraire portfolio dissection open market}
	\rho^{(k, n)} := (\widetilde{c}^{k, n})^{\dagger}\widetilde{\alpha}^{k, n}, \qquad \text{(modulo null portfolio)}	
	\end{equation}
	satisfies the conditions $(i)-(iii)$ of Proposition~\ref{prop : structural conditions open market}, and $\rho^{(k, n)}$ attains the maximal growth rate $\widetilde{g}^{k, n}$ of \eqref{def : maximal growth rate open market}, which is equal to $(1/2) (\widetilde{\alpha}^{k, n})^{\top} (\widetilde{c}^{k, n})^{\dagger} \widetilde{\alpha}^{k, n}$. Then, we have the similar representation of $\widetilde{g}^{k, n}$ as in \eqref{eq : g dissection} and the supermartingale num\'eraire portfolio $\rho$ among the top $m$ stocks, can be constructed from dissection $\rho^{(k, n)}$ of \eqref{def : numeraire portfolio dissection open market}.
	
	We finally present the associated definitions and results of Section~\ref{subsec : NA1} in the context of top $m$ open markets.
	
	\smallskip
	
	\begin{defn} [Arbitrage of the first kind in the top $m$ open market]	\label{Def : NA1 open market}
		An \textit{arbitrage of the first kind} for horizon $T > 0$ in the top $m$ open market is an $\mathcal{F}_{T}$-measurable random variable $h$ satisfying $\mathbb{P}(h \ge 0) = 1$, $\mathbb{P}(h > 0) > 0$, such that for every $x > 0$, there exists an admissible investment strategy $\vartheta$ among the top $m$ stocks, satisfying $X(\cdot; x, \vartheta) \ge 0$ and $X(T; x, \vartheta) \ge h$. If there are no arbitrages of the first kind, we say $\textit{NA}_1$ holds in the top $m$ open market.
	\end{defn}
	
	\smallskip
	
	\begin{defn} [Local martingale deflator in the top $m$ open market]	\label{Def : deflator open market}
		An adapted, positive process $Y$ is called \textit{local martingale deflator among the top $m$ stocks}, if $Y(0)=1$ and $YX_{\pi}$ is a local martingale for every portfolio $\pi \in \mathcal{L}(R) \cap \mathcal{T}(m)$ among the top $m$ stocks.
	\end{defn}
	
	\smallskip
	
	\begin{defn} [Market viability of the top $m$ open market]	\label{Def : viability open market}
		For a given nondecreasing, adapted, and right-continuous process $K$ with $K(0) = 0$, we define \textit{financing capital} associated with $K$ in the top $m$ open market by
		\begin{equation}
		x^m(K) := \inf_{x \ge 0} \{ \exists \, \vartheta \in \mathcal{L}_0(S)\cap\mathcal{T}(m) \text{ such that } X(\cdot ; x, \vartheta) \ge K\}. 
		\end{equation}
		We say that the top $m$ open market is \textit{viable}, if $x^m(K) = 0$ implies $K \equiv 0$.
	\end{defn}
	
	Proposition~\ref{prop : NA1} can be immediately extended to the top $m$ open market using the above Definitions~\ref{Def : NA1 open market}-\ref{Def : viability open market}, when the condition \eqref{con : bounded in prob} is substituted for
	\begin{equation*}
	\lim_{m \rightarrow \infty} \sup_{\pi \in \mathcal{L}(R) \cap \mathcal{T}(m)} \mathbb{P} \big[ X_{\pi}(T) > m \big] = 0 \text{ holds for any } T \ge 0.
	\end{equation*}
	
	Therefore, we present the fundamental theorem for the top $m$ open market.
	
	\smallskip
	
	\begin{thm}	\label{thm : fundamental theorem open market}
		The following statements are equivalent:
		\begin{enumerate} [(i)]
			\item There exists a supermartingale~(also a local martingale) num\'eraire portfolio among the top $m$ stocks.
			\item There exists a portfolio $\rho$ satisfying $\widetilde{A}^{k, n}_i = \widetilde{C}^{k, n}_{i\rho}$ for every $i \in [n]$ and $(k, n) \in \N^2$. 
			\item Each $(k, n)$-dissected top $m$ open market has locally finite growth, i.e., $\widetilde{G}^{k, n}(T) < \infty$ for every $T \ge 0$ and $(k, n) \in \N^2$.
			\item $\text{NA}_1$ holds in the top $m$ open market.
			\item There exists a local martingale deflator among the top $m$ stocks.
			\item The top $m$ open market is viable.
		\end{enumerate}
	\end{thm}
	
	It is sufficient to prove the implication $(vi) \Longrightarrow (iii)$, but the same proof of Theorem~\ref{thm : fundamental theorem} can be used by replacing the symbols $\alpha^{\ell, p}, c^{\ell, p}, R^{\ell, p}, A^{\ell, p}$, and $M^{\ell, p}$ with the corresponding tilde symbols.
	
	\smallskip
	
	We finally present the optional decomposition theorem in the framework of open market. First, let us denote $\mathcal{Y}^m$ the collection of local martingale deflators among the top $m$ stocks, and $\mathcal{M}^{\bot}_{loc}(\widetilde{M})$ the collection of scalar local martingales $L$ with RCLL paths, satisfying $L(0)=0$ and the strong orthogonality condition $[L, \widetilde{M}^{k, n}_i] = 0$ for every $(k, n) \in \N^2$ and $i \in [n]$. It is then easy to modify the proof of Proposition~\ref{prop : form of deflators} to derive the characterization
	\begin{equation}	\label{eq : Ym}
	\mathcal{Y}^m = \bigg\{ \frac{\mathcal{E}(L)}{X_{\rho}} \, \Big\vert \, L \in \mathcal{M}^{\bot}_{loc}(\widetilde{M}), \quad \Delta L > -1 \bigg\}.
	\end{equation}
	
	Moreover, we have the following version of Lemma~\ref{lem : X / X rho representation}. Denoting $\rho$ the num\'eraire portfolio among the top $m$ stocks and $X(\cdot) \equiv X(\cdot; x, \vartheta)$ for any given $x \in \R$ and investment strategy $\vartheta \in \mathcal{L}_0(S) \cap \mathcal{T}(m)$ among the top $m$ stocks, there exists a $\U$-valued process $\eta$ such that each $\eta^{(k, n)}$ is $\widetilde{M}^{k, n}$-integrable, $\eta^{(k, n)}_i(\cdot) \mathbbm{1}_{\{ u^{k, n}_i(\cdot) \le m \}} = \eta^{(k, n)}_i(\cdot)$ holds for every $(k, n) \in \N^2$, and satisfies
	\begin{equation}	\label{eq : X / X rho representation2}
	\frac{X}{X_{\rho}} = x + \sum_{k=1}^{\infty} \sum_{n=1}^{\infty} \int_0^{\cdot} \sum_{i=1}^n \eta^{(k, n)}_i(u) \, d\widetilde{M}^{k, n}_i(u).
	\end{equation}
	Indeed, defining
	\begin{equation*}
	\eta^{(k, n)}_i(t) := \frac{S_i(t) \vartheta^{(k, n)}_i(t) - X(t) \rho^{(k, n)}_i(t)}{X_{\rho}(t)}, \qquad \forall \, (k, n) \in \N^2,
	\end{equation*}
	as in the proof of Lemma~\ref{lem : X / X rho representation} shows the existence of such process $\eta$.
	
	Now that we have the representations \eqref{eq : Ym} and \eqref{eq : X / X rho representation2} in hand, the following optional decomposition results can be obtained by the same arguments in Section~\ref{subsec : ODT} with the tilde symbols replacing the regular ones.
	
	\smallskip
	
	\begin{thm}	\label{thm : optional decomposition open market}
		Assume that the top $m$ open market is viable. For a nonnegative, adapted process $X$ with $X(0)= x \ge 0$, the following statements are equivalent:
		\begin{enumerate} [(i)]
			\item For every local martingale deflator $Y \in \mathcal{Y}^m$ among the top $m$ stocks, the process $YX$ is a supermartingale.
			\item There exist an investment strategy $\vartheta \in \mathcal{L}_0(S) \cap \mathcal{T}(m)$ among the top $m$ stocks and an adapted, nondecreasing process $K$ satisfying $K(0)=0$ with right-continuous paths, such that $X = x + \vartheta \cdot S - K$ holds.
		\end{enumerate}
		Furthermore, the following statements are also equivalent:
		\begin{enumerate}
			\item [(iii)] For every local martingale deflator $Y \in \mathcal{Y}^m$ among the top $m$ stocks, the process $YX$ is a local martingale.
			\item [(iv)] There exists an investment strategy $\vartheta \in \mathcal{L}_0(S) \cap \mathcal{T}(m)$ among the top $m$ stocks satisfying $X \equiv X(\cdot; x, \vartheta)$, that is, $X = x + \vartheta \cdot S$.
		\end{enumerate}
	\end{thm}

        \smallskip

        We conclude this subsection with the following remark.
        
        \begin{rem}
            In this subsection (and in the work of \cite{Karatzas:Kim2}), the essential idea of handling the open market is to censor the return process by the event $\{u^{k, n}_i(s) \le m\}$ in Definition~\ref{Def: censored R} to keep the dimension of the open market equal to $m$. However, this censoring technique also works for any $\mathcal{F}_s$-measurable event $E_i(s)$, in other words, we can consider instead
            \begin{equation*}
		      \widetilde{R}^{k, n}_i(t) := \int_0^t \mathbbm{1}_{E_i(s)} dR^{k, n}_i(s), \qquad i \in [n], \quad k, n \in \N, \quad t \ge 0,
		  \end{equation*}
            in Definition \ref{Def: censored R}. This process $\widetilde{R}^{k, n}_i$ now represents the cumulative return of the $i$-th stock of the $(k, n)$-dissected market, censored according to the (collection of) events $\{E_i(s)\}_{s \ge 0}$. Then, the corresponding portfolio censored by the events $\{E_i(s)\}_{s \ge 0}$ should satisfy
            \begin{equation*}	
		      \pi^{(k, n)}_i(t) \mathbbm{1}_{(E_i(s))^c} = 0,
		  \end{equation*}
            instead of \eqref{def : top m portfolio} in Definition~\ref{Def: portfolio in open market}. If we modify all the other definitions accordingly (replacing $\{u^{k, n}_i(s) \le m\}$ with $E_i(s)$), all results in this subsection~(Theorems \ref{thm : fundamental theorem open market} and \ref{thm : optional decomposition open market}) remain to hold.

            Now that the dimension of the entire market changes over time, we can also consider the open market of stochastic dimension. For example, instead of choosing $m$ largest stocks by their rank at all times, we can select large stocks whose capitalization exceeds a certain threshold $\epsilon > 0$ by setting $E_i(s) = \{S_i(s) \ge \epsilon\}$ to construct another open market composed of large capitalization stocks. Then, this open market has a dimension process equal to $M(t):= |\{i : S_i(t) \ge \epsilon\}|$, and the portfolios in this open market can only invest in the stocks with capitalization greater than or equal to $\epsilon$ at all times. 
        \end{rem}
 
	\bigskip
	
	\section{Market with general price process}	\label{sec : RCLL price}

	We now relax the assumption on the continuity and strict positivity of the price process between the dimensional changes. Precisely, we shall consider a piecewise RCLL semimartingale $S$ to represent the price process of assets in the market. This general setting requires different proof techniques for the results presented in the previous section.
	
	\medskip
	
	\subsection{Preliminaries}  \label{subsec : preliminaries RCLL}
	
	First, we give a more general definition of the price process than the one given in Definition~\ref{Def : price process}. Some preliminary results, which will be frequently used in the proofs in the following subsections, are then presented.
	
	\smallskip
	
	\begin{defn} [Price process]	\label{Def : price process RCLL}
		A $\U$-valued process $S$ is called a \textit{price process}, if it is a piecewise RCLL semimartingale with a reset sequence $(\tau_k)_{k \ge 0}$.
	\end{defn}
	
	In Section \ref{sec : continuous price}, we considered a concept of investment strategy $\vartheta$ and its corresponding portfolio $\pi$ such that they are connected via the equations \eqref{def : corresponding pi dissection} and \eqref{def : corresponding vartheta dissection}. Their wealth process $X_\pi(\cdot) \equiv X(\cdot;1, \vartheta)$ in \eqref{eq : corresponding wealth of portfolio} is a strictly positive process as a stochastic exponential and $\vartheta$ has to be strictly admissible in order to define the portfolio $\pi$. However, since the concept of the portfolio is no longer needed in this section, we can allow the wealth process $X(\cdot;x, \vartheta)$ of an investment strategy $\vartheta$ to hit zero at some time. From this perspective, we do not assume strict positivity, or even nonnegativity, of the price process $S$ in Definition~\ref{Def : price process RCLL}. Since we shall only consider admissible investment strategies as in the following definition, only the nonnegative condition on the wealth process is important, not on the price process. We refer to \cite[Section~4.8]{KK} for a more detailed remark regarding the positivity condition of the price process.
	
	\smallskip
	
	\begin{defn} [Investment strategy] \label{Def : general investment strategy}
		For a given price process $S$ of Definition~\ref{Def : price process RCLL}, an investment strategy and its wealth process are defined as in Definition~\ref{Def: investment strategy}. The collection of all wealth processes of admissible investment strategies $\vartheta$ with initial capital $x > 0$ is denoted by
		\begin{equation*}
		\mathcal{W} := \{ X \equiv X(\cdot; x, \vartheta) \, \vert \, \vartheta \in \mathcal{L}_0(S), ~ x > 0, ~ X \geq 0\}.
		\end{equation*}
		We denote the subset $\mathcal{X} \subset \mathcal{W}$ of all wealth processes of strictly admissible investment strategies
		\begin{equation*}
		\mathcal{X} := \{X \in \mathcal{W} : X > 0\}.
		\end{equation*}
	\end{defn}
	
	We also need to modify the definition of local martingale deflators accordingly.
	
	\smallskip
	
	\begin{defn} [Local martingale deflator]  \label{Def : general deflator}
		An adapted strictly positive process $Y$ satisfying $Y(0) = 1$ is called \textit{local martingale deflator}, if $YX$ is a local martingale for every $X \in \mathcal{W}$. We denote $\mathcal{Y}$ the set of all local martingale deflators. Moreover, for a given price process $S$ and any $(k, n) \in \N^2$, an adapted, positive process $Y$ satisfying $Y(0)=1$ is called \textit{$(k, n)$-local martingale deflator}, if the product $Y(x + \vartheta \cdot S^{k,n})$ is a local martingale for every $n$-dimensional $S^{k,n}$-integrable process $\vartheta$ and $x>0$, satisfying $x + \vartheta \cdot S^{k,n} \geq 0$. We denote $\mathcal{Y}^{k, n}$ the set of all $(k, n)$-local martingale deflators for every $(k, n) \in \N^2$.
	\end{defn}
	
	Observe that $\mathcal{Y} \subset \mathcal{Y}^{k,n}$ for every $(k,n) \in \N^2$, since for an arbitrary $n$-dimensional $S^{k,n}$-integrable process $\vartheta$, we can construct a $\U$-valued process $\vartheta'$ by setting $\vartheta' := \vartheta$ on the $(k, n)$-dissection set and $\vartheta' := 0^{(m)}$ on any other $(\ell, m)$-dissection sets for $(\ell, m) \neq (k, n)$ such that $\vartheta' \cdot S = \vartheta \cdot S^{k,n}$ holds in view of the definition of stochastic integral in \eqref{def : stochastic integral}. We also note that any local martingale deflator $Y$ and any $(k, n)$-local martingale deflator $Y^{k, n}$ are themselves local martingales by taking $\vartheta(\cdot) \equiv 0^{(N_{\cdot})}$~(a $\U$-valued zero vector process with the same dimension process as $S$). 
	
	Furthermore, for any $(k, n) \in \N^2$, we note from Proposition~2.5 of \cite{Karatzas:Ruf:2017} that there is an equivalent, but seemingly weaker characterization
	\begin{equation}    \label{con : equivalent local martingale deflator}
	\mathcal{Y}^{k, n} = \mathcal{Z}(S^{k, n})
	\end{equation}
	of the set $\mathcal{Y}^{k, n}$ in Definition~\ref{Def : general deflator}, where $\mathcal{Z}(S^{k, n})$ is the collection of all positive local martingales $Z$ with $Z(0) = 1$ such that $ZS^{k, n}$ is a local martingale.
	
	We end this subsection by providing some lemmas.
	
	\smallskip
	
	\begin{lem} \label{lem : newlocsup}
		Let $Z$ be a local supermartingale (resp. local martingale) and $\tau$ a stopping time such that $Z = 0$ on $\llbracket 0, \tau \rrbracket$. If $X$ is an adapted process, then $X^{\tau}Z$ is a local supermartingale (resp. local martingale).
	\end{lem}
	
	\begin{proof}
		By the Doob-Meyer decomposition, we have $Z = M - A$ for a local martingale $M$ and a predictable, nondecreasing process $A$. Since $Z=0$ on $\llbracket 0, \tau \rrbracket$, we also have $A=M=0$ on $\llbracket 0, \tau \rrbracket$.
		Observe that we can write
		\begin{equation}
		X^{\tau}Z 
		= X^{\tau}M - X^{\tau}A 
		= \int (X(\tau) \mathbbm{1}_{\rrbracket \tau, \infty \llbracket}) \, dM - X(\tau)A.
		\end{equation}
		Since $X(\tau) \mathbbm{1}_{\rrbracket \tau, \infty \llbracket}$ is locally bounded and predictable, $\int (X(\tau) \mathbbm{1}_{\rrbracket T, \infty \llbracket}) \, dM$ is a local martingale and thus $X^{\tau}Z$ is a local supermartingale.
	\end{proof}
	
	In addition to Lemma~\ref{lem : newlocsup}, we shall need the following easy extension of Lemma~2.12 of \cite{Strong2}.
	
	\smallskip
	
	\begin{lem}     \label{lem : partition martingale}
		Suppose that $\eta$ is a stopping time, $(C_j)_{j \in \N}$ is an $\mathcal{F}_{\eta}$-measurable partition of $\Omega$, and $Y$ is an $\R$-valued semimartingale equal to $0$ on the interval $\llbracket 0, \eta \rrbracket$.
		\begin{enumerate} [(i)]
			\item If $Y\mathbbm{1}_{C_j}$ is a local martingale for all $j \in \N$, then $Y$ is a local martingale.
			\item If $Y$ is nonnegative and $Y\mathbbm{1}_{C_j}$ is a supermartingale for all $j \in \N$, then $Y$ is a supermartingale.
		\end{enumerate}
	\end{lem}
	
	\begin{proof}
		The assertion (i) is just Lemma~2.12 (2) of \cite{Strong2}. For (ii), Fatou's lemma yields for $0 \le s \le t$
		\begin{equation*}
		\E[Y_t \vert \mathcal{F}_s] = \E \Big[\sum_{j=1}^{\infty} Y_t\mathbbm{1}_{C_j} \big\vert \mathcal{F}_s \Big] \le \sum_{j=1}^{\infty} \E[ Y_t\mathbbm{1}_{C_j} \vert \mathcal{F}_s] \le \sum_{j=1}^{\infty} Y_s\mathbbm{1}_{C_j} = Y_s.
		\end{equation*}
	\end{proof}
	
	\medskip
	
	\subsection{The optional decomposition theorem} \label{subsec : ODT RCLL}
	
	In this subsection, we state and prove the optional decomposition theorem for a general piecewise semimartingale price process $S$ with RCLL paths. This will be the cornerstone for proving the following results, including the existence of an investment strategy with the supermartingale num\'eraire property and the superhedging duality.
	
	Let us mention that the proof of the optional decomposition theorem in Section~\ref{subsec : ODT} was more direct and illustrative leading to a rather explicit representation of the investment strategy~(see the equation \eqref{def : vartheta ODT}) in the optional decomposition. However, as we mentioned in the introduction, the same proof technique does not extend to the case when jumps are present in the dynamics of asset prices~(see \cite{KK:OD}). In contrast, the proof that we present in this subsection relies on the classical optional decomposition theorem for RCLL semimartingales of fixed dimension from \cite{Stricker:Yan}.
	
	\smallskip
	
	\begin{thm}	\label{thm : general optional decomposition}
		Suppose that $\mathcal{Y} \neq \emptyset$, i.e., a local martingale deflator exists. For a nonnegative, adapted process $X$ with $X(0)= x \ge 0$, the following statements are equivalent:
		\begin{enumerate} [(i)]
			\item For every local martingale deflator $Y \in \mathcal{Y}$, the process $YX$ is a supermartingale.
			\item There exist an investment strategy $\vartheta \in \mathcal{L}_0(S)$ and an adapted, nondecreasing process $K$ satisfying $K(0)=0$ with right-continuous paths, such that
			\begin{equation}    \label{eq : general optional decomposition}	
			X = x + \vartheta \cdot S - K = x + \sum_{k=1}^{\infty} \sum_{n=1}^{\infty} \int_0^{\cdot} \sum_{i=1}^n \vartheta^{k, n}_i(t) \, dS^{k, n}_i(t) - K.
			\end{equation}
		\end{enumerate}
	\end{thm}
	
	\begin{proof}
		For the implication, $(ii) \Longrightarrow (i)$, the integration by parts gives $YK = K_- \cdot Y + Y \cdot K$ for $Y \in \mathcal{Y}$. Since $K_- \cdot Y$ is a local martingale and $Y \cdot K$ is a nondecreasing process, $YK$ is a local submartingale. Moreover, $Y(x + \vartheta \cdot S)$ is a local martingale, which implies that $YX$ is a nonnegative local supermartingale and hence a true supermartingale.
		
		For the proof of the implication $(i) \Longrightarrow (ii)$, let us denote $X^k := X^{\tau_k} - X^{\tau_{k-1}}$ and $X^{k,n} := \mathbbm{1}_{\Omega^{k,n}}X^k$ to obtain the representation $X = X(0) + \sum_{k=1}^{\infty} \sum_{n=1}^{\infty} X^{k,n}$. Our goal is to show that each $X^{k,n}$ admits the decomposition
		\begin{equation}    \label{eq : X k, n decomp}
		X^{k,n} = \vartheta^{k,n} \cdot S^{k,n} - K^{k,n} 
		\end{equation}
		for some $n$-dimensional $S^{k,n}$-integrable process $\vartheta^{k,n}$ and an adapted nondecreasing process $K^{k,n}$ with $K^{k,n}(0) = 0$.
		
		Thanks to the assumption $\mathcal{Y} \neq \emptyset$ and the observation $\mathcal{Y} \subset \mathcal{Y}^{k, n}$ following Definition~\ref{Def : general deflator}, we choose an arbitrary $(k, n)$-local martingale deflator $Y^{k, n}$ in $\mathcal{Y}^{k, n}$ for every $(k, n) \in \N^2$, to construct the set $\{Y^{k,n}\}_{(k,n) \in \N^2}$. 
		
		We now define
		\begin{equation*}
		Z := \prod_{k = 1}^{\infty} Z^k, \qquad \text{where} \quad
		Z^k := \mathbbm{1}_{\{\tau_{k-1} = \infty\}} + \sum_{n = 1}^{\infty} \mathbbm{1}_{\Omega^{k,n}} \bigg(\frac{(Y^{k,n})^{\tau_k}}{(Y^{k,n})^{\tau_{k-1}}}\bigg),
		\end{equation*}
		then we have $Z \in \mathcal{Y}$ from the proof of Theorem 3.5 of \cite{Strong2}. By the hypothesis $(i)$, $ZX$ is a supermartingale. The Doob-Meyer decomposition yields $ZX = X(0) + M - A$ for a local martingale $M$ and a nondecreasing predictable process $A$ with $M(0) = A(0) = 0$. This implies that $(ZX)^{\tau_k} - (ZX)^{\tau_{k-1}} = M^{\tau_k} - M^{\tau_{k-1}} - (A^{\tau_k} - A^{\tau_{k-1}})$ is a local supermartingale. Now let us write
		\begin{equation*}
		Z^{\tau_k}(X^{\tau_k} - X^{\tau_{k-1}}) 
		= (ZX)^{\tau_k} - (ZX)^{\tau_{k-1}} - (Z^{\tau_k} - Z^{\tau_{k-1}})X^{\tau_{k-1}}.
		\end{equation*}
		Lemma \ref{lem : newlocsup} implies that $(Z^{\tau_k} - Z^{\tau_{k-1}})X^{\tau_{k-1}}$ is a local martingale, so $Z^{\tau_k}(X^{\tau_k} - X^{\tau_{k-1}})$ is a local supermartingale. We derive that
		\begin{equation*}
		\mathbbm{1}_{\Omega^{k,n}} \mathbbm{1}_{\rrbracket \tau_{k-1}, \infty \llbracket} \frac{Z^{\tau_k}} {Z^{\tau_{k-1}}} (X^{\tau_k} - X^{\tau_{k-1}}) = \mathbbm{1}_{\Omega^{k,n}}Z^k (X^{\tau_k} - X^{\tau_{k-1}})
		= \frac{(Y^{k, n})^{\tau_k}}{(Y^{k,n})^{\tau_{k-1}}} X^{k,n},
		\end{equation*}
		and the last expression is a local supermartingale, again on the strength of Lemma~\ref{lem : newlocsup}. Another application of Lemma~\ref{lem : newlocsup} implies that $(Y^{k,n})^{\tau_k} X^{k,n}$ is a local supermartingale. We write $Y^{k,n} X^{k,n} = (Y^{k,n} - (Y^{k,n})^{\tau_k})X^{k,n} + (Y^{k,n})^{\tau_k} X^{k,n}$, and yet another application of Lemma~\ref{lem : newlocsup} with the identity $X^{k, n} \equiv (X^{k, n})^{\tau_k}$ yields that $(Y^{k,n} - (Y^{k,n})^{\tau_k})X^{k,n}$ is a local martingale, thus $Y^{k,n} X^{k,n}$ is a local supermartingale. Keeping in mind that $X^{k,n} + X(\tau_{k-1}) \mathbbm{1}_{\Omega^{k,n}} \mathbbm{1}_{\rrbracket \tau_{k-1}, \infty \llbracket}$ is nonnegative, we consider an $\F(\tau_{k-1})$-measurable partition of $\Omega$ given by $ \cup_{j \in \N} A_j \cup \{\tau_{k-1} = \infty\}$ where $A_j = \{\tau_{k-1} < \infty, ~ j-1 \leq X(\tau_{k-1}) < j\}$. Hence, $\mathbbm{1}_{A_j} \mathbbm{1}_{\rrbracket \tau_{k-1}, \infty \llbracket} Y^{k,n} X^{k,n}$ is an adapted process bounded from below for each $j \in \N$.
		
		On the other hand, the Doob-Meyer decomposition implies that $Y^{k,n} X^{k,n} = N^{k,n} - B^{k,n}$ for a local martingale $N^{k,n}$ and a nondecreasing predictable process $B^{k,n}$, both of which satisfy $N^{k,n} = B^{k,n} = 0$ on $\llbracket 0, \tau_{k-1} \rrbracket$, as $X^{k,n} = 0$ on $\llbracket 0, \tau_{k-1} \rrbracket$. Since $\mathbbm{1}_{A_j} \mathbbm{1}_{\rrbracket \tau_{k-1}, \infty \llbracket}$ is locally bounded and simple predictable, we can write 
		\begin{align*}
		\mathbbm{1}_{A_j} \mathbbm{1}_{\rrbracket \tau_{k-1}, \infty \llbracket} Y^{k,n} X^{k,n} &= \mathbbm{1}_{A_j} \mathbbm{1}_{\rrbracket \tau_{k-1}, \infty \llbracket} N^{k,n} - \mathbbm{1}_{A_j} \mathbbm{1}_{\rrbracket \tau_{k-1}, \infty \llbracket} B^{k,n}
		\\
		& = \int \mathbbm{1}_{A_j} \mathbbm{1}_{\rrbracket \tau_{k-1}, \infty \llbracket} dN^{k,n} - \mathbbm{1}_{A_j} \mathbbm{1}_{\rrbracket \tau_{k-1}, \infty \llbracket} B^{k,n},
		\end{align*}
		where $\int \mathbbm{1}_{A_j} \mathbbm{1}_{\rrbracket \tau_{k-1}, \infty \llbracket} dN^{k,n}$ is a local martingale and $\mathbbm{1}_{A_j} \mathbbm{1}_{\rrbracket \tau_{k-1}, \infty \llbracket} B^{k,n}$ is a predictable nondecreasing process. Consequently, we obtain that $\mathbbm{1}_{A_j} \mathbbm{1}_{\rrbracket \tau_{k-1}, \infty \llbracket} Y^{k,n} X^{k,n} = \mathbbm{1}_{A_j} Y^{k,n} X^{k,n}$ is a local supermartingale bounded from below, hence a true supermartingale for each $j \in \N$.
		
		Let us recall the equivalent formulation \eqref{con : equivalent local martingale deflator} of $\mathcal{Y}^{k, n}$. Since the process $\mathbbm{1}_{A_j} X^{k,n} Y^{k,n}$ is a supermartingale for any $Y^{k,n} \in \mathcal{Y}^{k,n} = \mathcal{Z}(S^{k, n})$, we apply a version of optional decomposition theorem in \cite{Stricker:Yan}, to obtain the decomposition for each $j \in \N$
		\begin{equation}    \label{eq : U k, n}
		\mathbbm{1}_{A_j} X^{k,n} = \vartheta^{k,n,j} \cdot S^{k,n} - K^{k,n,j}
		\end{equation}
		for some $S^{k, n}$-integrable process $\vartheta^{k,n,j}$ and an adapted nondecreasing process $K^{k,n,j}$ satisfying $K^{k,n,j}(0) = 0$. By setting $\vartheta^{k,n} := \sum_{j=1}^{\infty} \mathbbm{1}_{A_j} \vartheta^{k,n,j}$ and $K^{k,n} := \sum_{j=1}^{\infty} \mathbbm{1}_{A_j} K^{k,n,j}$, we obtain the desired decomposition \eqref{eq : X k, n decomp}.
		
		Finally, taking $\vartheta := 0^{(N_0)} + \sum_{k = 1}^{\infty} \sum_{n = 1}^{\infty} \vartheta^{k,n} \hat{\mathbbm{1}}_{\rrbracket \tau_{k-1}, \tau_k \rrbracket \cap (\R_+ \times \Omega^{k,n})}$ and $K := \sum_{k = 1}^{\infty} \sum_{n = 1}^{\infty} K^{k,n}$ yields the result \eqref{eq : general optional decomposition} in view of the definition of stochastic integral in \eqref{def : stochastic integral}.
	\end{proof}
	
	The following result is the local martingale version of the optional decomposition theorem as in Corollary~\ref{cor : optional decomposition}. It can be proved in the same manner as Corollary~\ref{cor : optional decomposition}, except that here we can choose any $Y \in \mathcal{Y}$, instead of the reciprocal $1/X_{\rho}$ of the local martingale num\'eraire there.
	
	\smallskip
	
	\begin{cor}	\label{cor : general optional decomposition}
		Suppose that $\mathcal{Y} \neq \emptyset$. For a nonnegative, adapted process $X$ with $X(0)= x \ge 0$, the following statements are equivalent:
		\begin{enumerate} [(i)]
			\item For every local martingale deflator $Y \in \mathcal{Y}$, the process $YX$ is a local martingale.
			\item There exists an investment strategy $\vartheta \in \mathcal{L}_0(S)$ satisfying $X \equiv X(\cdot; x, \vartheta)$.
		\end{enumerate}
	\end{cor}
	
	\medskip
	
	\subsection{The fundamental theorem} \label{subsec: general fundamental theorem}
	
	As the first application of the optional decomposition theorem, we prove that the existence of a local martingale deflator implies the existence of an investment strategy, in the general piecewise semimartingale market, such that the corresponding wealth process has the supermartingale num\'eraire property. As it was previously mentioned in the introduction, \cite{kabanov2016no} noted that in the case of a finite dimensional RCLL semimartingale market~(of a fixed dimension), the supermartingale num\'eraire portfolio and the local martingale num\'eraire portfolio may not coincide. In the presence of jumps, the local martingale num\'eraire portfolio may not even exist under the stronger assumption of the existence of an equivalent martingale measure~(see \cite{Takaoka}).
	
	Once we have related the existences of a local martingale deflator to that of the supermartingale num\'eraire investment strategy, we provide the fundamental theorem by connecting them to the other concepts such as $NA_1$ and market viability.
	
	In Definition~\ref{Def : general investment strategy}, for any given $X \equiv X(\cdot;x, \vartheta)$ in $\mathcal{W}$, a simple normalization leads to define another admissible investment strategy $\tilde{\vartheta} := \vartheta/x$ satisfying $X(\cdot; 1, \tilde{\vartheta}) = X(\cdot; x, \vartheta)/x$, which also belongs to $\mathcal{W}$. Therefore, in what follows, we shall assume without loss of generality that every wealth process in $\mathcal{W}$ has the initial wealth equal to $x=1$.
	
	\smallskip
	
	\begin{defn} [Investment strategy with the supermartingale num\'eraire property]
		A strictly admissible investment strategy $\vartheta_*$ with the wealth process $X_* \equiv X(\cdot; 1, \vartheta_*) \in \mathcal{X}$ is said to have the \textit{supermartingale num\'eraire property}, if the ratio $X/X_*$ is a supermartingale for every $X \in \mathcal{X}$.
	\end{defn}
	
	\smallskip
	
	\begin{thm} \label{thm : general supermartingale numeraire}
		Suppose that $\mathcal{Y} \neq \emptyset$. Then there exists an admissible investment strategy $\vartheta_*$ having the supermartingale num\'eraire property.
	\end{thm}
	
	\begin{proof}
		We begin the proof by first noting that $\mathcal{Y}^{k,n} \neq \emptyset$ for every $(k,n) \in \N^2$ from the argument below of Definition~\ref{Def : general deflator}. This implies from the result of \cite{KK} that for every $(k, n) \in \N^2$ there exists a wealth process, denoted by $X_*^{k,n}$, which has the supermartingale num\'eraire property for each $n$-dimensional market where the prices process is given by $S^{k,n}$. Namely, for each $(k,n) \in \N^2$, the ratio $X^{k,n}/X_*^{k,n}$ is a supermartingale for every $X^{k,n}$ of the form $X^{k,n} = 1 + \vartheta^{(k, n)} \cdot S^{k,n} > 0$, for some $S^{k, n}$-integrable process $\vartheta^{(k, n)}$.
		
		We construct the process $X_*$ by the following recipe: 
		\begin{equation} \label{def: X-star}
		X_*^k := \mathbbm{1}_{\{\tau_{k-1} = \infty\}} + \sum_{n=1}^{\infty} \mathbbm{1}_{\Omega^{k,n}}(X_*^{k,n})^{\tau_k},
		\qquad X_* := \prod_{k=1}^{\infty} X_*^k.
		\end{equation} 
		Let us first prove that $YX_*$ is a local martingale for every $Y \in \mathcal{Y}$. For that purpose, we shall show that the product $(YX_*)^{\tau_m}$ is a local martingale for each $m \in \N$.
		
		For a fixed arbitrary $Y \in \mathcal{Y}$, we also have $Y \in \mathcal{Y}^{k,n}$, thus $YX_*^{k,n}$ is a local martingale for every $(k,n) \in \N^2$. By denoting $\nu_*^{(k,n)}$ the investment strategy corresponding to the num\'eraire wealth process $X_*^{k, n}$ in each $(k, n)$-dissected market~(of fixed dimension), we can write $X_*^{k,n} = 1 + \nu_*^{(k,n)} \cdot S^{k,n}$. Then, $Y(X_*^{k, n}-1) = Y(\nu_*^{(k, n)} \cdot S^{k, n})$ is also a local martingale for each $(k, n) \in \N^2$. Moreover, we have $X_*^k = 1 + \sum_{n=1}^{\infty} \nu_*^{(k,n)} \cdot S^{k,n}$ and consider an $\mathcal{F}_{\tau_{k-1}}$-measurable partition
		\begin{equation}    \label{eq : k-1 partition of Omega} 
		\bigcup_{n=1}^{\infty}\{\Omega^{k, n}\} \cup \{\tau_{k-1} = \infty\}
		\end{equation}
		of $\Omega$. From Lemma~\ref{lem : partition martingale}, we obtain that $Y(X_*^k-1)$, and consequently $YX_*^k$, are local martingales for every $k \in \N$. Furthermore, $(YX_*^k)^{\tau_k} - Y^{\tau_{k-1}}$ is a local martingale as well, and an application of Lemma~\ref{lem : newlocsup} with the identity $(X_*^k)^{\tau_{k-1}} = 1$ yields the local martingale property of $(YX_*^k)^{\tau_k}/Y^{\tau_{k-1}} - 1$, and hence of $(YX_*^k)^{\tau_k}/Y^{\tau_{k-1}}$.
		
		In what follows, we denote $Y^k := Y^{\tau_k}/Y^{\tau_{k-1}}$ for $k \in \N$. For a fixed $m \in \N$, we have $Y^{\tau_m} = \prod_{i=1}^m Y^i$ and $Y^i X_*^i$ is a local martingale for every $i \in [m]$. Let us take a common localizing sequence $(\sigma_{\ell})_{\ell \in \N}$ for $\{Y^i X_*^i\}_{i \in [m]}$. We now prove that $(YX_*)^{\tau_m \wedge \sigma_{\ell}}$ is a martingale for every $\ell \in \N$. Since we have $Y^i X_*^i(s \vee \tau_{i-1}) = Y^i X_*^i(s)$ for each $i \in \N$, we obtain for any $s < t$
		\begin{align*}
		\E \Big[(YX_*)^{\tau_m \wedge \sigma_{\ell}}(t) \big| \F_s \Big] 
		&= \E\Bigg[ \;  \prod_{i=1}^m (Y^i X_*^i)^{\sigma_{\ell}}(t) \; \bigg| \F_s \Bigg] \\
		& = \E \Bigg[ \; \prod_{i=1}^{m-1} (Y^i X_*^i)^{\sigma_{\ell}}(t) \; \E \Big[ (Y^m X_*^m)^{\sigma_{\ell}}(t) \Big| \F_{s \vee \tau_{m-1}}  \Big] \bigg| \F_s \Bigg] \\
		& = (Y^m X_*^m)^{\sigma_{\ell}}(s) \; \E \Bigg[ \; \prod_{i=1}^{m-1} (Y^i V_*^i)^{\sigma_{\ell}}(t) \; \Bigg| \F_s \Bigg].
		\end{align*}
		By successively conditioning on $\F_{s \vee \tau_{m-2}}, \F_{s \vee \tau_{m-3}}, \cdots , \F_{s \vee \tau_1}$, we derive that the last expression is equal to
		\begin{equation*}
		\prod_{i=1}^m (Y^i X_*^i)^{\sigma_{\ell}} (s) = (YX_*)^{\tau_m \wedge \sigma_{\ell}}(s), 
		\end{equation*}
		proving the martingale property of $(YX_*)^{\tau_m \wedge \sigma_{\ell}}$. Consequently, $(YX_*)^{\tau_m}$ for all $m \in \N$, hence also $YX_*$, are local martingales for any $Y \in \mathcal{Y}$. 
		
		Thanks to the optional decomposition theorem~(Corollary~\ref{cor : general optional decomposition}), we obtain that the existence of an admissible investment strategy $\vartheta_*$ such that $X_* = 1 + \vartheta_* \cdot S$ holds. What is now left to prove is the supermartingale num\'eraire property of the process $X_*$. For a fixed $X \equiv X(\cdot; 1, \vartheta) \in \mathcal{X}$, we shall show by induction that $(X/X_*)^{\tau_m}$ is a supermartingale for every $m \in \N$; then $X/X_*$ is also a supermartingale as every nonnegative local supermartingale is a supermartingale.
		
		Let us recall the representation $X = 1 + \sum_{k=1}^{\infty} \sum_{n=1}^{\infty} \vartheta^{(k,n)} \cdot S^{k,n}$, then the expression
		\begin{equation*}
		\Big(\frac{X^{\tau_1}}{X_*^{\tau_1}} - 1 \Big) \mathbbm{1}_{\Omega^{1, n}} = \frac{1 + \vartheta^{(1,n)} \cdot S^{1,n}}{(X_*^{1,n})^{\tau_1}} 
		- 1 = \frac{1 + \vartheta^{(1,n)} \cdot S^{1,n}}{(X_*^{1,n})} - 1 
		\end{equation*}
		is a supermartingale for every $n \in \N$ from the first paragraph of this proof. Thanks to Lemma~\ref{lem : partition martingale}, $(X/X_*)^{\tau_1} - 1$, and thus $(X/X_*)^{\tau_1}$, are supermartingales.
		
		Assuming $(X/X_*)^{\tau_m}$ is a supermartingale, we shall show that $(X/X_*)^{\tau_{m+1}}$ is a supermartingale. Since $X_*^{\tau_{m+1}} = X_*^{\tau_m} X_*^{m+1}$ and $X^{\tau_{m+1}} = X^{\tau_m} + \sum_{n=1}^{\infty} \vartheta^{(m+1,n)} \cdot S^{m+1,n} > 0$, we have for $0 \le s \le t$
		\begin{align*}
		&\E\bigg[\Big(\frac{X}{X_*}\Big)^{\tau_{m+1}} (t) \Big| \F_s\bigg] 
		= \E \Bigg[ \Big(\frac{X}{X_*}\Big)^{\tau_m}(t) \, \E \bigg[ \frac{1 + \frac{\sum_{n=1}^{\infty} \int_0^t \vartheta^{(m+1,n)}(u) dS^{m+1, n}(u)}{X^{\tau_m}(t)}}  {X_*^{m+1}(t)} \, \bigg| \, \F_{s \vee \tau_m} \bigg] \Bigg| \F_s \Bigg]
		\\
		& \qquad \qquad =\E \Bigg[ \Big(\frac{X}{X_*}\Big)^{\tau_m}(t) \, \E \bigg[ \frac{1 + \frac{\sum_{n=1}^{\infty} \int_0^t \vartheta^{(m+1,n)}(u) dS^{m+1, n}(u)}{X(\tau_m)} \mathbbm{1}_{\rrbracket \tau_m, \infty \llbracket}(t)}  {X_*^{m+1}(t)} \, \bigg| \, \F_{s \vee \tau_m} \bigg] \Bigg| \F_s \Bigg]
		\\
		& \qquad \qquad =\E \Bigg[ \Big(\frac{X}{X_*}\Big)^{\tau_m}(t) \, \E \bigg[ \frac{1 + \sum_{n=1}^{\infty} \int_0^t \frac{\vartheta^{(m+1,n)}(u)}{X(\tau_m)} dS^{m+1, n}(u)}  {X_*^{m+1}(t)} \, \bigg| \, \F_{s \vee \tau_m} \bigg] \Bigg| \F_s \Bigg].
		\end{align*}
		Here, the last two identities use the fact that the integrals $\int_0^t \vartheta^{(m+1, n)}(u)dS^{m+1, n}(u)$ take nonzero values only when $t > \tau_m$. We denote $\zeta^{m+1}(t)$ the expression in the inner expectation, i.e., 
		\begin{equation*}
		\zeta^{m+1} := \frac{1 + \sum_{n=1}^{\infty} \big(\vartheta^{(m+1,n)} / X(\tau_m)\big) \cdot S^{m+1,n}}{X_*^{m+1}} 
		= \frac{X^{\tau_{m+1}}/X^{\tau_m}}{X_*^{m+1}} > 0.
		\end{equation*}
		We note that $\zeta^{m+1} - 1$ is equal to $0$ on $\llbracket 0, \tau_m \rrbracket$ and for every $n \in \N$
		\begin{equation*}
		\mathbbm{1}_{\Omega^{m+1, n}} (\zeta^{m+1} - 1) =  \frac{1 +  \big(\vartheta^{(m+1,n)} / X(\tau_m)\big) \cdot S^{m+1,n}}{X_*^{m+1, n}} - 1
		\end{equation*}
		is a supermartingale from the first paragraph. Lemma \ref{lem : partition martingale} again implies that $\zeta^{m+1} - 1$, hence also $\zeta^{m+1}$, are supermartingales. From the identity $\zeta^{m+1}(s \vee \tau_m) = \zeta^{m+1}(s)$ and the supermartingale property of $(X/X_*)^{\tau_m}$, we obtain for $0 \le s \le t$
		\begin{align*}
		\E\bigg[\Big(\frac{X}{X_*}\Big)^{\tau_{m+1}} (t) \Big| \F_s\bigg] 
		&= \E\bigg[\Big(\frac{X}{X_*}\Big)^{\tau_{m}} (t) \,\E \Big[ \zeta^{m+1}(t) \big| \F_{s \vee \tau_m} \Big] \bigg]
		\le \E\bigg[\Big(\frac{X}{X_*}\Big)^{\tau_{m}} (t) \, \zeta^{m+1}(s) \Big\vert \F_s \bigg]
		\\
		&\le \Big(\frac{X}{X_*}\Big)^{\tau_{m}}(s) \, \zeta^{m+1}(s)
		= \Big(\frac{X}{X_*}\Big)^{\tau_{m+1}}(s).
		\end{align*}
		This proves the supermartingale num\'eraire property of $X_*$ and completes the proof.

	\end{proof}
	
	We are now ready to re-state the fundamental theorem in the general case of piecewise semimartingale market with RCLL paths. As we pointed out in the first paragraph of this subsection, there is no structural condition in this general market, which nicely characterizes the supermartingale num\'eraire strategy in terms of the local rates of the market. Thus, the conditions $(ii)$ and $(iii)$ in Theorem~\ref{thm : fundamental theorem}, which were derived from the structural condition, do not appear in the following theorem. The definitions of market viability and no arbitrage of the first kind remain the same as in Definitions~\ref{def: NA1} and \ref{Def : market viability} in the present context.
	
	\smallskip
	
	\begin{thm} \label{Thm : general fundamental}
		The following statements are equivalent:
		\begin{enumerate} [(i)]
			\item The market is viable.
			\item $NA_1$ holds.
			\item There exists a local martingale deflator.
			\item An investment strategy $\vartheta_*$ having the supermartingale num\'eraire property exists, and its wealth process $X_*(T) = X(T;1, \vartheta_*)$ is finite almost everywhere for any $T \ge 0$.
			\item The collection $\mathcal{X}$ of wealth processes of strictly positive investment strategies is bounded in probability, i.e.,
			\begin{equation*}
			\lim_{m \to \infty} \sup_{X \in \mathcal{X}} \P \big[ X(T) > m \big] = 0 \text{ holds for any } T \ge 0.
			\end{equation*}
		\end{enumerate}
	\end{thm}
	
	\begin{proof}
		As in the proof of Proposition~\ref{prop : NA1}, we refer to Exercise 2.21, Proposition 2.22 of \cite{KK2} and Theorem~3.5 of \cite{Strong2} for the equivalence between $(i), (ii), (iii)$ and $(v)$. For the implication $(iii) \Longrightarrow (iv)$, the existence of $\vartheta_*$ with the supermartingale num\'eraire property follows from Theorem~\ref{thm : general supermartingale numeraire}. Moreover, if $\P \big[ X_*(T) = \infty \big] > 0$ holds for some $T \ge 0$, then an $\mathcal{F}_T$-measurable random variable $h = xX_*(T)$ for every initial capital $x > 0$ is an arbitrage of the first kind for horizon $T$, such that $\varphi := x\vartheta_*$ is an admissible strategy satisfying $X(T;x,\varphi) = xX(T;1, \vartheta_*) = h$. This violates the condition $(ii)$, hence $X_*(T)$ should be finite almost everywhere for any $T \ge 0$.
		
		Assuming $(iv)$, we now prove $(v)$. Let us denote $\vartheta^*$ the strategy having the supermartingale num\'eraire property and $X_* \equiv X(\cdot; 1, \vartheta^*)$ its wealth process. For any $T \ge 0$, $X \in \mathcal{X}$, and $m, \ell \in \N$, Markov's inequality and the supermartingale property of $X/X_*$ yield
		\begin{align*}
		\P \big[ X(T) > m \big] 
		&= \P \bigg[ \frac{X(T)}{X_*(T)} > \frac{m}{X_*(T)} \bigg]
		\le \P \big[ X_*(T) > \ell \big] + \P \bigg[ \frac{X(T)}{X_*(T)} > \frac{m}{X_*(T)}, ~ X_*(T) \le \ell \bigg]
		\\
		& \le \P \big[ X_*(T) > \ell \big] + \P \bigg[ \frac{X(T)}{X_*(T)} > \frac{m}{\ell}\bigg]
		\\
		&\le \P \big[ X_*(T) > \ell \big] + \frac{\ell}{m} \E \bigg[ \frac{X(T)}{X_*(T)}\bigg]
		\le \P \big[ X_*(T) > \ell \big] + \frac{\ell}{m}.
		\end{align*}
		Therefore, by taking the supremum over all $X \in \mathcal{X}$, we obtain for every $m, \ell \in \N$
		\begin{equation*}
		\sup_{X \in \mathcal{X}} \P \big[ X(T) > m \big] \le \P \big[ X_*(T) > \ell \big] + \frac{\ell}{m}.
		\end{equation*}
		For any $\epsilon > 0$, we can choose large enough $\ell \in \N$ such that $\P \big[ X_*(T) > \ell \big] < \epsilon / 2$. Then, we take large enough $m$ satisfying $\ell/m < \epsilon/2$, to deduce $\sup_{X \in \mathcal{X}} \P \big[ X(T) > m \big] < \epsilon$. This establishes $(v)$ and completes the proof.
	\end{proof}
	
	\medskip
	
	\subsection{Superhedging duality and the market completeness}   \label{subsec: superhedging duality}
	
	As another application of the optional decomposition theorem, we first present in this subsection the superhedging duality. We recall the defintions of cumulative withdrawal stream and its superhedging capital from Definition~\ref{Def : market viability}. For the results in this subsection, we shall make the assumption $\F_0 = \{\emptyset, \Omega\}$ throughout.
	
	\smallskip
	
	\begin{thm} \label{thm: general superhedging duality}
		Suppose that $\mathcal{Y} \neq \emptyset$. Let $K$ be a cumulative withdrawal stream and 
		\begin{equation*}
		x(K) := \inf \{x \ge 0 \, \vert \, \exists \, X(\cdot; x, \vartheta) \in \mathcal{W} \text{ such that } X(t;x, \vartheta) \ge K(t) \text{ for all } t \geq 0\},
		\end{equation*}
		the superhedging capital associated with $K$. Then, the following superhedging duality holds:
		\begin{equation*}
		x(K) = \sup_{Y \in \mathcal{Y}} \E\bigg[\int_0^{\infty} Y(s) \, dK(s)\bigg].
		\end{equation*}
	\end{thm} 
	
	\begin{proof}
		Let us denote the right-hand side by $w_K := \sup_{Y \in \mathcal{Y}} \E[\int_0^{\infty} Y(s) dK(s)]$. We first show the inequality $x(K) \le w_K$. If $w_K = \infty$, there is nothing to prove. Thus, we assume $w_K < \infty$ and consider the process
		\begin{equation}    \label{def : minimal financing}
		X(t) := \esssup_{Y \in \mathcal{Y}} \E \bigg[ \int_t^{\infty} \frac{Y(u)}{Y(t)} \, dK(u) \Big| \F(t) \bigg]
		\end{equation}
		with $X(0) = w_K$ (as $\F_0$ is trivial). We shall show that for any local martingale deflator $Z \in \mathcal{Y}$, the process
		\begin{equation*}
		\zeta_Z(t) := Z(t)X(t) + \int_0^t Z(u) \, dK(u), \qquad t \ge 0,
		\end{equation*}
		is a supermartingale. To this purpose, let us define 
		\begin{equation*}
		\mathcal{Y}^Z_t := \{Y \in \mathcal{Y} \, \vert \, Y(u) = Z(u) \text{ for all } u \in [0, t]\}
		\end{equation*}
		and denote
		\begin{equation*}
		\widetilde{\zeta}_Z(t) := \esssup_{Y \in \mathcal{Y}^Z_t} \E\bigg[ \int_0^{\infty} Y(u) \, dK(u) \Big| \F(t) \bigg].
		\end{equation*}
		We now argue $\zeta_Z(\cdot) \equiv \widetilde{\zeta}_Z(\cdot)$. For a fixed $t \ge 0$, we clearly have $\widetilde{\zeta}_Z(t) \le \zeta_Z(t)$. For the reverse inequality, for any $Y \in \mathcal{Y}$, we construct another local martingale deflator $Y_Z \in \mathcal{Y}^Z_t$ by concatenation:
		\begin{align*}
		Y_Z(u) := \begin{cases}
		~ Z(u) &\text{for } 0 \le u \le t, \\
		\frac{Z(t)}{Y(t)} Y(u) &\text{for } u > t.
		\end{cases}
		\end{align*}
		The process $Y_Z$ is a local martingale deflator (see Excercise~2.26 of \cite{KK2}) with the property that $Y_Z(u)/Y_Z(t) = Y(u)/Y(t)$ holds for every $u \ge t$. Denoting the collection of all such processes $Y_Z$ by $\mathcal{Z}^Z_t$, then $\mathcal{Z}^Z_t \subset \mathcal{Y}^Z_t$, and we can write
		\begin{equation*}
		X(t) = \esssup_{Y_Z \in \mathcal{Z}^Z_t} \E\bigg[ \int_t^{\infty} \frac{Y_Z(u)}{Y_Z(t)} \, dK(u) \Big| \F(t)\bigg].
		\end{equation*}
		This shows the reverse inequality $\widetilde{\zeta}_Z(t) \ge \zeta_Z(t)$ for an arbitrary $t \ge 0$, hence $\zeta_Z(\cdot) \equiv \widetilde{\zeta}_Z(\cdot)$.
		
		For any two elements $Y_1, Y_2 \in \mathcal{Y}^Z_t$, we construct another element $Y_3$ in $\mathcal{Y}_t^Z$, again from Exercise~2.26 of \cite{KK2}, as follows:
		\begin{align*}
		Y_3(u) = \begin{cases}
		Z(u) &\text{ for } 0 \leq u \leq t, \\
		Y_1(u) &\text{ for } u > t, ~ \text{on the set } \{\E[\int_0^{\infty} Y_1(s) dK(s) | \F(t)] \geq \E[\int_0^{\infty} Y_2(s) dK(s) | \F(t)]\}, \\
		Y_2(u) &\text{ for } u > t, ~ \text{on the set } \{\E[\int_0^{\infty} Y_1(s) dK(s) | \F(t)] < \E[\int_0^{\infty} Y_2(s) dK(s) | \F(t)]\}.
		\end{cases}
		\end{align*}
		Then, we have $\E[\int_0^{\infty} Y_3(s) dK(s) | \F(t)] \geq \E[\int_0^{\infty} Y_i(s) dK(s) | \F(t)]$ for $i = 1,2$, which implies that the collection $\{ \E[\int_0^{\infty}Y(s) \, dK(s) | \F(t)] : Y \in \mathcal{Y}^Z_t\}$ has the so-called ``directed upwards'' property (see Theorem~A.32 of \cite{FoellmerSchied}), and we can find a sequence $(Y_m)_{m \in \N} \subset \mathcal{Y}^Z_t$ such that the sequence $(\E[\int_0^{\infty}Y_m(s) \, dK(s) | \F(t)])_{m \in \N}$ is increasing and 
		\begin{equation*}
			\zeta_Z(t) = \lim_{m \to \infty} \E[ \int_0^{\infty} Y_m(s) \, dK(s) | \F(t) ].
		\end{equation*}		
		By the Monotone convergence theorem, we have
		\begin{equation*}
		\E \big[\zeta_Z(t) | \F(s)\big] = \lim_{m \rightarrow \infty} \E\bigg[\int_0^{\infty} Y_m(u) \, dK(u) \Big| \F(s)\bigg].
		\end{equation*}
		Now for $s \leq t$, due to the inclusion $\mathcal{Y}^Z_t \subset \mathcal{Y}^Z_s$, we have
		\begin{align*}
		\zeta_Z(s) = \widetilde{\zeta}_Z(s) 
		&= \esssup_{Y \in \mathcal{Y}^Z_s} \E\bigg[ \int_0^{\infty} Y(u) \, dK(u) \Big| \F(s) \bigg]
		\\
		&\ge \esssup_{Y \in \mathcal{Y}^Z_t} \E\bigg[ \int_0^{\infty} Y(u) \, dK(u) \Big| \F(s) \bigg]
		\ge \E \big[\zeta_Z(t) | \F(s)\big],
		\end{align*}
		proving the supermartingale property of $\zeta_Z$.
		
		From integration by parts, we have
		\begin{equation*}
		Z(X + K) = \zeta_Z + \int_0^{\cdot} K(u-) \, dZ(u).
		\end{equation*}
		The process $Z(X+K)$ is then a local supermartingale, hence a supermartingale, as it is nonnegative, for all $Z \in \mathcal{Y}$. Thanks to Theorem~\ref{thm : general optional decomposition}, there exist an admissible strategy $\vartheta \in \mathcal{L}_0(S)$ and a cumulative withdrawal stream $F$ such that $X + K = w_K + \vartheta \cdot S - F$ holds. This proves $x(K) \leq w_K$. 
		
		For the reverse inequality, consider a wealth process $X \in \mathcal{W}$ satisfying $X(\cdot) \ge K(\cdot)$. Since $YX$ is a local martingale for every $Y \in \mathcal{Y}$, the process
		\begin{equation*}
		YX - \int_0^{\cdot} K(u-) \, dY(u) = Y(X - K) + \int_0^{\cdot} Y(u) \, dK(u)
		\end{equation*}
		is a nonnegative local martingale, hence a supermartingale. Therefore, we have the inequality
		\begin{equation*}
		X(0) \ge \E \bigg[ \int_0^{\infty} Y(u) \, dK(u) \bigg].
		\end{equation*}
		Taking supremum over all $Y \in \mathcal{Y}$ and infimum over all such initial capitals $X(0)$ financing $K$, yields the reverse inequality $x(K) \geq w_K$.
	\end{proof}
	
	The process $X+K$ in the proof of Theorem~\ref{thm: general superhedging duality} where $X$ is of \eqref{def : minimal financing}, which finances the given cumulative withdrawal stream $K$, is actually the \textit{minimal} one, as we present in the following result.
	
	\smallskip
	
	\begin{thm}
		Suppose that $\mathcal{Y} \neq \emptyset$. Let $K$ be a cumulative withdrawal stream and define
		\begin{equation*}
		\widetilde{X}(t) := K(t) + \esssup_{Y \in \mathcal{Y}} \E\bigg[ \int_t^{\infty} \frac{Y(u)}{Y(t)} \, dK(u) \Big| \F(t)\bigg], \qquad t \ge 0.
		\end{equation*}
		Then, $\widetilde{X}$ minimally finances $K$; for any adapted nonnegative process $X$ such that $YX$ is a supermartingale for every $Y \in \mathcal{Y}$ and $X(\cdot) \ge K(\cdot)$, we have $X(\cdot) \ge \widetilde{X}(\cdot)$. 
	\end{thm}
	
	\begin{proof}
		Since $YX$ is a supermartingale for every $Y \in \mathcal{Y}$, the process
		\begin{equation*}
		YX - \int_0^{\cdot} K(u-) \, dY(u) = Y(X - K) + \int_0^{\cdot} Y(u) \, dK(u)
		\end{equation*}
		is a nonnegative local supermartingale, thus a supermartingale. Hence, we obtain for $t \ge 0$
		\begin{equation*}
		Y(t) \big(X(t) - K(t)\big) + \int_0^t Y(u) \, dK(u) \ge \E\bigg[\int_0^{\infty} Y(u) \, dK(u) \Big| \mathcal{F}(t) \bigg].
		\end{equation*}
		Rearraging the inequality yields for any $Y \in \mathcal{Y}$
		\begin{equation*}
		X(t) \ge K(t) + \E \bigg[ \int_t^{\infty} \frac{Y(u)}{Y(t)} \, dK(u) \Big| \F(t) \bigg],
		\end{equation*}
		which proves the result.    
	\end{proof}
	
	As a corollary to Theorem \ref{thm: general superhedging duality}, we derive the following superhedging duality for European contingent claims. 
	
	\smallskip
	
	\begin{defn} [European contingent claim]
		A pair $(T, \xi_T)$ is called a \textit{European contingent claim}, if $T$ is a finite stopping time and $\xi_{T}$ is a nonnegative $\F_T$-measurable random variable. Here, $T$ is the maturity of the claim, and $\xi_T$ is the payoff at the maturity.
	\end{defn}

    \begin{example}
        In addition to the classical examples of European contingent claims, whose payoffs depend on individual asset price at $T$, such as call and put ($\xi_T = (S_i(T)-K)^+$ and $(K-S_i(T))^+$, respectively, for some $K > 0$), we can consider a new type of European contingent claim depending on the number of assets in the market at time $T$, e.g., $\xi_T = \mathbbm{1}_{\{N(T) > M\}}$ for some $M \in \mathbb{N}$.
    \end{example}
	
	Let us remark that any European contingent claim $(T, \xi_T)$ can be expressed as a cumulative withdrawal stream $K^{(T, \xi_T)} = \xi_T \mathbbm{1}_{\llbracket T, \infty \llbracket}$ and the following result immediately follows.
	
	\smallskip
	
	\begin{cor}
		Suppose that $\mathcal{Y} \neq \emptyset$. Let $(T, \xi_T)$ be a European contingent claim and denote
		\begin{equation*}
		x^{(T, \xi_T)} := \inf \{x > 0 : \exists \; X(\cdot; x, \vartheta) \in \mathcal{W} \text{ such that } X(T) \geq \xi_T \}
		\end{equation*}
		the smallest initial capital starting from which the claim $(T, \xi_T)$ can be financed. Then, the following superhedging duality holds:
		\begin{align} \label{eq: Eu claim superhedging duality} 
		x^{(T, \xi_T)} = \sup_{Y \in \mathcal{Y}} \E \big[ Y(T) \, \xi_T \big].
		\end{align}
		If the quantity of \eqref{eq: Eu claim superhedging duality} is finite, the minimal hedging process for the European contingent claim $(T, \xi_T)$ is given by
		\begin{equation} \label{eq: Eu claim minimal hedge}
		X^{(T, \xi_T)}(t) = \esssup_{Y \in \mathcal{Y}} \E\Bigg[ \frac{Y(T) \xi_T}{Y(t)} \Bigg| \F(t) \Bigg] = x^{(T, \xi_T)} + \vartheta \cdot S - K
		\end{equation}
		for some $\vartheta \in \mathcal{L}_0(S)$ and a cumulative withdrawal stream $K$.
	\end{cor}
	
	In what follows we discuss the notions of replicability of European claims and market completeness in the present context.
	
	\smallskip
	
	\begin{defn} [Replicability] \label{def: replicability}
		Suppose that $\mathcal{Y} \neq \emptyset$. A European contingent claim $(T, \xi_T)$ such that the superhedging capital of \eqref{eq: Eu claim superhedging duality} is finite, is said to be \textit{replicable}, if the minimal hedging process $X^{(T, \xi_T)}$ of \eqref{eq: Eu claim minimal hedge} does not have any capital withdrawals, i.e., $K \equiv 0$.
	\end{defn}
	
	\smallskip
	
	\begin{defn} [Market completeness]
		Suppose that $\mathcal{Y} \neq \emptyset$. The market is said to be complete, if every \textit{European contingent claim} $(T, \xi_T)$ with the finite superhedging capital of \eqref{eq: Eu claim superhedging duality} is replicable.
	\end{defn}
	
	In view of the optional decomposition theorem~(Corollary~\ref{cor : general optional decomposition}), a European contingent claim $(T, \xi_T)$ is replicable if and only if $YX^{(T, \xi_T)}$ is a local martingale for every $Y \in \mathcal{Y}$; the market is complete if and only if $YX^{(T, \xi_T)}$ is a local martingale for every $Y \in \mathcal{Y}$ and for a minimal hedging process $X^{(T, \xi_T)}$ of every European contingent claim with a finite superhedging capital.
	
	In the following result, known as the second fundamental theorem of asset pricing, we prove that the completeness of the market is equivalent to the uniqueness of the local martingale deflator in the market.
	
	\smallskip
	
	\begin{thm}    \label{thm : market completeness}
		Suppose that $\mathcal{Y} \neq \emptyset$. The market is complete if and only if $\mathcal{Y}$ is a singleton. 
	\end{thm}
	
	\begin{proof}
		Let us assume that $\mathcal{Y} = \{Y\}$ is a singleton. For any European contingent claim $(T, \xi_T)$ such that the quantity of \eqref{eq: Eu claim superhedging duality} is finite, we have 
		\begin{equation*}
		X^{(T, \xi_T)}(t) =  \E\bigg[ \frac{Y(T) \xi_T}{Y(t)} \Big| \F(t) \bigg]. 
		\end{equation*}
		Then $YX^{(T, \xi_T)}$ is a uniformly integrable martingale, thus a direct application of the optional decomposition theorem~(Corollary~\ref{cor : general optional decomposition}) yields the replicability of $(T, \xi_T)$ and the completeness of the market.
		
		For the reverse implication, we assume that the market is complete. Let $Y, Y' \in \mathcal{Y}$ and take a common localizing sequence $(\sigma_n)_{n \in \N}$ such that $Y^{\sigma_n}$ and $(Y')^{\sigma_n}$ are uniformly integrable martingales for $n \in \N$. We define two probability measures $Q_n$ and $Q_n'$ on $\F_{\sigma_n}$ by setting
		\begin{equation}
		\frac{dQ_n}{d\P} = Y(\sigma_n) \qquad \text{and} \qquad \frac{dQ_n'}{d\P} = Y'(\sigma_n) \qquad \text{for each } n \in \N.
		\end{equation}
		For any $A \in \F_{\sigma_n}$, a pair $(\sigma_n, \mathbbm{1}_A)$ is a European claim such that its superhedging capital of \eqref{eq: Eu claim superhedging duality} is finite. Since $YX^{(\sigma_n, \mathbbm{1}_A)}$ and $Y'X^{(\sigma_n, \mathbbm{1}_A)}$ are local martingales, $(X^{(\sigma_n, \mathbbm{1}_A)})^{\sigma_n}$ is a local martingale under $Q_n$ and $Q_n'$ (see Lemma 11, Chapter 2 of \cite{jarrow2018continuous}). Moreover, since $X^{(\sigma_n, \mathbbm{1}_A)} \leq 1$, $(X^{(\sigma_n, \mathbbm{1}_A)})^{\sigma_n}$ is a martingale under $Q_n$ and $Q_n'$, hence we have $x^{(\sigma_n, \mathbbm{1}_A)} = \E^{Q_n}[X^{(\sigma_n, \mathbbm{1}_A )}(\sigma_n)]  =  \E^{Q_n'}[X^{(\sigma_n, \mathbbm{1}_A)}(\sigma_n)] = Q_n(A) = Q_n'(A)$ for all $A \in \F_{\sigma_n}$. This implies that $Y^{\sigma_n} = (Y')^{\sigma_n}$ for all $n \in \N$, thus $Y \equiv Y'$. This completes the proof.
	\end{proof}

    Recalling Definition \ref{Def : general deflator} and Theorem~\ref{thm : market completeness}, the completeness of $(k, n)$-dissected market is equivalent to the uniqueness of $(k, n)$-local martingale deflator on the $(k, n)$-dissection set, i.e., $Y_1 = Y_2$ on $\rrbracket \tau_{k-1}, \tau_k \rrbracket \cap (\R_+ \times \Omega^{k,n})$ for any $Y_1, Y_2 \in \mathcal{Y}^{k,n}$. Moreover, for given $(k, n)$-local martingale deflators $Y^{k, n} \in \mathcal{Y}^{k, n}$ for all pairs $(k, n) \in \mathbb{N}^2$, we can construct a local martingale deflator $Z \in \mathcal{Y}$ via recipe in the proof of Theorem 3.5 of \cite{Strong2}:
    \begin{equation}    \label{def : entire local martingale deflator}
        Z := \prod_{k = 1}^{\infty} Z^k, \qquad \text{where} \quad
        Z^k := \mathbbm{1}_{\{\tau_{k-1} = \infty\}} + \sum_{n = 1}^{\infty} \mathbbm{1}_{\Omega^{k,n}} \bigg(\frac{(Y^{k,n})^{\tau_k}}{(Y^{k,n})^{\tau_{k-1}}}\bigg).
    \end{equation}
    We now have the following corollary stating the completeness of the entire market in terms of the completeness of every dissected market.
    
    \begin{cor}
        Suppose that $\mathcal{Y} \neq \emptyset$. The entire market is complete if and only if each $(k, n)$-dissected market is complete for every $(k, n) \in \mathbb{N}^2$.
    \end{cor}

    \begin{proof}
        Suppose that the entire market is complete but there exists a pair $(k',n')$ such that $(k', n')$-dissected market is not complete. This implies that there exist $Y_1, Y_2 \in \mathcal{Y}^{k',n'}$ such that $Y_1 \neq Y_2$ on the $(k', n')$-dissection set. From the completeness of the entire market, we have $\mathcal{Y} = \{ Y \}$ and $Y \in \mathcal{Y}^{k, n}$ for every $(k, n) \in \mathbb{N}^2$. Thus, using the recipe \eqref{def : entire local martingale deflator}, we can construct two different local martingale deflators $Z_1, Z_2 \in \mathcal{Y}$, where we set $Y^{k, n} = Y$ for all other pairs $(k, n) \neq (k', n')$ but we choose $Y_1$ and $Y_2$, respectively, for $Y^{k', n'}$. This contradicts the completeness of the entire market.
           
        For the other implication, assume that there exist $Y_1, Y_2 \in \mathcal{Y}$ such that $Y_1 \neq Y_2$. Then, there exists a $(k,n)$-dissection set such that $Y_1 \neq Y_2$ on that set. Since $\mathcal{Y} \subset \mathcal{Y}^{k,n}$ for all $(k,n) \in \N^2$, we have $Y_1, Y_2 \in \mathcal{Y}^{k,n}$, which contradicts the completeness of $(k, n)$-dissected market.
    \end{proof}

	
	
	
	\bigskip
	
	\subsection*{Acknowledgments}
	We are deeply grateful to Ioannis Karatzas for detailed reading and suggestions which improved this paper.
	
	\bigskip
	
	\subsection*{Funding}
	E. Bayraktar is supported in part by the National Science Foundation under grant DMS-2106556 and by the Susan M. Smith Professorship.

	\bigskip
	
	\bigskip
	
	\renewcommand{\bibname}{References}
	\bibliography{aa_bib}
\end{document}